\documentclass{jfm}
\usepackage{graphicx}
\usepackage{float}
\usepackage{epstopdf, epsfig}
\usepackage{subfig}
\usepackage{siunitx}
\usepackage{color}
\usepackage{lineno,hyperref}
\usepackage{overpic}
\hypersetup{
     colorlinks   = true,
     linkcolor = blue,
     citecolor    = blue,
     urlcolor = blue
}

\shorttitle{Imbibition in dual-permeability pore network}
\shortauthor{Q. Gu, H. Liu and L. Wu}

\title{Preferential imbibition in a dual-permeability pore network}

\author{Qingqing Gu\aff{1},
  Haihu Liu\aff{2}
 \corresp{\email{haihu.liu@mail.xjtu.edu.cn}}
  \and Lei Wu\aff{1}
  \corresp{\email{wul@sustech.edu.cn}}
 }

\affiliation{\aff{1}Department of Mechanics and Aerospace Engineering, Southern University of Science and Technology, Shenzhen 518055, China
\aff{2}School of Energy and Power Engineering, Xi'an Jiaotong University, 28 West Xianning Road, Xi'an 710049, China}
\makeatletter 
\newcommand*{\rom}[1]{\expandafter\@slowromancap\romannumeral #1@} 
\makeatother 
\begin{document}
\maketitle

\begin{abstract}
A deep understanding of two-phase displacement in  porous media with permeability contrast is essential for the design and optimisation of enhanced oil recovery processes. In this paper, we investigate the forced imbibition behaviour in two dual-permeability geometries that are of equal permeability contrast. First, a mathematical model is developed for the imbibition in a pore doublet, which shows that the imbibition dynamics can be fully described by the viscosity ratio $\lambda$ and capillary number $Ca_m$ which creatively incorporates the influence of channel width and length. Through the finite difference solution of the mathematical model, a $\lambda-Ca_m$ phase diagram is established to characterise the imbibition preference in the pore doublet. We then investigate the imbibition process in a dual-permeability pore network using a well-established lattice Boltzmann method, focusing on the competition between the viscous and capillary forces. Like in the pore doublet, the preferential imbibition occurs in high permeability zone at high $Ca_{m}$ but in low permeability zone at low $Ca_{m}$. When $Ca_m$ is not sufficiently high, an oblique advancing pattern is observed which is attributed to non-trivial interfacial tension. Thanks to the newly defined capillary number, the critical $Ca_{m}$ curve on which the breakthrough simultaneously occurs in both permeability zones,  is found to match perfectly with that from the pore doublet and it is the optimal condition for maximising the imbibition efficiency in the entire pore network.
\end{abstract}

\section{Introduction}
Immiscible two-phase displacement in permeable media has drawn extensive research attention due to its importance in secondary and tertiary oil recovery processes~\citep{lakeEnhancedOilRecovery1989}. However, many petroleum-bearing underground geological formations exist in the form of layers, which poses great technical challenges for the economical recovery of original oil due to early breakthrough~\citep{shengEnhancedOilRecovery2013,bahadoriFundamentalsEnhancedOil2018}.
Injecting gas (e.g. carbon dioxide) or liquid (e.g. water) into the subsurface system with permeability variations often leads to the preference of the injected flow into one of the layers, and whether high or low permeability depends on the fluid properties such as viscosity, interfacial tension, density, buoyancy and solubility, the properties of porous media such as surface wettability, porosity and permeability, and the operational conditions such as the injection rate. In order to optimise the gas or liquid flooding operations and thus improve the oil recovery, it is crucial to understand the fundamentals of two-phase displacement in porous media with permeability contrast.

Extensive works have been devoted to understanding the two-phase displacement mechanisms from experimental~\citep{lenormandNumericalModelsExperiments1988,zhangInfluenceViscousCapillary2011,
zhaoWettabilityControlMultiphase2016,huExperimentalStudyDisplacement2020}, theoretical~\citep{chatzisDynamicImmiscibleDisplacement1983,
laidlawTheoreticalExperimentalInvestigation1983,sorbieExtendedWashburnEquation1995,
al-housseinyControlInterfacialInstabilities2012,al-housseinyPreferentialFlowPenetration2014,
zhengControllingViscousFingering2015,zhengViscousFluidInjection2015} and numerical~\citep{liuPoreScaleSimulationsGas2013,sunMicromodelExperimentsPore2016,
chenInertialEffectsProcess2019} perspectives. \citet{lenormandNumericalModelsExperiments1988} experimentally studied a non-wetting fluid displacing a wetting fluid (i.e. drainage) in a micromodel and found that the competition between capillary and viscous forces creates the instability of the advancing front, leading to three different displacement regimes, namely viscous fingering, capillary fingering and stable displacement, which are mapped on a phase diagram of viscosity ratio versus capillary number. Later, the phase diagram was improved by~\citet{zhangInfluenceViscousCapillary2011} and was extended to three-dimensional porous media by~\citet{huExperimentalStudyDisplacement2020} with the aid of fast development in precise microfabrication, fluid saturation visualization and image analysis. Unlike the single permeability system, there are only a few experimental studies concerning the multiphase flows in porous media with permeability contrast. For instance,~\citet{zhangLiquidCO2Displacement2011} studied the drainage process in a dual-permeability pore network, demonstrating the influence of injection rate on displacement mechanisms.~\cite{maVisualizationImprovedSweep2012} demonstrated the use of foam to realize the flow diversion from high permeable to low permeable regions in a dual-permeability micromodel with aligned solid posts.~\citet{nijjerStableUnstableMiscible2019} investigated the effect of permeability contrast and viscosity variations on miscible displacement in layered porous media.

Theoretical study of the two-phase displacement with variable permeabilities is limited to a pore doublet model (PDM)~\citep{mooreEffectViscosityCapillarity1956}, which is a simple network with two connected capillaries.~\citet{chatzisDynamicImmiscibleDisplacement1983} derived the explicit formulation of velocity in each capillary when the wetting and non-wetting fluids are of the same viscosity, and they provided a semi-quantitative understanding of a relatively long string of pore doublets.~\citet{laidlawTheoreticalExperimentalInvestigation1983} studied the simultaneous arrival of the interfaces at the downstream end of pore doublet under a controlled pressure drop, and concluded that the effectiveness of pressure drop in controlling trapping is dependent on the  scale of the pore doublet system. Nevertheless, their analysis cannot be extended to porous media as the pressure drop between two adjacent nodal pores within the porous media is hardly controllable.~\citet{sorbieExtendedWashburnEquation1995} developed an extended pore doublet model by incorporating an inertial term into the energy balance equation. Recently, \citet{al-housseinyPreferentialFlowPenetration2014} conducted a drainage study in a pore doublet, and discovered the possible existence of preferential flow in two identical daughter channels that vary in size along the flow direction. Inspired by their quantitative description of the meniscus movement under given flow rate, we will carry out theoretical analysis of the pore doublet consisting of two unequal-sized branch channels and focus on the forced imbibition with an injection velocity.

As a complement to theoretical and experimental studies, numerical simulations have developed into a useful tool to provide insights into the two-phase flow phenomena that occurs during immiscible displacement. Among them, pore-scale simulations are becoming increasingly popular with the advent of advanced algorithms and parallel computing. Simulations at the pore scale are of great importance since (1) pore-scale phenomenon such as trapping has a significant impact on the larger scale~\citep{juanesImpactRelativePermeability2006,cinarExperimentalStudyCO22009,
 soulainePorescaleModellingMultiphase2018}; (2) they are able to capture heterogeneity, interconnectivity and non-uniform flow behaviour (e.g. various fingerings) and provide local information on fluid distribution and velocity for the construction of constitutive equations at macroscopic scales~\citep{liuMultiphaseLatticeBoltzmann2015}. Several approaches have been developed to simulate multiphase flows at pore scale, which mainly include pore-network models, lattice Boltzmann method (LBM) and the conventional computational fluid dynamics (CFD) methods such as the volume-of-fluid (VOF) method~\citep{raeiniDirectSimulationsTwophase2014,
yinDirectSimulationsTwophase2018}, level-set (LS) method~\citep{prodanovicLevelSetMethod2006}, and the phase-field (PF) method~\citep{badalassiComputationMultiphaseSystems2003,akhlaghiamiriEvaluationLevelSet2013}. Pore-network models~\citep{joekar-niasarNonequilibriumEffectsCapillarity2010,
kibbeyPoreNetworkModel2012,fagbemiCouplingPoreNetwork2020} simulate fluid flow through an idealized network of pores connected by throats. Although this approach is well-tailored for studying capillary-controlled displacement that provides infinite resolution in network elements, a number of approximations are made concerning the pore space geometry, which may result in loss of geometric and topological information. Conventional CFD methods rely on the evolution of an indicator to track the fluid interface in addition to solving the macroscopic equations for fluid flow.  Due to the lack of versatility of implementing the boundary conditions for arbitrary grain shapes, it remains a challenge to apply these methods for the simulation of two-phase flow in complex porous media~\citep{mukherjeePorescaleModelingTwophase2011}.

We will concentrate on the LBM simulation of multiphase flow, in which the simplified kinetic models are used to capture microscopic or mesoscopic flow physics while the macroscopic averaged quantities satisfy the desired macroscopic equations. Compared to the pore-network models, LBM allows for better representing the pore morphology of the actual porous medium~\citep{rothmanMacroscopicLawsImmiscible1990,panPorescaleModelingSaturated2001,
porterLatticeBoltzmannSimulationsCapillary2009,boekLatticeBoltzmannStudiesFluid2010}. In addition, due to its kinetic nature and local dynamics, LBM has several advantages over the conventional CFD methods, especially in dealing with complex boundaries, incorporation of microscopic interactions, flexible reproduction of the interface between different fluids, and parallelisation of the algorithm. Despite plenty of literatures on the pore scale flow behaviour in single permeability porous system~\citep{ramstadRelativePermeabilityCalculations2012,chenLatticeBoltzmannSimulations2018,
azizPorescaleInsightsTransport2018, huPhaseDiagramQuasistatic2019, akaiPoreScaleNumericalSimulation2020}, the imbibition dynamics in a dual-permeability porous system is not well understood. In this work, we present a systematic study of the imbibition dynamics in two dual-permeability geometries, which are of equal permeability contrast. We start from the simple pore doublet model, and for the first time use the theoretical predictions along with the LBM validations to quantify the meniscus filling behaviour. In particular, a new capillary number is introduced to characterise the preferential penetration in two unequal-sized branch channels. The validated LBM is then used to simulate the imbibition process in a dual-permeability pore network for varying capillary numbers and viscosity ratios, and the obtained results are compared with those obtained previously from the pore doublet.

\section{Lattice Boltzmann method for immiscible two-phase flow}\label{sec:section-2-num-method}
Direct numerical simulation of the two-phase flow in two-dimensional pore-spaces is performed using a state-of-the-art colour-gradient lattice Boltzmann model~\citep{xuLatticeBoltzmannSimulation2017}. In this model, the distribution functions $f_{i}^{R}$ and $f_{i}^{B}$ are used to represent the red and blue fluids, where the subscript $i$ is the lattice velocity direction and ranges from 0 to 8 for the two-dimensional nine-velocity (D2Q9) lattice model used in this work. $f_{i}(\boldsymbol{x},t)$ is the total distribution function at position $\boldsymbol x$ and time $t$, and is defined as $f_{i}=f_{i}^{R}+f_{i}^{B}$. Conservation of mass for each fluid and total momentum conservation require
\begin{equation}\label{mass-conservation}
\rho^{k}=\sum_{i} f_{i}^{k}, \qquad \rho \boldsymbol{u}=\sum_{i}f_{i}\boldsymbol{c}_{i},\quad k=R \text{~or~}B,
\end{equation}
where $\rho=\rho^{R}+\rho^{B}$ is the total density with the superscripts `R' and `B' referred to as the red and blue fluids respectively, and $\boldsymbol u$ is the local fluid velocity. The lattice velocity $\boldsymbol c_i$ is defined as $\boldsymbol c_0=(0,0)$, $\boldsymbol c_{1,3}=(\pm c,0)$, $\boldsymbol c_{2,4}=(0,\pm c)$, $\boldsymbol c_{5,7}=(\pm c,\pm c)$, and $\boldsymbol c_{6,8}=(\mp c,\pm c)$, where $c=\delta_x/\delta_t$ is the lattice speed with $\delta_x$ being the lattice length and $\delta_t$ being the time step. The sound of speed is related to the lattice speed by $c_s=c/\sqrt{3}$. The evolution of $f_{i}^{R}$ and $f_{i}^{B}$ in time and space is described by
\begin{equation}\label{two-colour-lb-equation}
 f_{i}^{k}(\boldsymbol{x}+\boldsymbol{c}_{i}\delta_t,t+\delta_t)=f_{i}^{k}(\boldsymbol{x},t)+{(\Omega_{i}^{k})}^{(3)}\left[{(\Omega_{i}^{k})}^{(1)}+{(\Omega_{i}^{k})}^{(2)}\right],
\end{equation}
where ${(\Omega_{i}^{k})}^{(1)}$ is the single-phase collision operator, ${(\Omega_{i}^{k})}^{(2)}$ is the perturbation operator, and ${(\Omega_{i}^{k})}^{(3)}$ is the recolouring operator to guarantee the immiscibility of both fluids. Note that the single-phase collision and perturbation operators are to recover the Navier-Stokes equations for the fluid mixture, and thus can be implemented via the total distribution function $f_i$. Using the multiple relaxation time (MRT) scheme~\citep{ginzburgMultireflectionBoundaryConditions2003}, the single-phase collision operator reads as
\begin{equation}
{(\Omega_{i})}^{(1)}=-(\mathsfbi{M}^{-1}\mathsfbi{S}\mathsfbi{M})_{ij}(f_{j}-f_{j}^{eq}),
\label{TRT-collision-term}
\end{equation}
where $f_{i}^{eq}$ is the equilibrium distribution function and is given by
\begin{equation}
f_{i}^{eq}(\rho, \boldsymbol{u})=\rho W_i\left[ 1+\frac{\boldsymbol{c}_i \cdot \boldsymbol{u}}{c_s^2}+\frac{(\boldsymbol{c}_i\cdot \boldsymbol{u})^2}{2c_s^4} -\frac{\boldsymbol{u}^2}{2c_s^2} \right].
\label{maxwell-Boltzmann-equi-distr}
\end{equation}
Herein, $W_i$ is the weight factor with $W_{0}=4/9$,~$W_{1-4}=1/9$ and $W_{5-8}=1/36$. The transformation matrix $\mathsfbi{M}$ is given by~\citep{lallemandTheoryLatticeBoltzmann2000}
\begin{equation}\label{transformation-matrix}
\mathsfbi{M} = \left[
  \begin{array}{ccccccccc}
		1&  1&  1&  1&  1&  1&  1&  1& 1\\
		-4&  -1&  -1&  -1&  -1&  2&  2&  2& 2\\
		4& -2&  -2&  -2&  -2& 1 &  1&  1& 1\\
		0&  1&  0&  -1&  0&  1&  -1&  -1& 1\\
		0& -2&  0&  2&  0&  1&  -1&  -1& 1\\
		0&  0&  1&  0&  -1&  1& 1 & -1 & -1\\
		0&  0&  -2&  0&  2&  1&  1&  -1& -1\\
		0&  1&  -1&  1&  -1&  0&  0&  0& 0\\
		0&  0&  0&  0&  0&  1&  -1&  1& -1
  \end{array}
\right].
\end{equation}

With the transformation matrix $\mathsfbi{M}$, the distribution function $f_i$ can be projected onto the moment space through $m_i = M_{ij}f_j$, and the resulting nine moments are
\begin{equation}\label{f-in-moment-space}
 \mathsfbi{m} = (\rho, e,\varepsilon,j_{x},q_{x},j_{y},q_{y},p_{xx},p_{xy})^{T},
\end{equation}
where $e$ and $\varepsilon$ are related to the total energy and the energy square, $j_{x}$ and $j_{y}$ are the $x$ and $y$ components of the momentum, $q_{x}$ and $q_{y}$ are the components of the energy flux, and $p_{xx}$ and $p_{xy}$ correspond to the diagonal and off-diagonal components of the viscous stress tensor. The values of the equilibrium moment are $ \mathsfbi{m}^{eq}=\rho(1,-2+3\boldsymbol{u}^2,1-3\boldsymbol{u}^2, u_x,-u_x, u^y,-u^y,
u_x^2-u_y^2,u_x u_y)^{T}$, which are obtained by $m_i^{eq} = M_{ij}f_j^{eq} $. The diagonal relaxation matrix $\mathsfbi{S}$ in (\ref{TRT-collision-term}) is given as $\mathsfbi{S}=diag(s_{\rho},s_{e},s_{\varepsilon},s_{j},s_{q},s_{j},s_{q},s_{p},s_{p} )$. $s_{\rho}$ and $s_{j}$ can take any values since they correspond to the conserved moments (density $\rho$ and momentum $\boldsymbol j$). $s_{e}$ and $s_{p}$ are related to the bulk and shear viscosities, while $s_{\varepsilon}$ and $s_{q}$ are the free parameters. To improve the numerical stability, we choose $s_{e}=s_{\varepsilon}=s_{p}=1/\tau $ and $s_{q}=8(2-s_{p})/(8-s_{p})$~\citep{panEvaluationLatticeBoltzmann2006} in our simulations, where the dimensionless relaxation time $\tau$ is related to the dynamic viscosity of the fluid mixture by $\eta=c_{s}^{2}\rho(\tau-0.5)\delta_t$. When both fluids have unequal viscosities, a harmonic mean is employed to determine the viscosity of the fluid mixture, i.e. $1/\eta =(1+\rho^{N})/(2\eta^{R})+(1-\rho^{N})/(2\eta^{B})$, where the phase field $\rho^N$ is defined as
\begin{equation}\label{fai-rhoN}
\rho^{N} (\boldsymbol{x},t)=\frac{\rho^{R}(\boldsymbol{x},t)-\rho^{B}(\boldsymbol{x},t)}{\rho^{R}(\boldsymbol{x},t)+\rho^{B}(\boldsymbol{x},t)}, \quad -1\leq \rho^N \leq 1.
\end{equation}

The perturbation operator that generates an interfacial force $\boldsymbol{F}_{s}$ is given by
\begin{equation}\label{surface-force-perturbation-operator}
(\Omega_{i})^{(2)} =\mathsfbi{M}^{-1}\left(\mathbf{I}-\frac{1}{2}\mathbf{S}\right)\boldsymbol F,
\end{equation}
with
\begin{eqnarray}
\boldsymbol{F}
(\boldsymbol x,t)& = & [0, 6(u_{x}F_{sx}+u_{y}F_{sy}), -6(u_{x}F_{sx}+u_{y}F_{sy}), \nonumber\\
&& F_{sx}, -F_{sx}, F_{sy}, -F_{sy},
 2(u_{x}F_{sx}-u_{y}F_{sy}), u_{x}F_{sy}+u_{y}F_{sx}]^{T},
\label{force-new}
\end{eqnarray}
where $\boldsymbol I$ is the second-order identity tensor, and $F_{sx}$ and $F_{sy}$ are the components of the interfacial force $\boldsymbol{F}_{s}$. The interfacial tension between two fluids is modelled as a spatially varying body force $\boldsymbol F_s$ based on the continuum surface force (CSF) concept~\citep{brackbillContinuumMethodModeling1992}, which is given by
\begin{equation}\label{eq:interfacial-force}
\boldsymbol{F}_{s} =\frac{1}{2}\sigma K \bnabla \rho ^{N},
\end{equation}
where $\sigma$ is the interfacial tension coefficient. $K$ is the local interface curvature related to the unit normal vector $\boldsymbol n$ by
\begin{equation}\label{eq:extension-curvature}
K=n_{x}n_{y}(\frac{\partial}{\partial y}n_{x}+\frac{\partial}{\partial x}n_{y})-n_{y}^2\frac{\partial}{\partial x}n_{x}-n_{x}^2\frac{\partial}{\partial y}n_{y},
\end{equation}
where $n_x$ and $n_y$ are the $x$ and $y$ components of $\boldsymbol n$ defined by $\boldsymbol n =\bnabla\rho^{N}/\left | \bnabla \rho ^{N} \right |$. In the calculations of interface curvature and normal vector, the partial derivatives of a variable $\psi$ are evaluated by
\begin{equation}\label{isotropic-partial-derivative}
\boldsymbol \nabla \psi (\boldsymbol{x},t)=\frac{1}{c_{s}^{2}}\sum_{i}W_{i}\psi(\boldsymbol{x}+\boldsymbol{c}_{i}\delta_{t},t)\boldsymbol{c}_{i}.
\end{equation}

In the presence of interfacial force, the fluid velocity should be redefined as~\citep{guoDiscreteLatticeEffects2002}
\begin{equation}\label{redefine-velocity-interface-region}
\rho \boldsymbol{u}=\sum_{i}f_{i}\boldsymbol{c}_{i}+\frac{1}{2}\boldsymbol{F}_{s}\delta_{t}
\end{equation}
to correctly recover the Navier-Stokes equations. To minimize the mixing and segregate the red and blue fluids, the recolouring operator proposed by~\citet{latva-kokkoDiffusionPropertiesGradientBased2005} is used,
\begin{equation}\label{recoloring-1}
(\Omega_{i}^{R})^{(3)}(f_i^R)=\frac{\rho^{R}}{\rho }f_{i}^{*}+\beta \frac{\rho^{R}\rho^{B}}{\rho}W_i \cos(\varphi_{i})\left | \boldsymbol{c}_{i} \right |,
\end{equation}
\begin{equation}\label{recoloring-2}
(\Omega_{i}^{B})^{(3)}(f_i^B)=\frac{\rho^{B}}{\rho }f_{i}^{*}-\beta \frac{\rho^{R}\rho^{B}}{\rho}W_i\cos(\varphi_{i})\left | \boldsymbol{c}_{i} \right |,
\end{equation}
where $f_{i}^{*}$ represents the total distribution function after the perturbation step. $\beta$ is a segregation parameter ranging from 0 to 1 and set to be 0.7 in order to maintain a narrow interface thickness and keep spurious velocities low~\citep{hallidayLatticeBoltzmannAlgorithm2007}. $\varphi_{i}$ is the angle between  $\boldsymbol{\nabla} \rho ^{N}$ and the lattice velocity $\boldsymbol{c}_{i}$.

On the solid surface, no-slip boundary condition is imposed using the halfway bounce-back scheme~\citep{laddNumericalSimulationsParticulate1994}, and a wetting boundary condition is needed to obtain the desired contact angle $\theta$.  Here,  the wetting boundary condition recently developed by~\citet{xuLatticeBoltzmannSimulation2017} is adopted, and its basic idea is to modify the orientation of the phase field gradient at three-phase contact lines so as to match the desired contact angle. Because of its high accuracy and the ability of dealing with arbitrarily complex geometries, this wetting boundary condition has been used many times in pore-scale two-phase simulations~\citep{guLatticeBoltzmannSimulation2018,guPorescaleStudyCountercurrent2019,
xuPredictionImmiscibleTwophase2018}, and has been recently extended to the three-dimensional case~\citep{akaiWettingBoundaryCondition2018}. For the details of the wetting boundary condition, interested readers are referred to~\citet{xuLatticeBoltzmannSimulation2017}.

\section{Mathematical model for forced imbibition in a pore doublet}
\label{imbibition}
\begin{figure}
   \centerline{\includegraphics[width=0.6\textwidth]{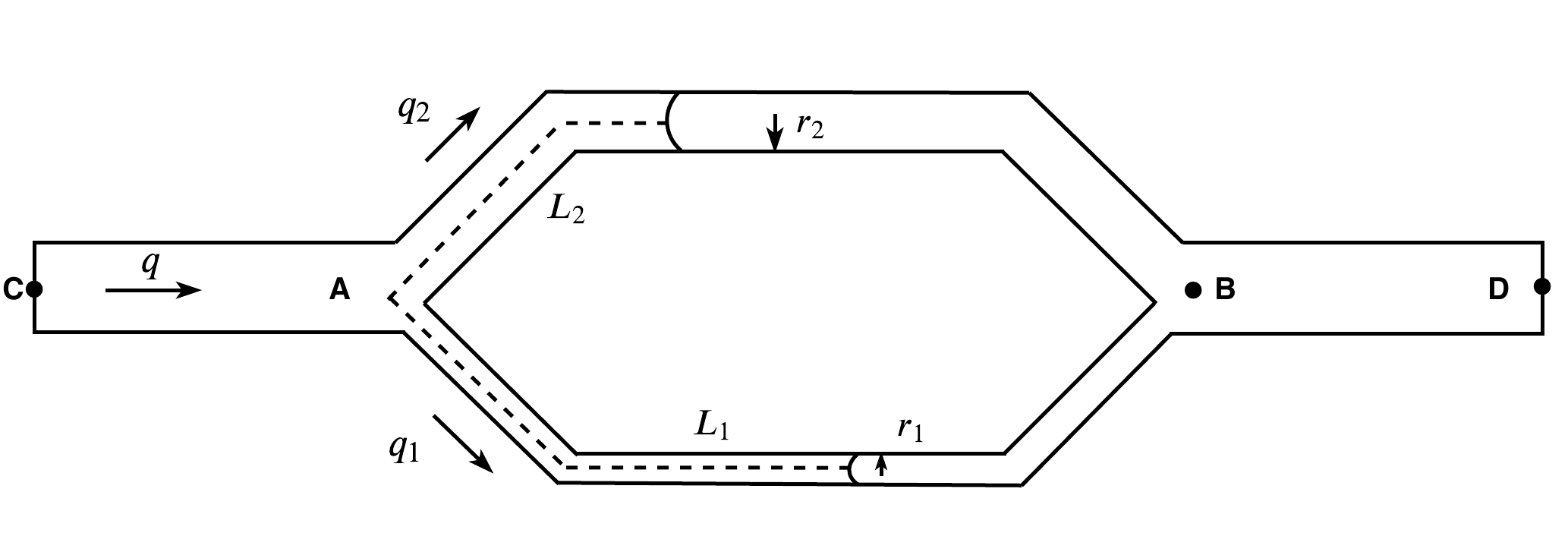}}
	\caption{Schematic diagram of the imbibition process in a pore doublet ($r_2=2r_1$).}
	\label{fig:dualDoublet-geometry}
\end{figure}

In order to understand the mechanism underlying the forced imbibition, we first consider a simple geometry known as the pore doublet model, which is sketched in figure~\ref{fig:dualDoublet-geometry}. The pore doublet consists of three parts: a feeding channel CA that supplies the wetting fluid, two capillary tubes that bifurcate from the point A and reunite downstream at the point B, and an exit channel BD. The branch channel at the bottom (capillary 1) has a narrower width $2r_1$ and the one at the top (capillary 2) has a wider width $2r_2$. The two branches are symmetric with the same length of $L$ along the flow direction, and the angle between the horizontal line and the centreline of each branch channel is $45^{\circ}$. Initially, the entire pore doublet is saturated with the non-wetting fluid. The wetting fluid is injected from the left inlet at a given flow rate $q$, while a constant pressure is assumed at the right outlet. The feeding and exit channels are of equal widths $h=2(r_1+r_2)$, and a constant contact angle of $\theta=30^{\circ}$ (measured from the wetting fluid side) is considered. In the following, we will present a theoretical modelling of the imbibition process based on the aforementioned pore doublet.

\subsection{Governing equations}
\label{governing-equations}
Assuming that the flow through the pore doublet is the steady laminar flow and the two-phase interface advances with a constant mean curvature, the pressure difference between point A and B can be written as:
\begin{equation}\label{pressure-difference-A-B-1}
\Delta p=p_A-p_B=\frac{3q_1}{2r_1^3}\left[\eta_w L_1 +\eta_n(L-L_1)\right]-\frac{\sigma \cos \theta}{r_1},
\end{equation}
\begin{equation}\label{pressure-difference-A-B-2}
\Delta p=p_A-p_B=\frac{3q_2}{2r_2^3}\left[\eta_w L_2 +\eta_n(L-L_2)\right]-\frac{\sigma \cos \theta}{r_2},
\end{equation}
where $p_A$ and $p_B$ are the pressures at the points A and B respectively, $q_1$ and $q_2$ are the volumetric flow rates in the capillaries 1 and 2, and $L_1$ and $L_2$ are the lengths that are occupied by the wetting fluid in the small and large capillaries. In the above equations, the total pressure drop includes the viscous pressure drop $\Delta p_{vis}$ and the capillary pressure drop $\Delta p_{cap}$, which are defined as
\begin{equation}\label{eq:vis-delta-p}
\Delta p_{vis,i}=\frac{3q_i}{2r_i^3}[\eta_w L_i +\eta_n(L-L_i)],\quad i=1,2,
\end{equation}
\begin{equation}\label{eq:capillary-pressure-p}
\Delta p_{cap,i}=-\frac{\sigma \cos \theta}{r_i},\quad i=1,2.
\end{equation}
\subsection{Non-dimensionalisation of governing equations}
\label{non-dimensional-governing-equations}
 In order to nondimensionalise the governing equations, the scaling parameters (denoted by the subscript $s$) for length, time and pressure are introduced,
\begin{equation}\label{eq:p-s-t-s}
l_s= r_1, \quad t_s=\frac{2r_1 L}{q}, \quad p_s=\frac{3\eta_n qL}{2r_1^3},
\end{equation}
where the subscript $n$ is referred to the non-wetting fluid. Substitution of (\ref{eq:p-s-t-s}) into (\ref{eq:vis-delta-p}) leads to
\begin{equation}\label{eq:delta-p-vis-non-4}
{\Delta \hat{p}_{vis,i}}=\left(\frac{\hat{u}_i}{\hat{L}}\right)\left[ \frac{\lambda  \cdot \left(\frac{\hat{L}_i }{\hat{L}}\right)}{\hat{r}_i^2}+\frac{\left(1-\frac{\hat{L}_i}{\hat{L}}\right) }{\hat{r}_i^2}\right ],\quad i=1,2,
\end{equation}
where $\lambda=\eta_w/\eta_n$ is the viscosity ratio of wetting to non-wetting fluid, and the hat over a variable means that the variable is nondimensional. Similarly, (\ref{eq:capillary-pressure-p}) can be written as
\begin{equation}\label{eq:delta-p-cap-non-2}
  {\Delta \hat{p}_{cap,i}}=\frac{\cos \theta}{Ca_m}\cdot \frac{1}{\hat{r}_i},\quad i=1,2,
  \end{equation}
where $Ca_m=3\eta_n qL/2r_1^2\sigma$ is the preferential capillary number. Note that this capillary number is different from the standard one, which takes into account the influence of pore length and size. Combining (\ref{eq:delta-p-vis-non-4}) and (\ref{eq:delta-p-cap-non-2}), one can obtain the total pressure drop as
\begin{equation}\label{eq:delta-p-non}
\Delta \hat{p}=\Delta \hat{p}_{vis,i}+\Delta \hat{p}_{cap,i}=\hat{r}_i\cdot \frac{\mathrm{d}\left(\frac{\hat{L}_i }{\hat{L}}\right) }{\mathrm{d} \hat{t}}\left[ \frac{\lambda  \left(\frac{\hat{L}_i }{\hat{L}}\right)}{\hat{r}_i^3}+\frac{ \left(1-\frac{\hat{L}_i }{\hat{L}}\right) }{\hat{r}_i^3}\right ]-\frac{\cos \theta}{Ca_m}\cdot \frac{1}{\hat{r}_i},\quad i=1,2.
\end{equation}
The mass conservation can also be written in dimensionless form as
\begin{equation}\label{eq:mass-conser-non-3}
\frac{\mathrm{d} \left(\frac{\hat L_1}{\hat L}\right)}{\mathrm{d}\hat t}\cdot \hat{r}_1+\frac{\mathrm{d} \left(\frac{\hat L_2}{\hat L}\right)}{\mathrm{d}\hat t}\cdot \hat{r}_2=1.
\end{equation}
To solve the interface movement, we write (\ref{eq:delta-p-non}) for each daughter channel. Equating the resulting two equations gives,
\begin{equation}\label{eq:coupled-ode-1}
 \hat{r}_1\cdot \frac{\mathrm{d}\left(\frac{\hat L_1}{\hat L}\right) }{\mathrm{d} \hat{t}}\left[ \frac{\lambda  \frac{\hat{L}_1}{\hat L} }{\hat{r}_1^3}+\frac{ \left(1-\frac{\hat{L}_1}{\hat L}\right) }{\hat{r}_1^3}\right ]-\frac{\cos \theta}{Ca_m}\cdot \frac{1}{\hat{r}_1}=\hat{r}_2\cdot \frac{\mathrm{d}\left(\frac{\hat{L}_2}{\hat L}\right)}{\mathrm{d} \hat{t}}\left[ \frac{\lambda  \frac{\hat{L}_2}{\hat L} }{\hat{r}_2^3}+\frac{ \left(1-\frac{\hat{L}_2}{\hat L}\right) }{\hat{r}_2^3}\right ]-\frac{\cos \theta}{Ca_m}\cdot \frac{1}{\hat{r}_2}.
\end{equation}
Substituting (\ref{eq:mass-conser-non-3}) into (\ref{eq:coupled-ode-1}), we obtain an ordinary differential equation (ODE) for $\hat{L}_i(t)$, i.e.
\begin{equation}\label{eq:coupled-ode-3}
\frac{\mathrm{d}\left(\frac{\hat{L}_1}{\hat L}\right) }{\mathrm{d} \hat{t}}=\frac{\frac{\cos \theta}{Ca_m}\cdot \left(\frac{1}{\hat{r}_1}-\frac{1}{\hat{r}_2}\right)+\phi \left(\frac{\hat{L}_2}{\hat L}\right)}{\hat{r}_1 \left[\phi \left(\frac{\hat{L}_1}{\hat L}\right)+\phi \left(\frac{\hat{L}_2}{\hat L}\right)\right]},
\end{equation}
where $\phi\left(\hat{L}_i/\hat L\right)= \left[\lambda \hat{L}_i/\hat L+\left(1-\hat{L}_i/\hat L\right)\right]/\hat{r}_i^3$. The ODE for $\hat{L}_2(t)$ can be obtained by exchanging subscripts 1 and 2, so
\begin{equation}\label{eq:coupled-ode-L2t}
\frac{\mathrm{d}\left(\frac{\hat{L}_2}{\hat L}\right) }{\mathrm{d} \hat{t}}=\frac{\frac{\cos \theta}{Ca_m}\cdot \left(\frac{1}{\hat{r}_2}-\frac{1}{\hat{r}_1}\right)+\phi\left(\frac{\hat{L}_1}{\hat L}\right)}{\hat{r}_2 \left[\phi\left(\frac{\hat{L}_1}{\hat L}\right)+\phi\left(\frac{\hat{L}_2}{\hat L}\right)\right]}.
\end{equation}
To prevent the flow in the branch channels from moving backward, the following constraints must be satisfied~\citep{al-housseinyPreferentialFlowPenetration2014}
\begin{equation}\label{eq:constraints-12}
0 \leq \frac{\mathrm{d} (\frac{\hat{L}_i}{\hat L})}{\mathrm{d} \hat{t}} \leq \frac{1}{\hat{r}_i}, \quad i = 1, 2.
\end{equation}

\subsection{Semi-analytical solutions}
\label{solved-ode-results}
\captionsetup[subfigure]{position=top,labelfont=rm,textfont=normalfont,singlelinecheck=off,justification=raggedright}
\begin{figure}
    \subfloat[]{\includegraphics[width=0.3\textwidth]{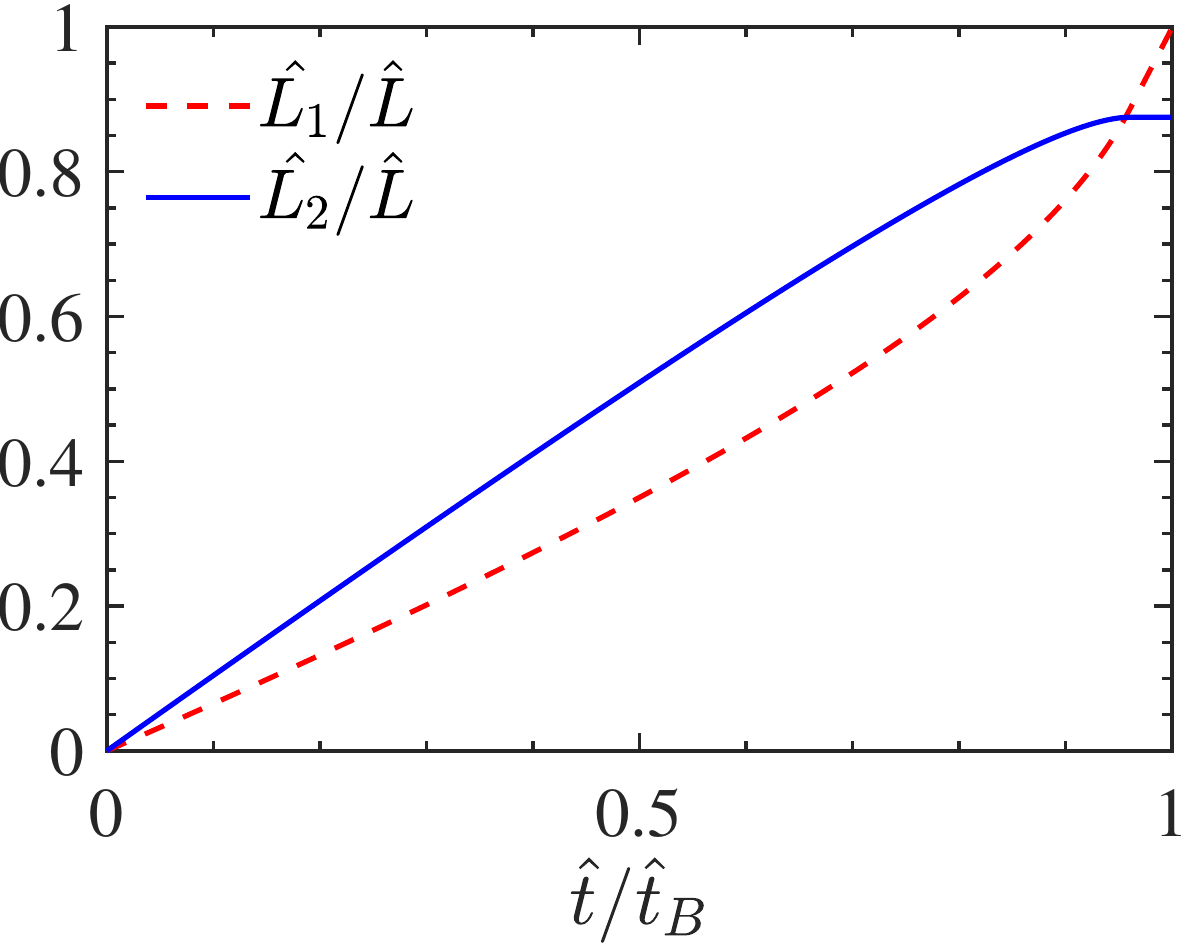}}~
    \subfloat[]{\includegraphics[width=0.3\textwidth]{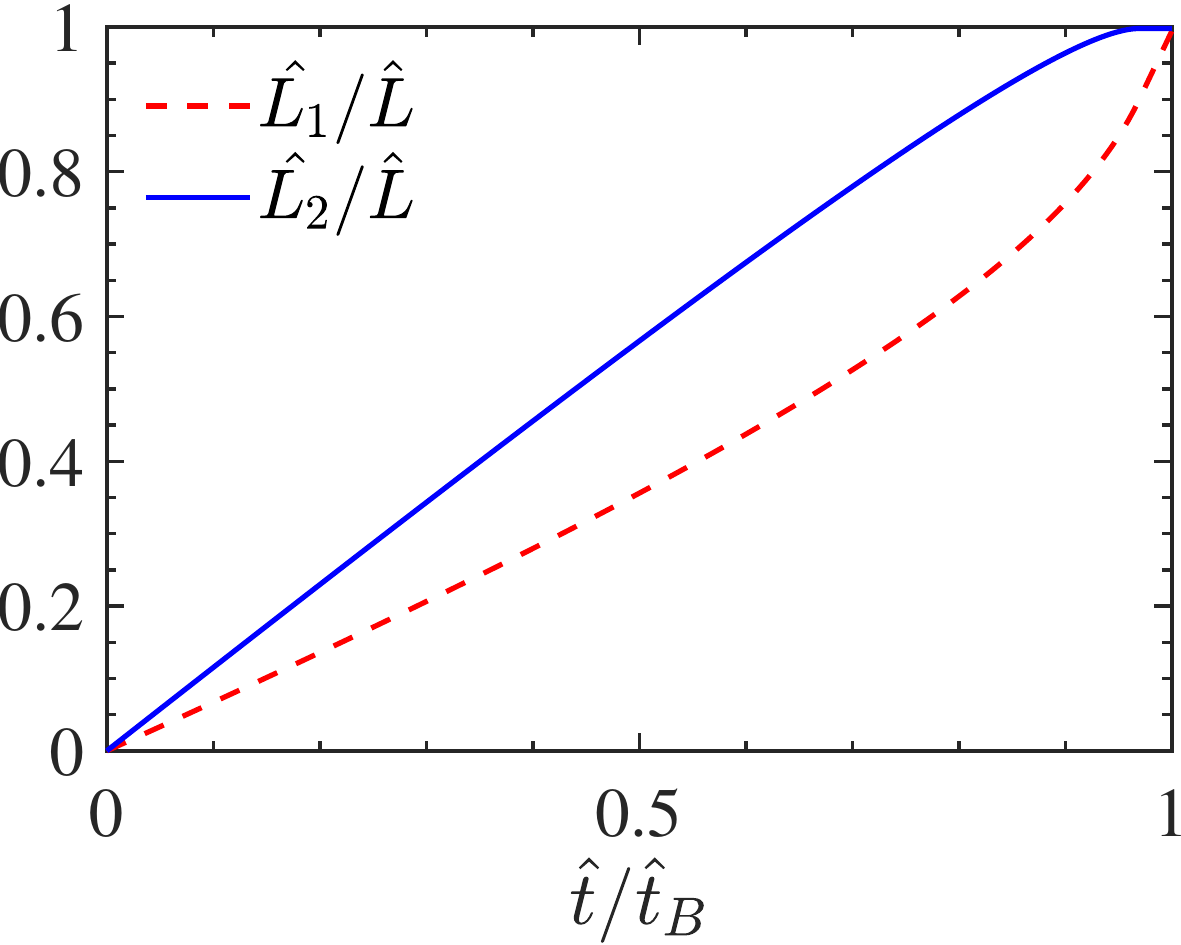}}~
    \subfloat[]{\includegraphics[width=0.3\textwidth]{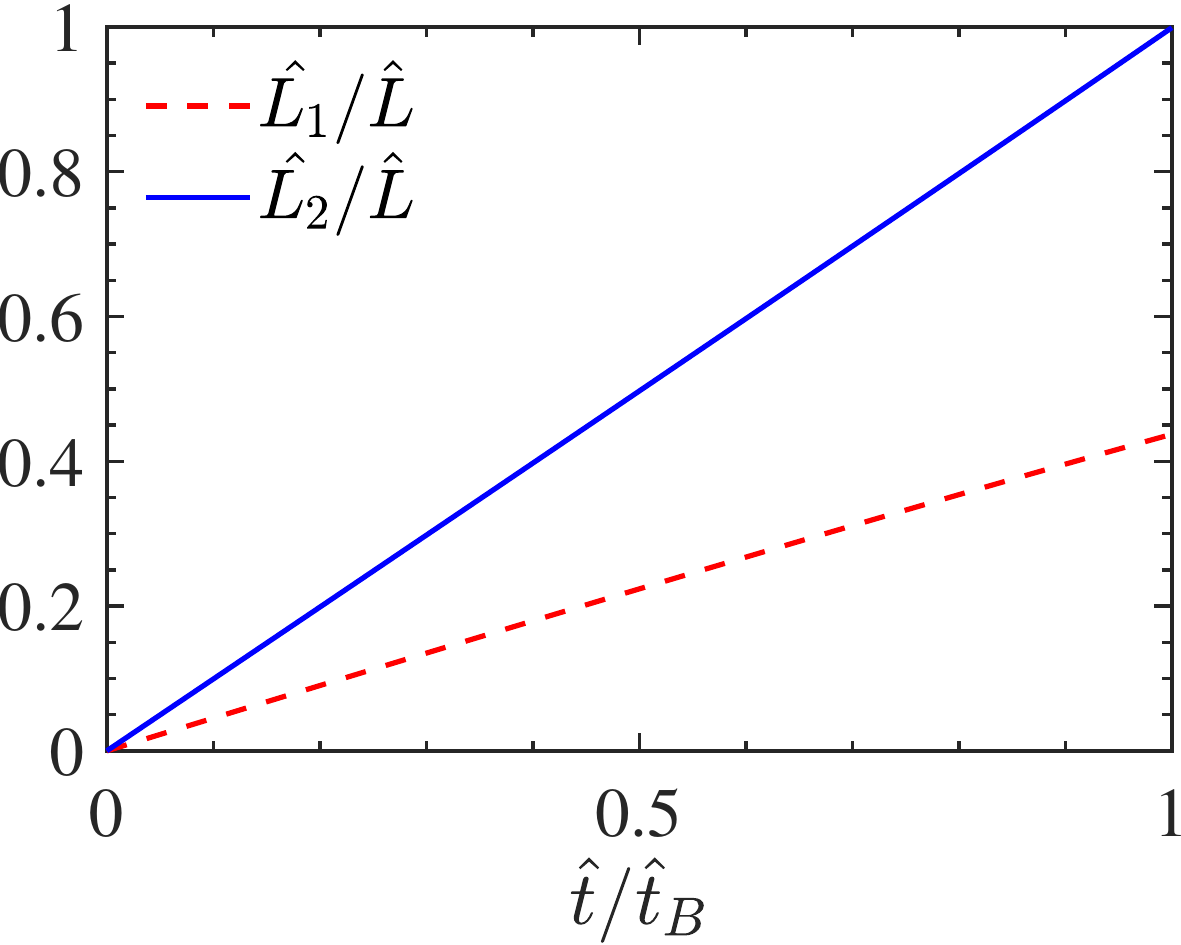}}\\
    \subfloat[]{\includegraphics[width=0.3\textwidth]{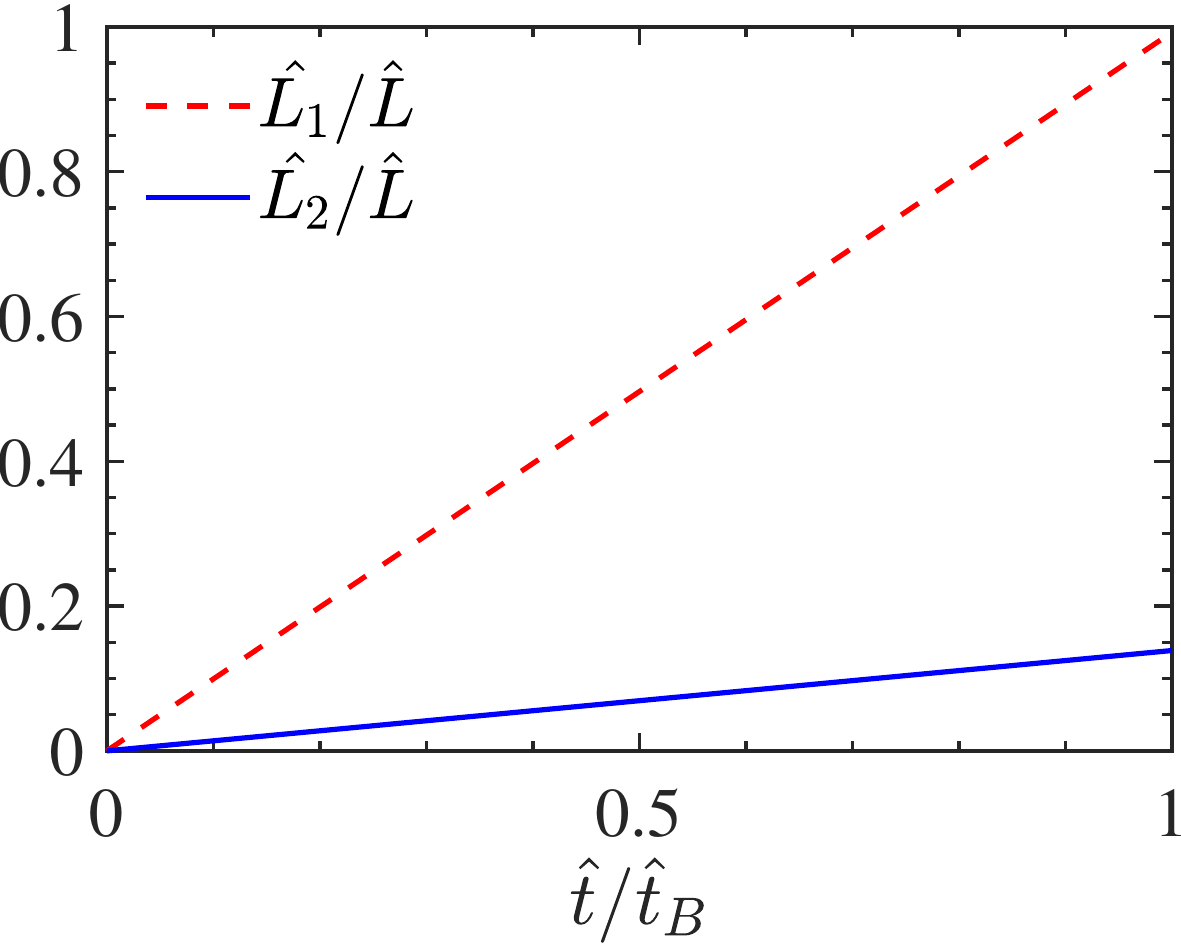}}~
    \subfloat[]{\includegraphics[width=0.3\textwidth]{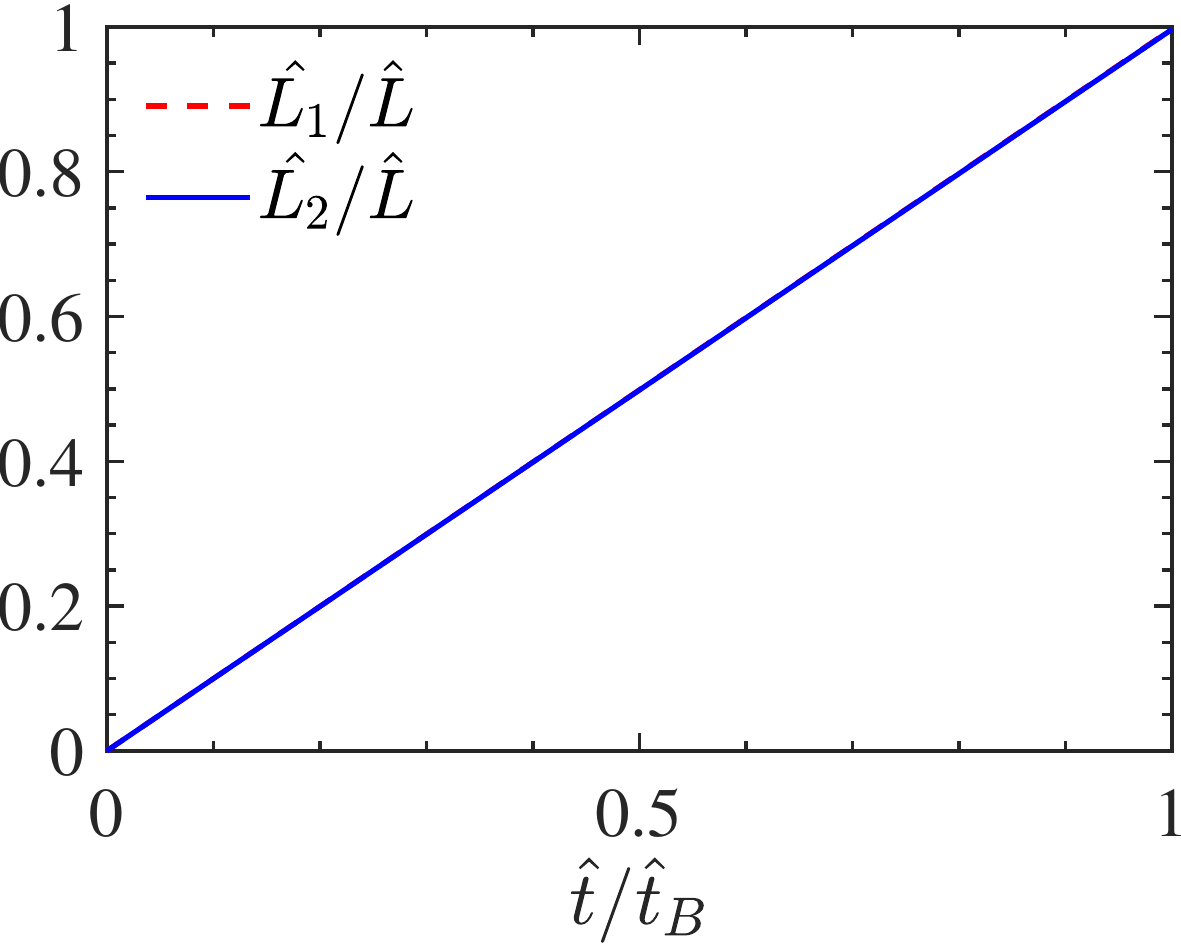}}~
    \subfloat[]{\includegraphics[width=0.3\textwidth]{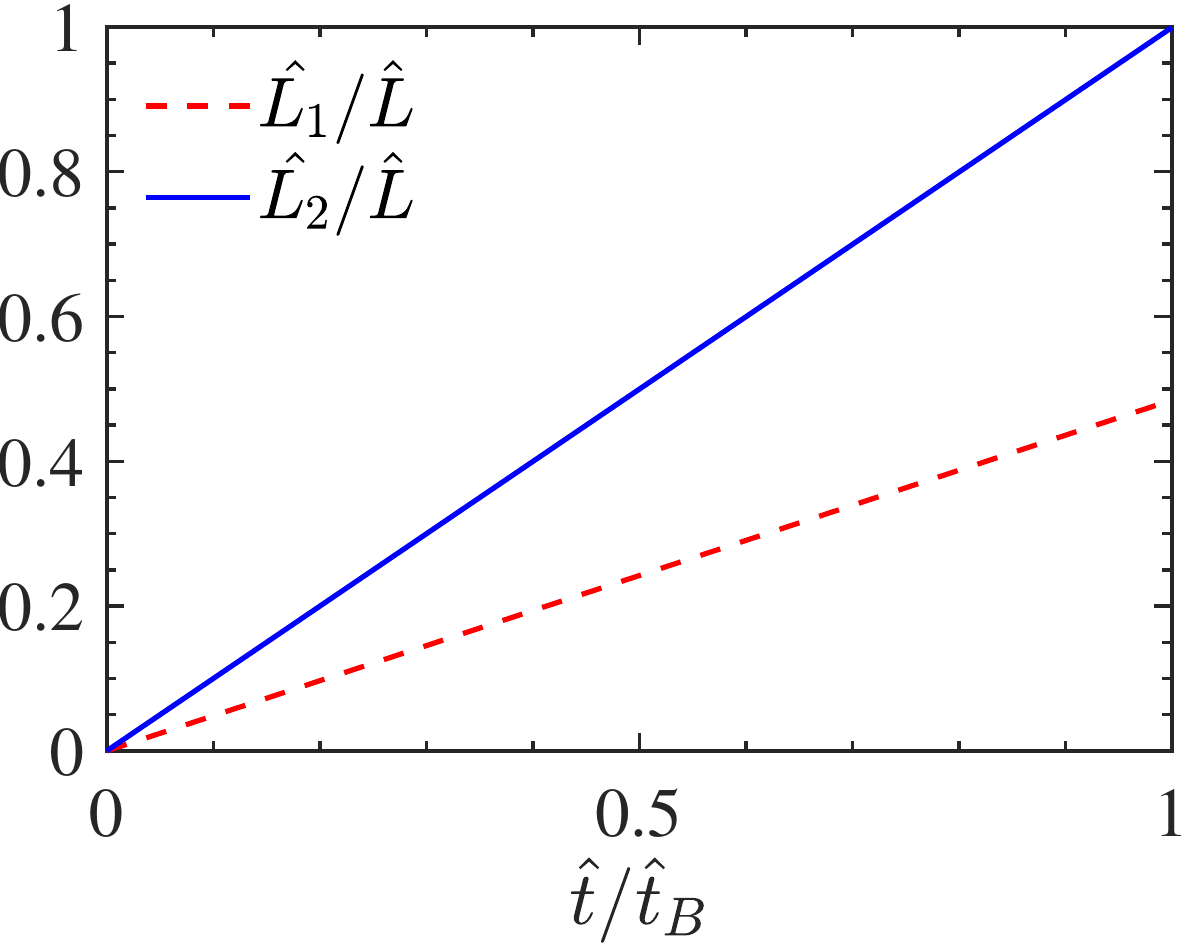}}\\
    \subfloat[]{\includegraphics[width=0.3\textwidth]{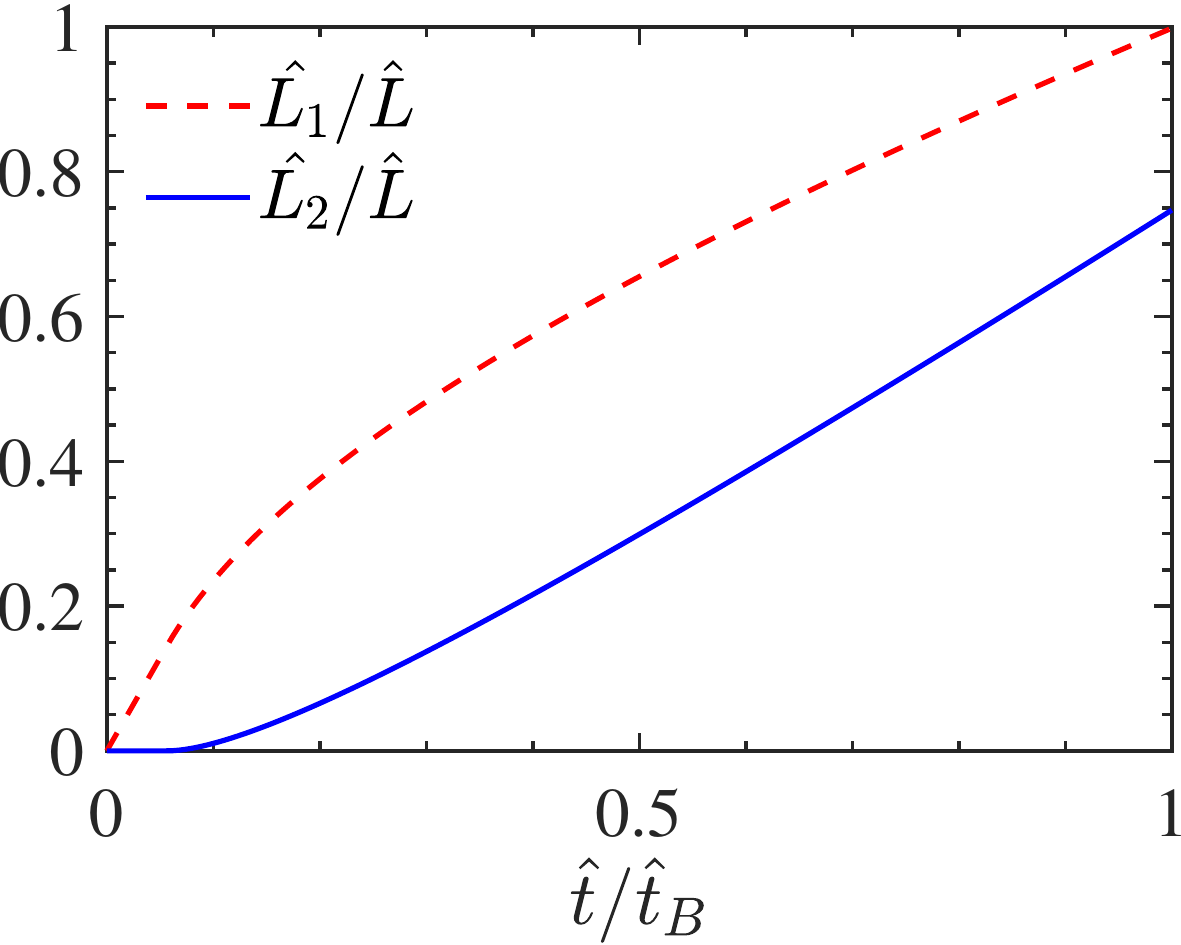}}~
    \subfloat[]{\includegraphics[width=0.3\textwidth]{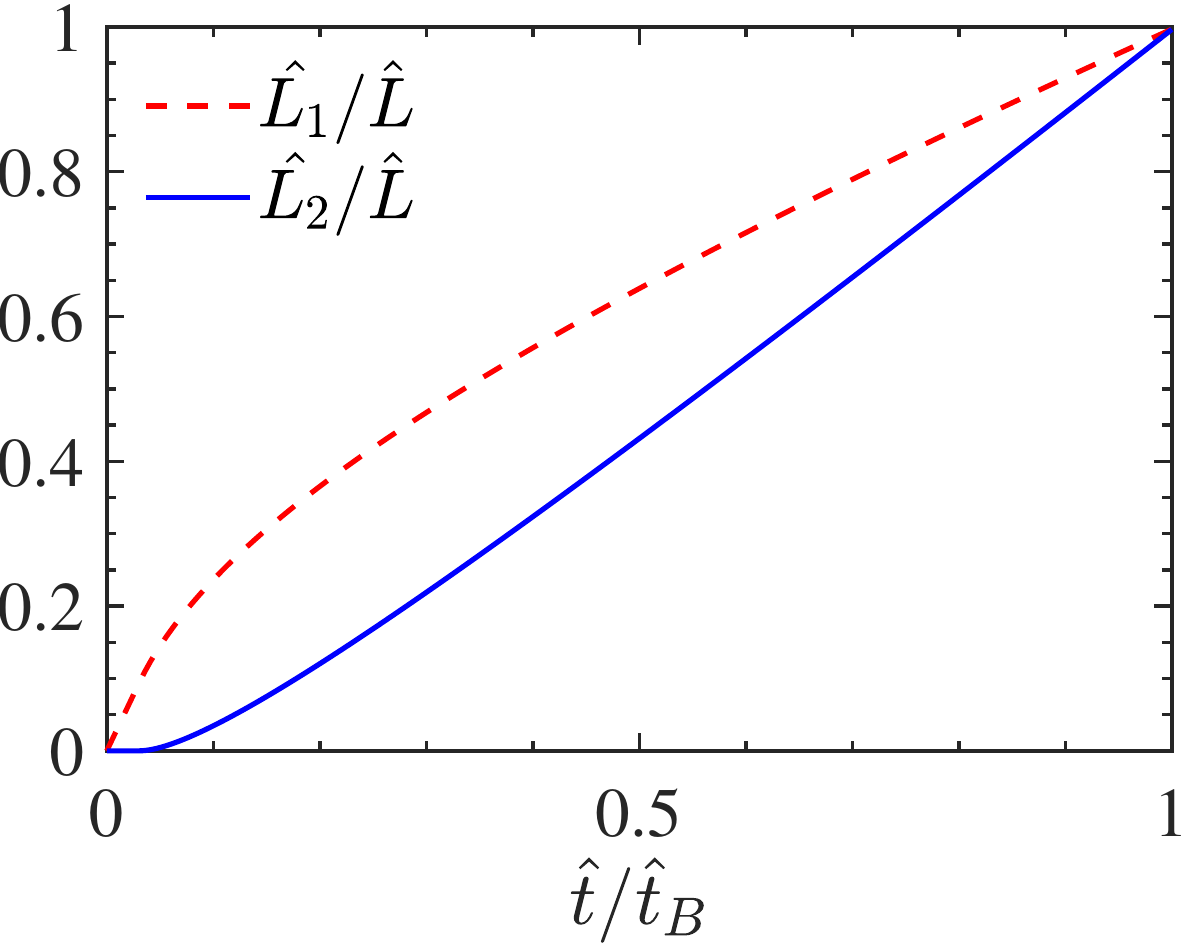}}~
    \subfloat[]{\includegraphics[width=0.3\textwidth]{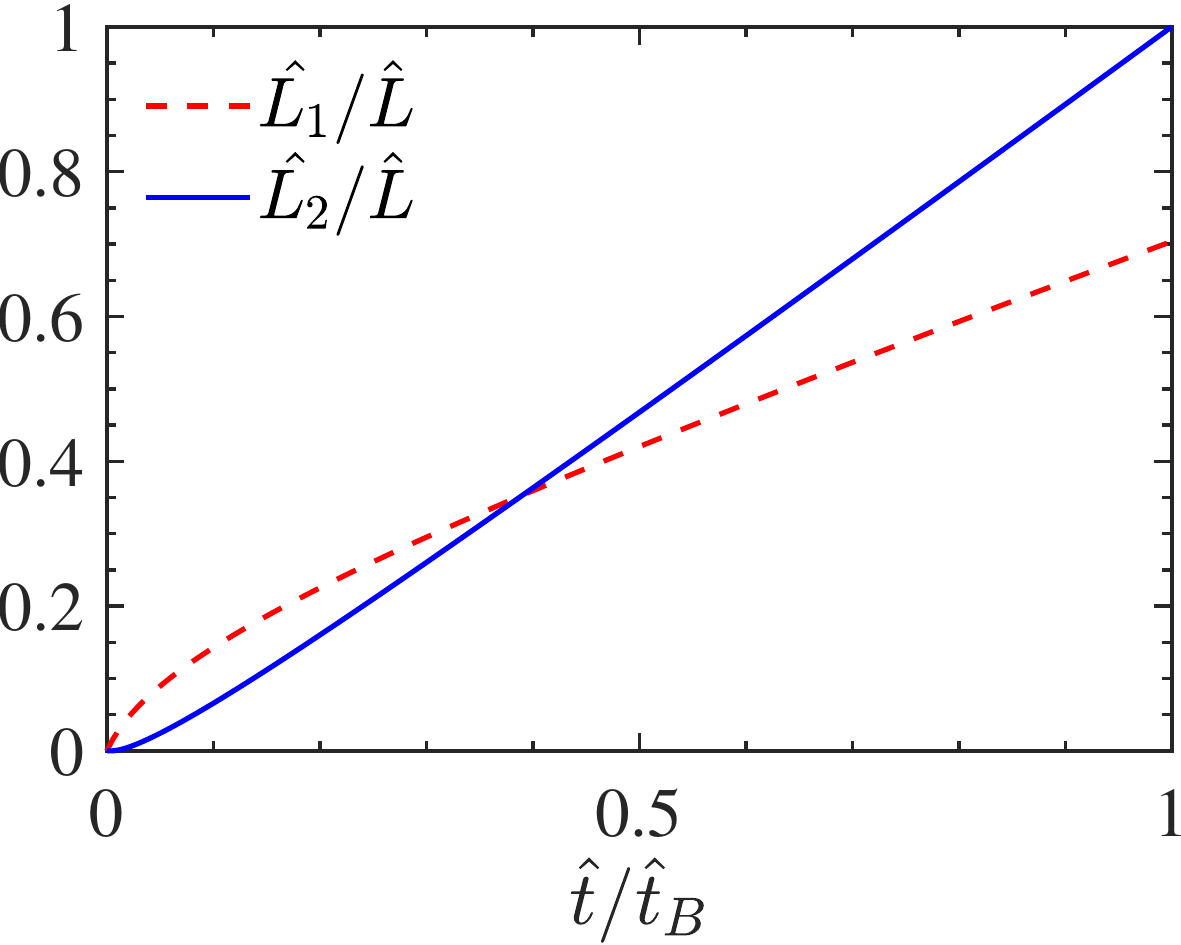}}
\caption{(Colour online) The lengths of the wetting fluid in the branch channels as a function of time  obtained by solving equations (\ref{eq:coupled-ode-3}) and (\ref{eq:coupled-ode-L2t}) at  $\lambda=0.025$ for (a) $Ca_m=\num{3.16}$, (b) $Ca_m=\num{3.48}$ and (c) $Ca_m=\num{5.05}$; at $\lambda=1$ for (d) $Ca_m=\num{0.606}$, (e) $Ca_m=\num{1.82}$ and (f) $Ca_m=\num{4.54}$; at $\lambda=20.0$ for (g) $Ca_m=\num{0.126}$, (h) $Ca_m=\num{0.173}$ and (i) $Ca_m=\num{0.423}$.}
   \label{fig:non-dimension-lambda-0025-1-20-num-solution}
\end{figure}
Consider a pore doublet geometry with $\hat{r}_1 =1$, $\hat{r}_2=2$, $\hat{L}=63.11$ and $\theta=\ang{30}$. To obtain the location of the meniscus in each capillary, we numerically solve the equations (\ref{eq:coupled-ode-3}) and (\ref{eq:coupled-ode-L2t}) subject to the constraint (\ref{eq:constraints-12}) using the first-order forward difference scheme for different values of $Ca_m$ and $\lambda$. Solutions are found with the initial condition that $[\hat{L}_1,\hat{L}_2]=[0,0]$ at $\hat{t}=0$.

Numerical results for several typical capillary numbers at $\lambda =0.025$, 1 and 20 are shown in figure~\ref{fig:non-dimension-lambda-0025-1-20-num-solution}, where the penetration lengths $\hat{L}_1$ and $\hat{L}_2$ are plotted as a function of time $\hat{t}$, normalized by the breakthrough time $\hat{t}_B$. For each viscosity ratio, at low $Ca_m$ (figure~\ref{fig:non-dimension-lambda-0025-1-20-num-solution}a,d,g), we can see that  $\hat{L}_1=\hat{L}>\hat{L}_2$ when the breakthrough occurs, although $\hat{L}_1$ lags behind $\hat{L}_2$ till $\hat{t}/\hat{t}_B=0.96$ in figure~\ref{fig:non-dimension-lambda-0025-1-20-num-solution}(a); at high $Ca_m$  (figure~\ref{fig:non-dimension-lambda-0025-1-20-num-solution}c,f,i), the meniscus in the channel 2 breaks through first, i.e. $\hat{L}_1<\hat{L}_2=\hat{L}$. This suggests that there exists a critical value of $Ca_m$ between the low and high $Ca_m$, known as the critical preferential capillary number ($Ca_{m,c}$), at which the breakthrough of wetting fluid occurs simultaneously in both branch channels. As shown in figure~\ref{fig:non-dimension-lambda-0025-1-20-num-solution}(b,e,h), the values of $Ca_{m,c}$ are 3.48, 1.82 and 0.173 for the viscosity ratios of 0.025, 1 and 20. Clearly, the critical preferential capillary number is strongly dependent on the viscosity ratio. In addition, we also interestingly find that for $\lambda=1$, the imbibition rates are a constant and exactly the same in both branch channels.

\begin{figure}
\centering
\begin{overpic}[scale=.55]{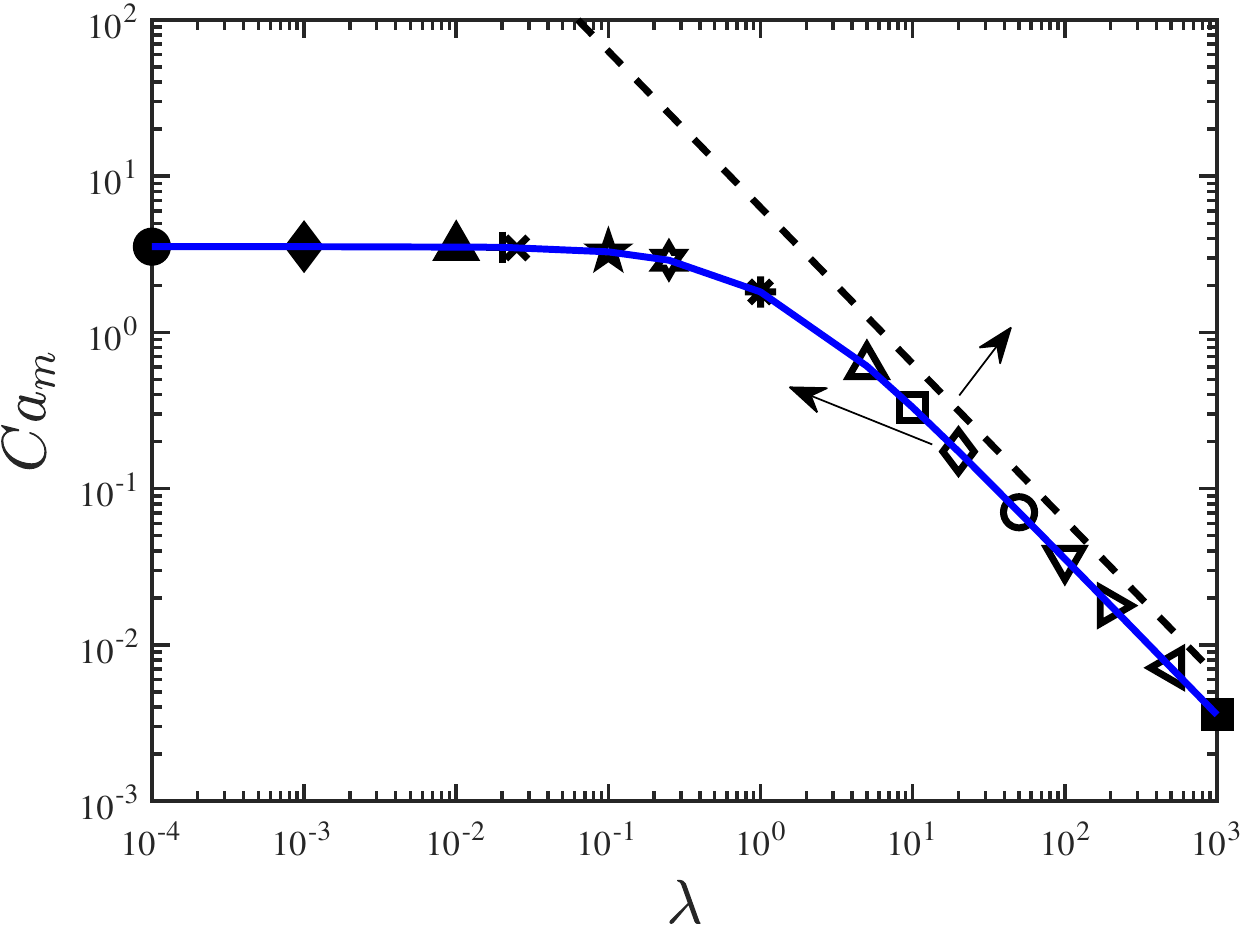}
\put(22,27){\color{black} (\rom{1})}
\put(28,25){\includegraphics[scale=.06]{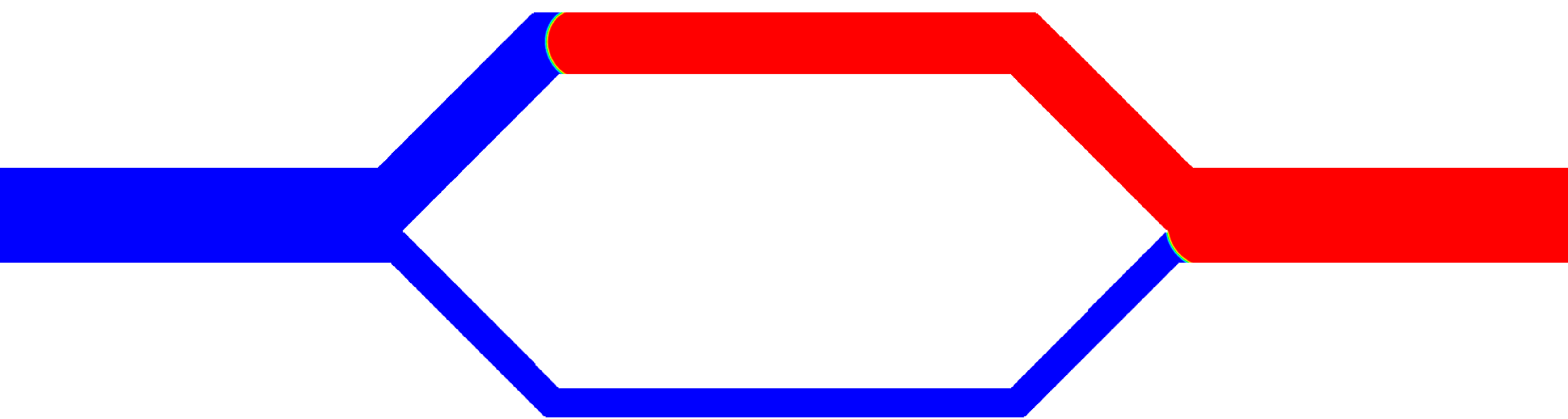}}
\put(27,42){\color{black} (\rom{3})}
\put(37,40){\includegraphics[scale=.06]{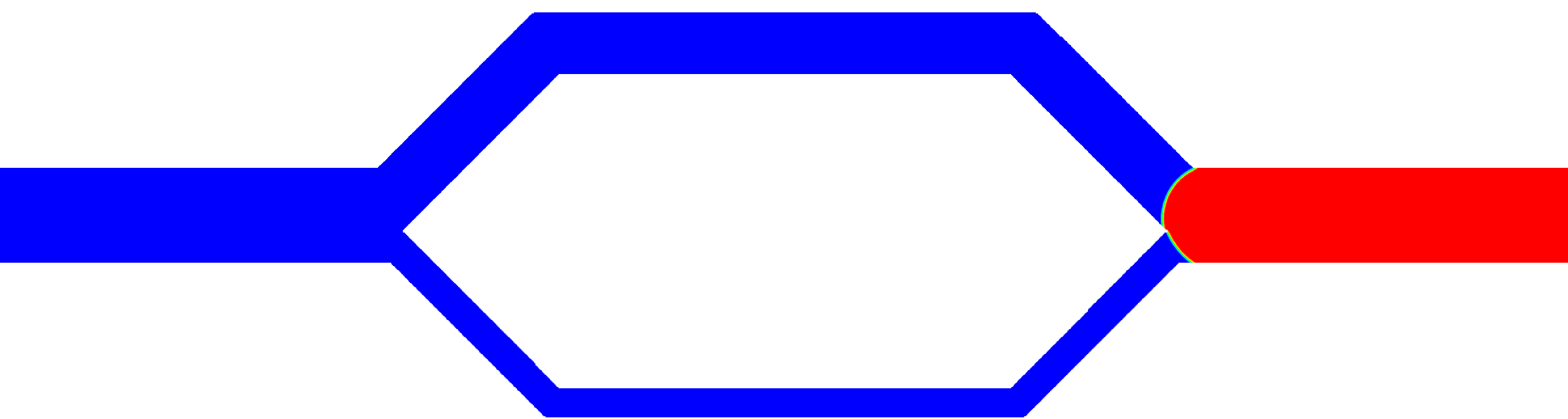}}
\put(57,64){\color{black} (\rom{2})}
\put(65,62){\includegraphics[scale=.06]{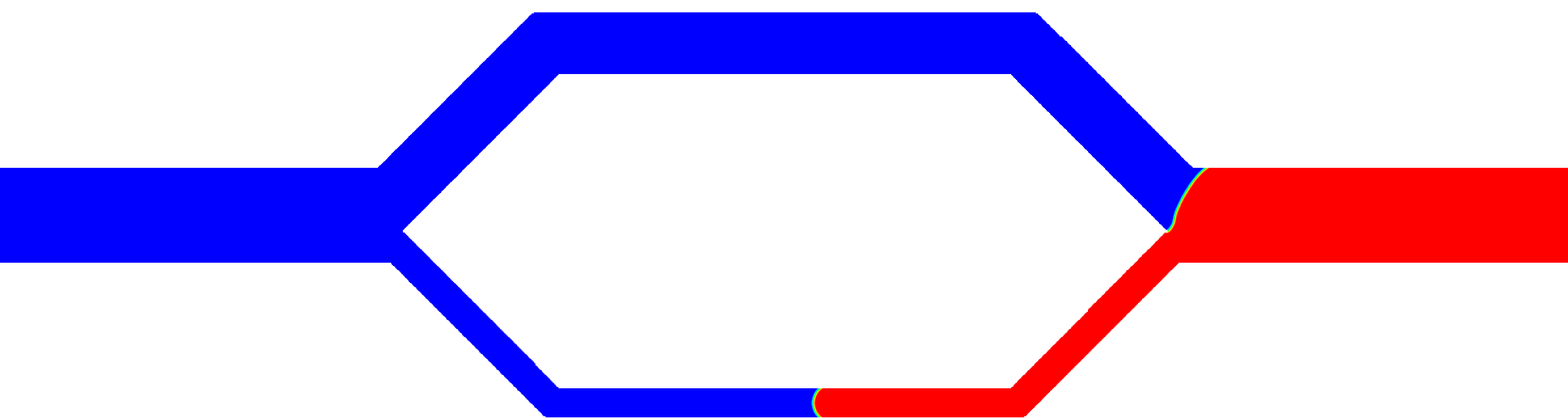}}
\put(82,48){\color{black}$ \lambda^{-1}$}
\end{overpic}
\caption{(Colour online) The $\lambda-Ca_m$ diagram showing the imbibition preference in a pore doublet. The discrete symbols of different shapes represent the cases where the simultaneous breakthrough occurs. Connecting these symbols gives the blue solid line which divides the plane into two regions, i.e.  (\rom{1}) and (\rom{2}). In (\rom{1}), the meniscus first breaks through the channel 1, whereas in (\rom{2}) the breakthrough first occurs in the channel 2. The border on which $\hat{L}_1=\hat{L}_2=\hat{L}$ at breakthrough is denoted as (\rom{3}), and it follows a scaling relation $Ca_{m,c}=3.011\lambda^{-1}$ for $\lambda\geq10$. The dashed line is added to show the proportional relationship between $Ca_{m,c}$ and $\lambda^{-1}$.}
\label{fig:poredoublet-cri-Ca-Tendency}
\end{figure}
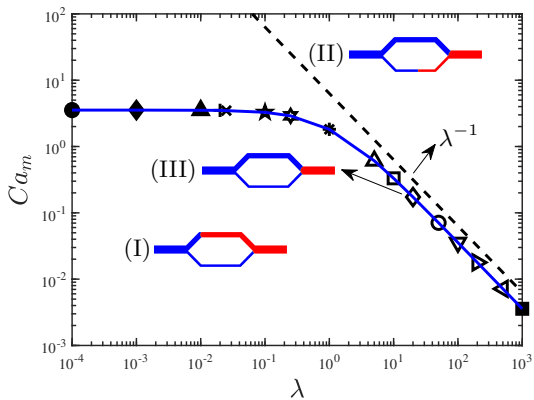

Different imbibition behaviours at low and high values of $Ca_m$ are attributed to the competition between the capillary pressure and the viscous resistance. At low flow rates ($Ca_m$), the viscous resistance is negligibly small while the capillary pressure is dominant, which acts a driving force for the wetting fluid to progress; since the capillary pressure is inversely proportional to the channel width, the penetration length in the channel 1 is larger than that in the channel 2, i.e. $\hat L_1>\hat L_2$, at breakthrough. However, at high flow rates the viscous force is dominant; because of the lower viscous resistance in the channel 2, the penetration length in the channel 2 would be larger than in the channel 1, i.e. $\hat L_1<\hat L_2$.

To understand the effect of the viscosity ratio on the imbibition process, the theoretical analysis is further conducted for a wide range of viscosity ratios, varying from \num{e-4} to \num{e3}. Figure \ref{fig:poredoublet-cri-Ca-Tendency} depicts the imbibition preference of the meniscus at breakthrough in the $\lambda-Ca_m$ diagram. Three typical regions are identified due to the competition between capillary and viscous forces: (\rom{1}) the region below the solid blue line, where the meniscus in the channel 1 outpaces that in the channel 2 at breakthrough, i.e. $\hat L_1=\hat L>\hat L_2$; (\rom{2}) the region above the solid blue line, where the meniscus in the channel 2 outpaces that in the channel 1 at breakthrough, i.e. $\hat L_1<\hat L_2=\hat L$; (\rom{3}) the border of the above-mentioned two regions, on which the menisci in the channels 1 and 2 arrive the downstream junction at the same time, i.e. $\hat L_1=\hat L_2=\hat L$ at breakthrough. It is noted that the border corresponds to the critical curve of $Ca_m$, i.e. the $Ca_{m,c}$ curve. We can observe that for $\lambda\geq10$, the critical capillary number $Ca_{m,c}$ obeys a scaling relation $Ca_{m,c}=3.011\lambda^{-1}$; whereas for $\lambda\leq0.1$, it tends to converge to a value of around 3.5. Through figure \ref{fig:poredoublet-cri-Ca-Tendency}, we are able to predict the filling order of the wetting fluid for varying viscosity ratio and $Ca_m$ in a pore doublet. In a previous work~\citep{sorbieExtendedWashburnEquation1995}, the existence of critical parameters for characterising the simultaneous filling of both branch channels has been discussed in terms of the aspect ratio ($r_i/L$) and the channel width ratio, and the influence of aspect ratio is explained as a result of the fluid inertia; however, the aspect ratio is incorporated into the definition of the preferential capillary number in the present study.
\subsection{Comparison between LBM simulations and semi-analytical solutions}
\label{lbm-pore-doublet}
\begin{figure}
    \subfloat[]{\includegraphics[width=0.33\textwidth]{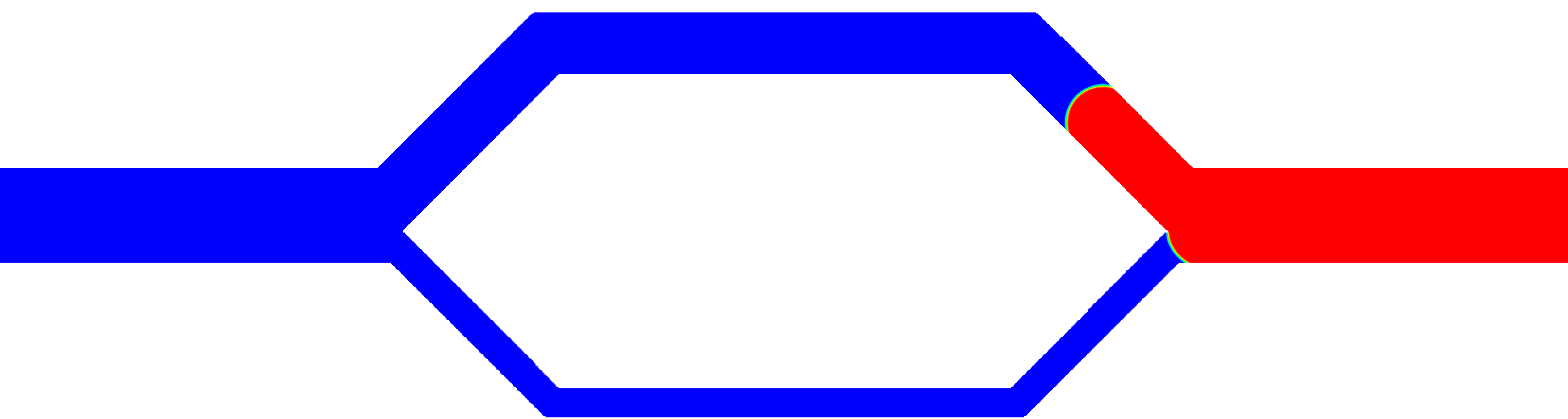}}~
    \hfill
    \subfloat[]{\includegraphics[width=0.33\textwidth]{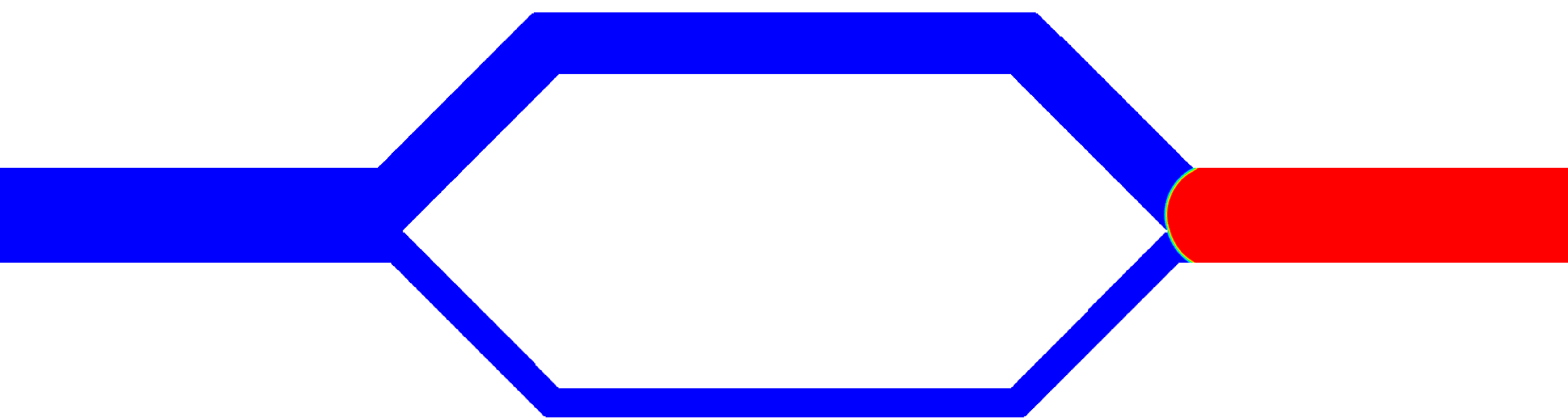}}~
    \hfill
    \subfloat[]{\includegraphics[width=0.33\textwidth]{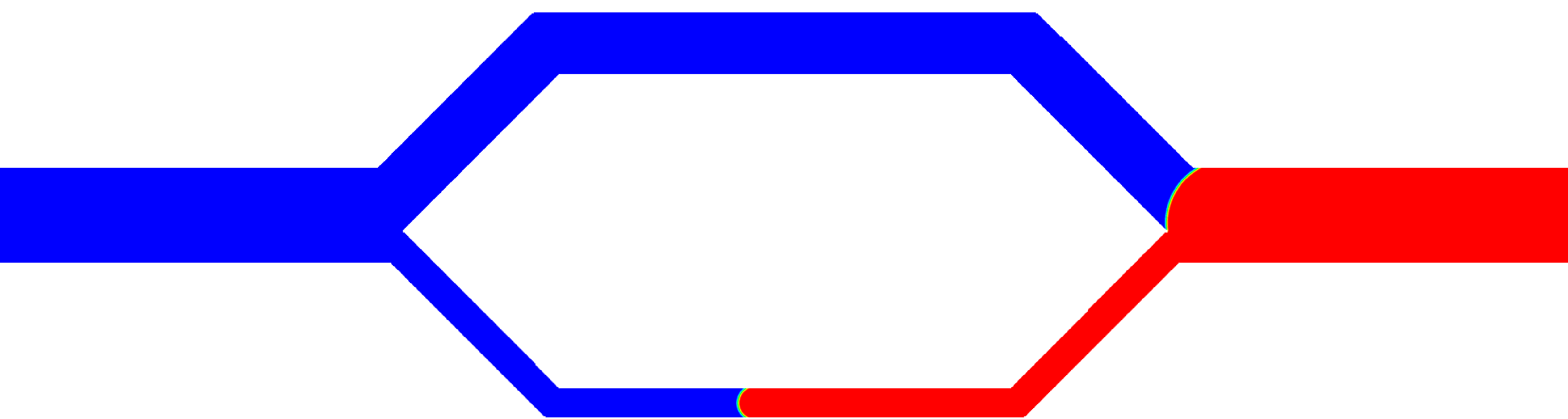}}\\
    \subfloat[]{\includegraphics[width=0.33\textwidth]{figures/Ca2E_3_Vi1_t420000.pdf}}~
    \hfill
    \subfloat[]{\includegraphics[width=0.33\textwidth]{figures/Ca6E_3_Vi1_t216000.pdf}}~
    \hfill
    \subfloat[]{\includegraphics[width=0.33\textwidth]{figures/Ca15E_2_Vi1_t77500.pdf}}\\
    \subfloat[]{\includegraphics[width=0.33\textwidth]{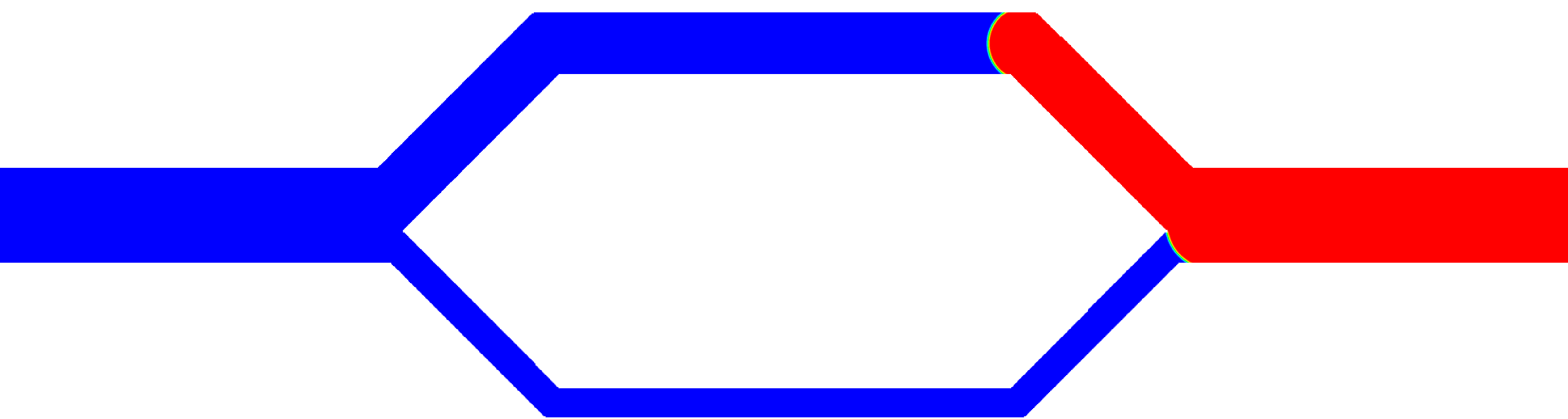}}~
    \hfill
    \subfloat[]{\includegraphics[width=0.33\textwidth]{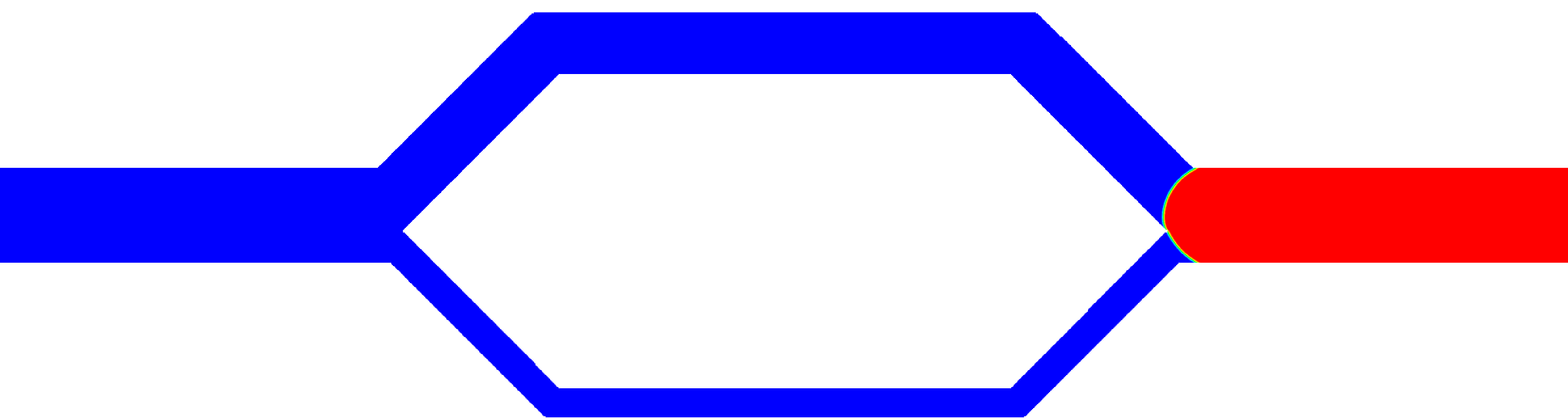}}~
    \hfill
    \subfloat[]{\includegraphics[width=0.33\textwidth]{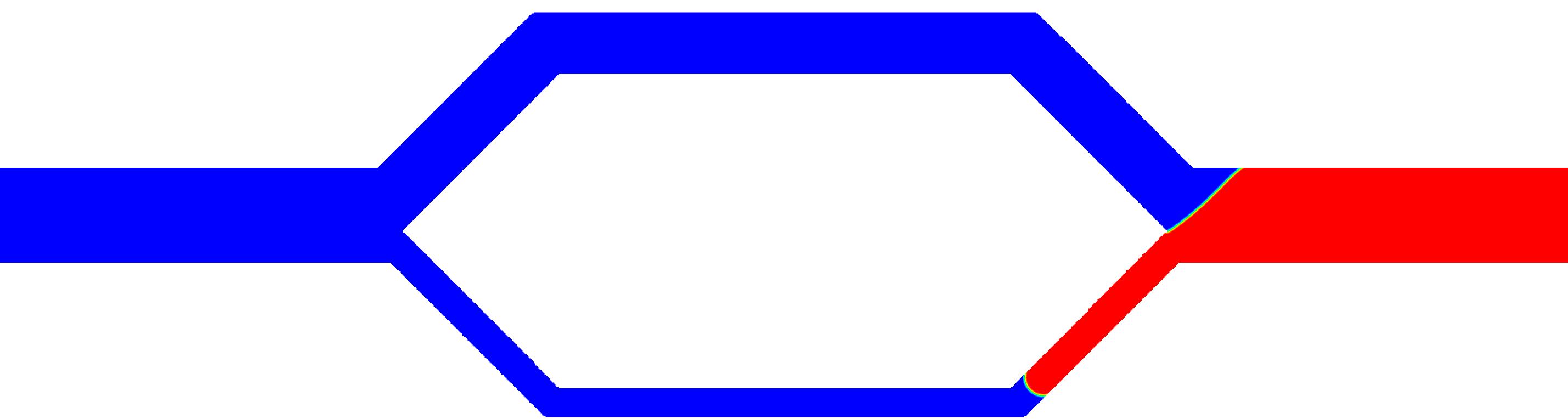}}
\caption{(Colour online) Fluid distributions at breakthrough obtained from the LBM simulations for the same parameters as those in figure~\ref{fig:non-dimension-lambda-0025-1-20-num-solution}. Specifically, the first row: $\lambda=0.025$ with: (a) $Ca_m=\num{3.16}$, (b) $Ca_m=\num{3.48}$ and (c) $Ca_m=\num{5.05}$; the second row: $\lambda=1$ with (d) $Ca_m=\num{0.606}$, (e) $Ca_m=\num{1.82}$ and (f) $Ca_m=\num{4.54}$; the third row: $\lambda=20.0$ with (g) $Ca_m=\num{0.126}$, (h) $Ca_m=\num{0.173}$ and (i) $Ca_m=\num{0.423}$. The non-wetting and wetting fluids are shown in red and blue, respectively.}
   \label{fig:lb-poredoublet-lambda-0025-1-20}
\end{figure}

In this section, the colour-gradient model is used to simulate the imbibition behaviour in a pore doublet and its capability is assessed by comparing with the semi-analytical solutions in section \ref{solved-ode-results}. The simulations are run in a $1575\times 409$ lattice domain with $r_1=15$ lattices and $r_2=30$ lattices, which are found fine enough to produce grid-independent results. Figure \ref{fig:lb-poredoublet-lambda-0025-1-20} shows the simulation results corresponding to the same values of $Ca_m$ and $\lambda$ in figure \ref{fig:non-dimension-lambda-0025-1-20-num-solution}. It is clear that the simulation results at breakthrough agree well with the semi-analytical solutions qualitatively, and for each $\lambda$, two menisci in branch channels are found to arrive at the downstream junction simultaneously at  $Ca_{m,c}$, consistent with the semi-analytical predictions in figure \ref{fig:non-dimension-lambda-0025-1-20-num-solution} as well.

\begin{figure}
\centering
\subfloat[]{\includegraphics[height=1.1cm]{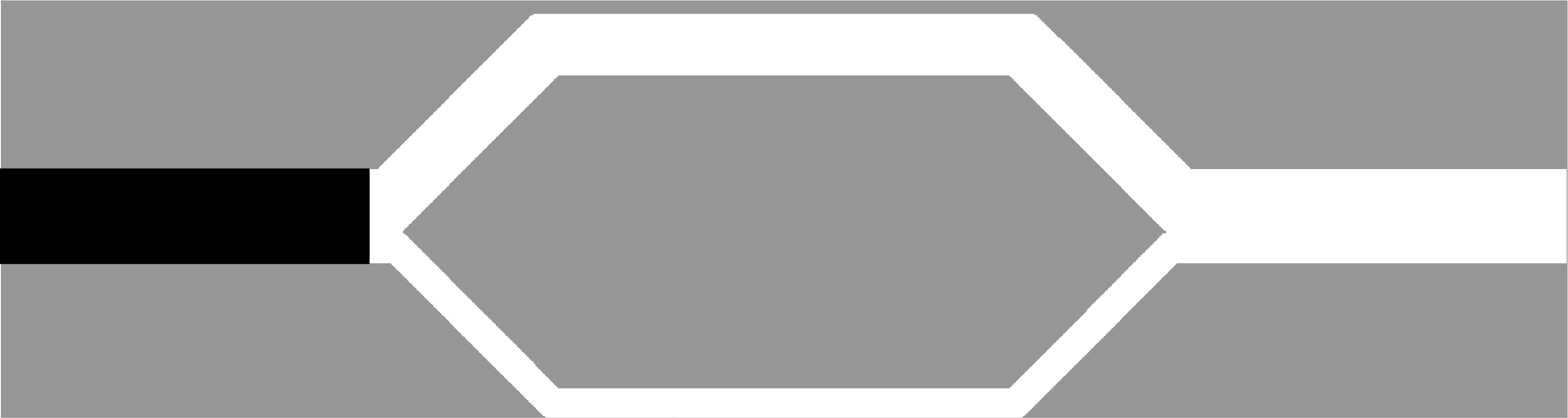}%
    \llap{\raisebox{1.4cm}{
      \includegraphics[height=1.1cm]{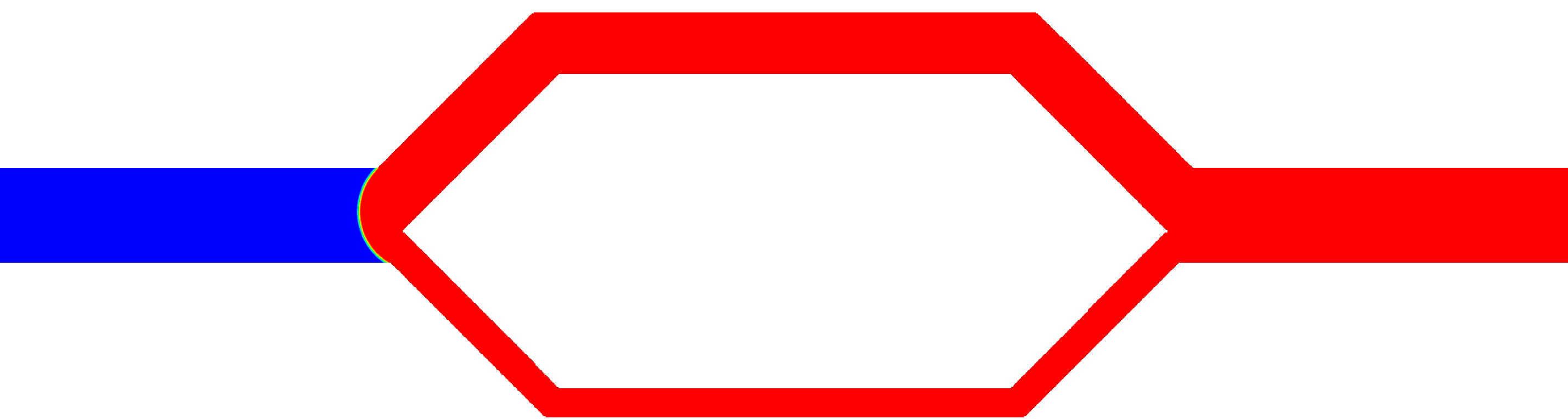}%
    }}
}~
\subfloat[]{\includegraphics[height=1.1cm]{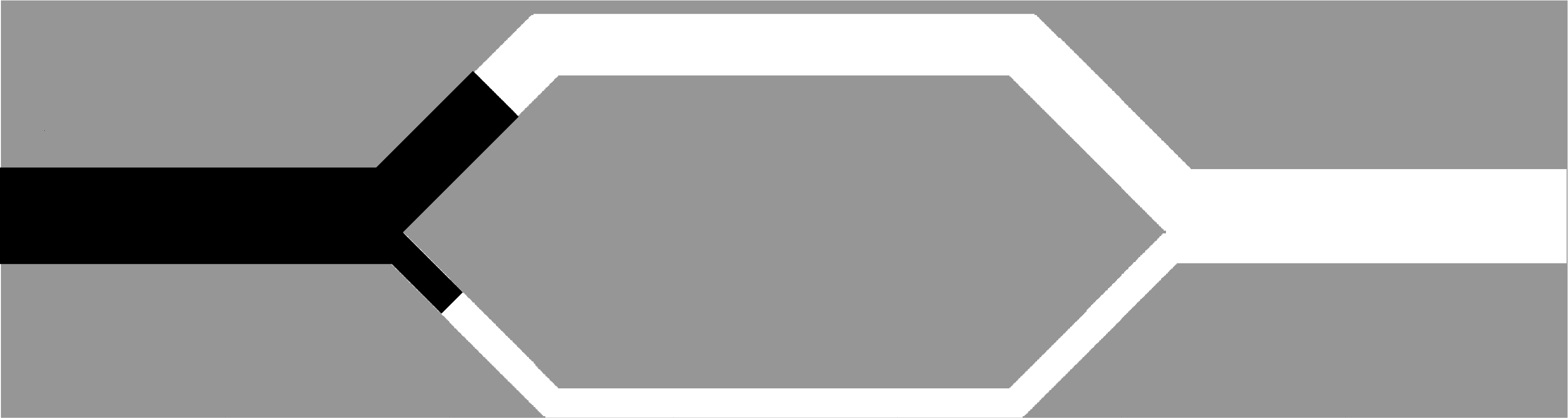}%
    \llap{\raisebox{1.4cm}{
      \includegraphics[height=1.1cm]{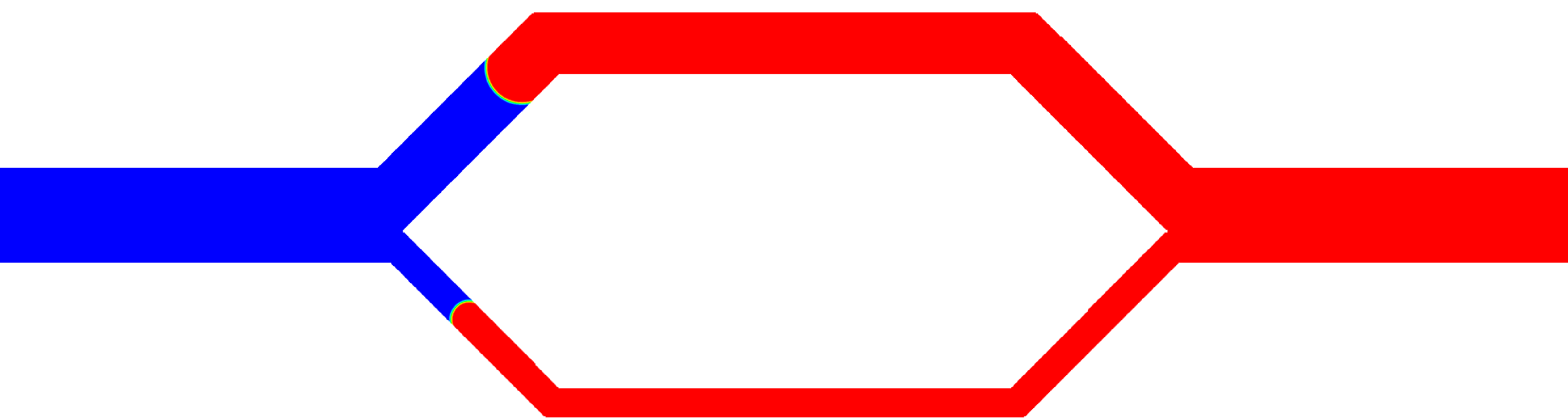}%
    }}
}~
\subfloat[]{\includegraphics[height=1.1cm]{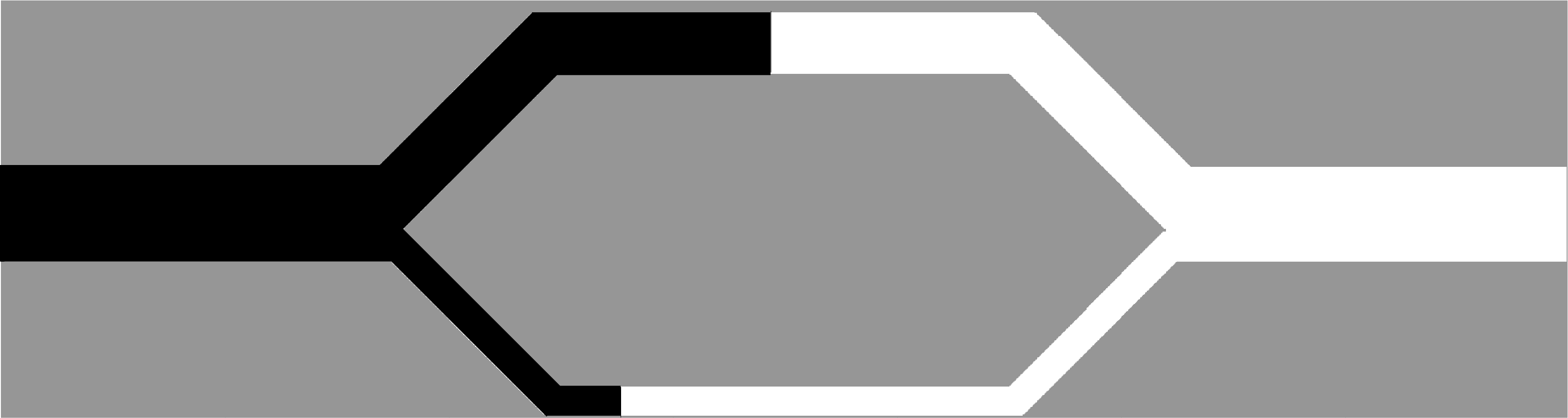}%
    \llap{\raisebox{1.4cm}{
      \includegraphics[height=1.1cm]{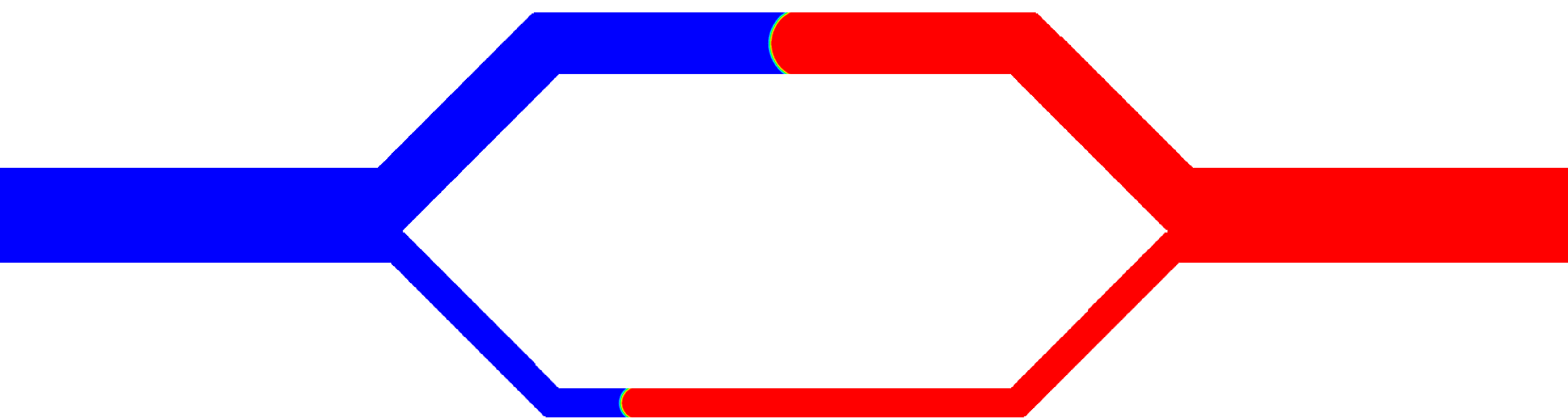}%
    }}
}\\
    \subfloat[]{\includegraphics[height=1.1cm]{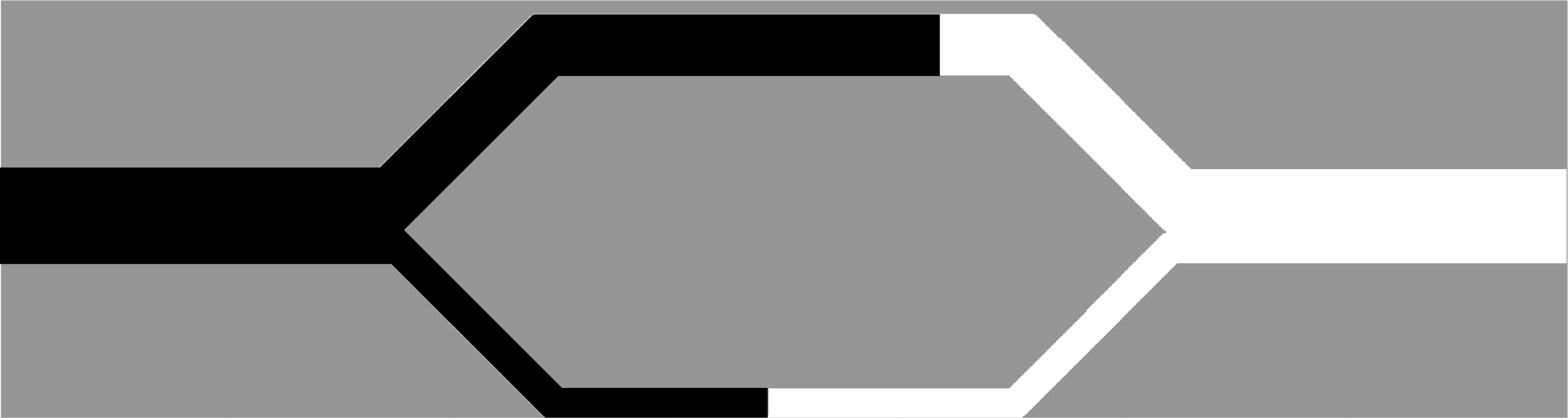}%
    \llap{\raisebox{1.4cm}{
      \includegraphics[height=1.1cm]{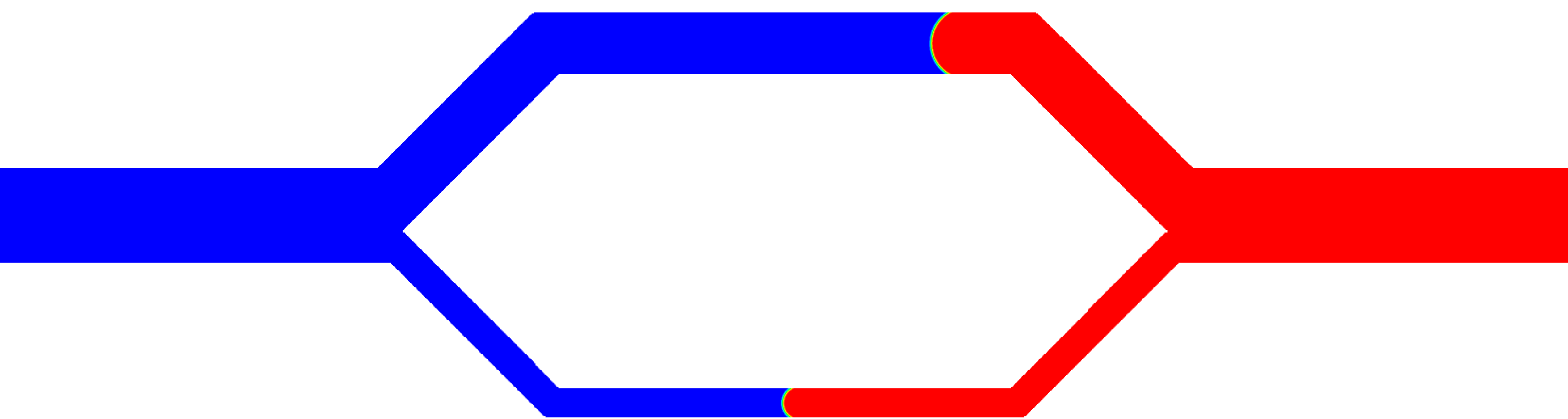}%
    }}
}~
\subfloat[]{\includegraphics[height=1.1cm]{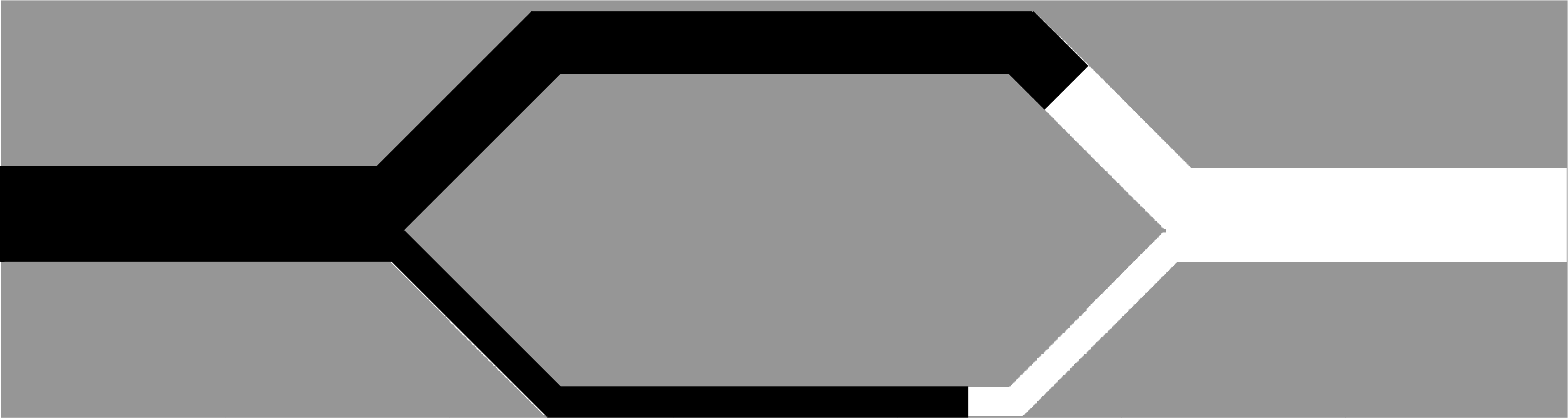}%
    \llap{\raisebox{1.4cm}{
      \includegraphics[height=1.1cm]{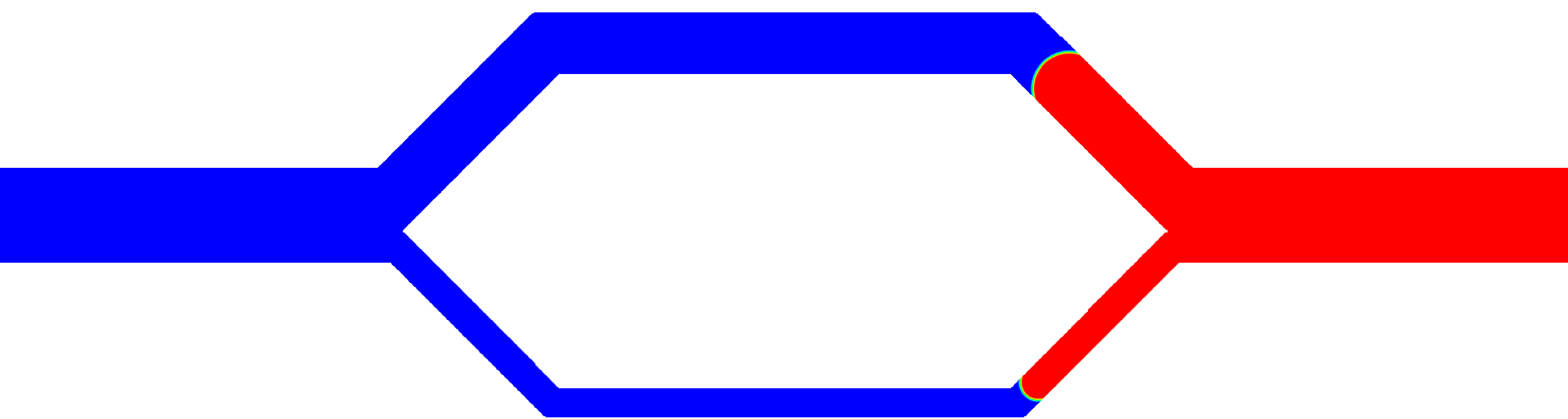}%
    }}
}~
    \subfloat[]{\includegraphics[height=1.1cm]{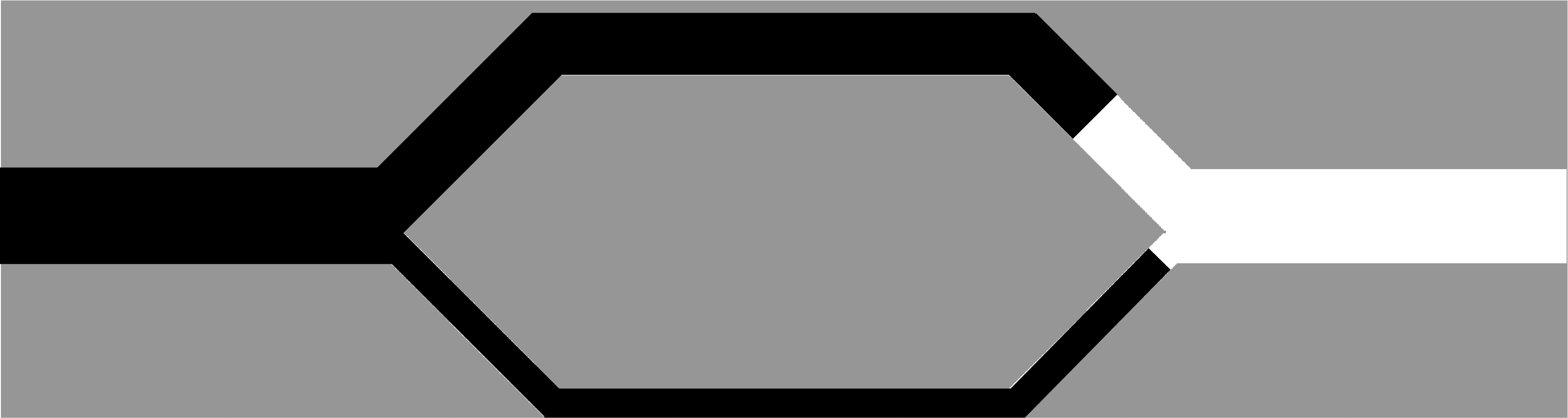}%
    \llap{\raisebox{1.4cm}{
      \includegraphics[height=1.1cm]{figures/Ca1point3E_3_Vi0025_t920000.pdf}%
    }}
}
\caption{(Colour online) Snapshots of imbibition process obtained by the LBM simulations (top) and the semi-analytical solutions (bottom) for $Ca_m=\num{3.16}$ and $\lambda=0.025$ at: (a) $\hat{t}/\hat{t}_B=0$, (b) $\hat{t}/\hat{t}_B=0.19$, (c) $\hat{t}/\hat{t}_B=0.5$, (d) $\hat{t}/\hat{t}_B=0.69$, (e) $\hat{t}/\hat{t}_B=0.88$ and (f) $\hat{t}/\hat{t}_B=1.0$. In the top images, the non-wetting and wetting fluids are shown in red and blue, respectively. In the bottom images, the non-wetting and wetting fluids are shown in white and black, respectively.}
   \label{fig:lb-poredoublet-lambda-0025-Ca5e-2}
\end{figure}

To assess the transient behaviour, as an example, we present the snapshots of imbibition process for $Ca_m=\num{3.16}$ and $\lambda=0.025$ in figure~\ref{fig:lb-poredoublet-lambda-0025-Ca5e-2}, where the upper and lower rows represent the simulation results and the semi-analytical predictions, respectively. Again, good agreement between the simulation results and semi-analytical predictions are obtained, although the LBM simulation a little overestimates $\hat L_1$ in figure~\ref{fig:lb-poredoublet-lambda-0025-Ca5e-2}(e). Having verified the colour-gradient LBM, we will use it to investigate the imbibition displacement in a dual-permeability pore network in the next section, where the theoretical predictions are not applicable due to the inherent complex geometry.

\section{Forced imbibition in a dual-permeability pore network}
\label{lb-dual-perm-pore-network}
In this section, we first describe the geometry setup of the problem along with the boundary conditions. Then the simulation results of imbibition displacement in the pore network are presented and compared with those previously obtained  from the pore doublet.

\begin{figure}
  \centering
    \subfloat[]{\includegraphics[width=0.5\textwidth]{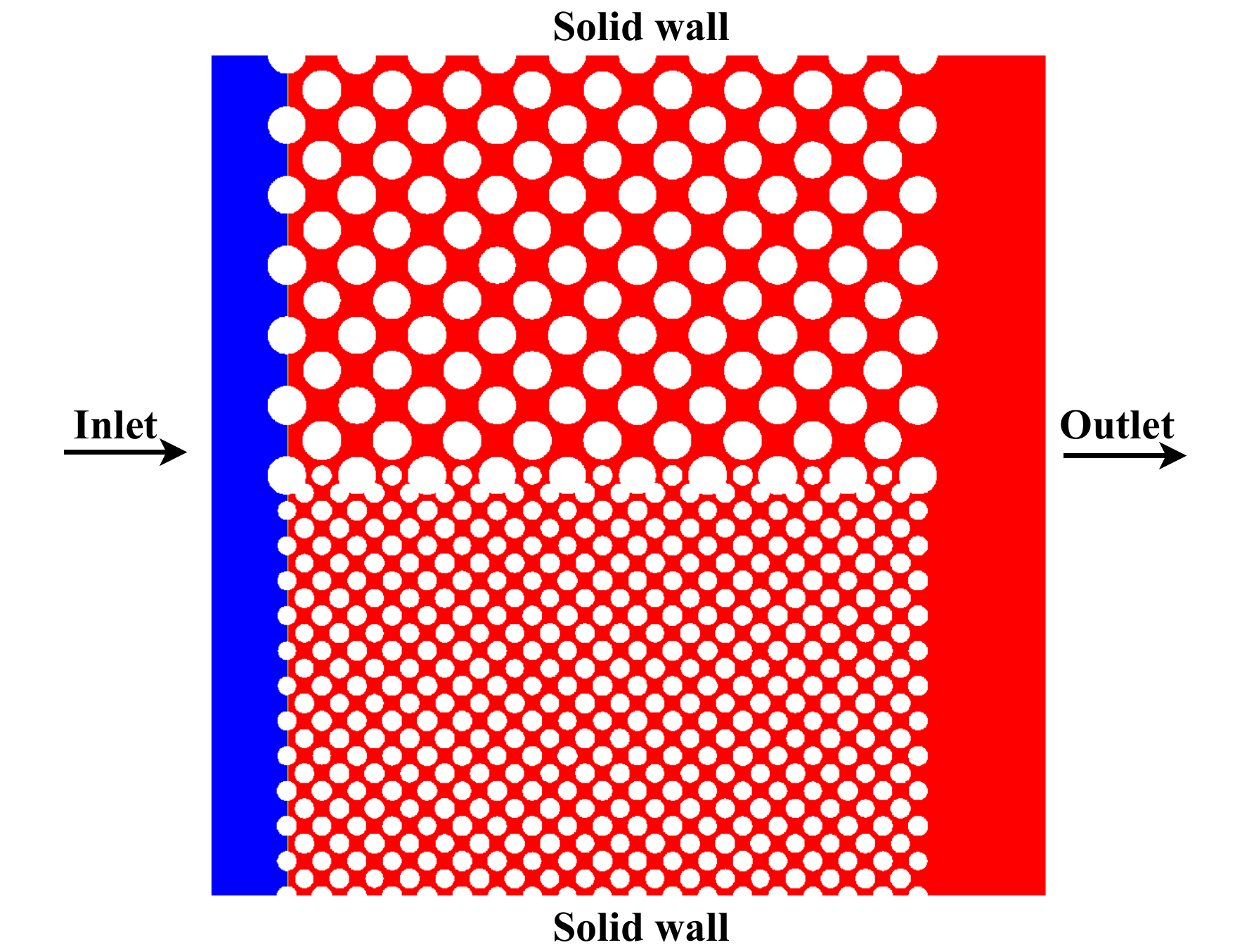}}~
    \subfloat[]{\includegraphics[width=0.3\textwidth]{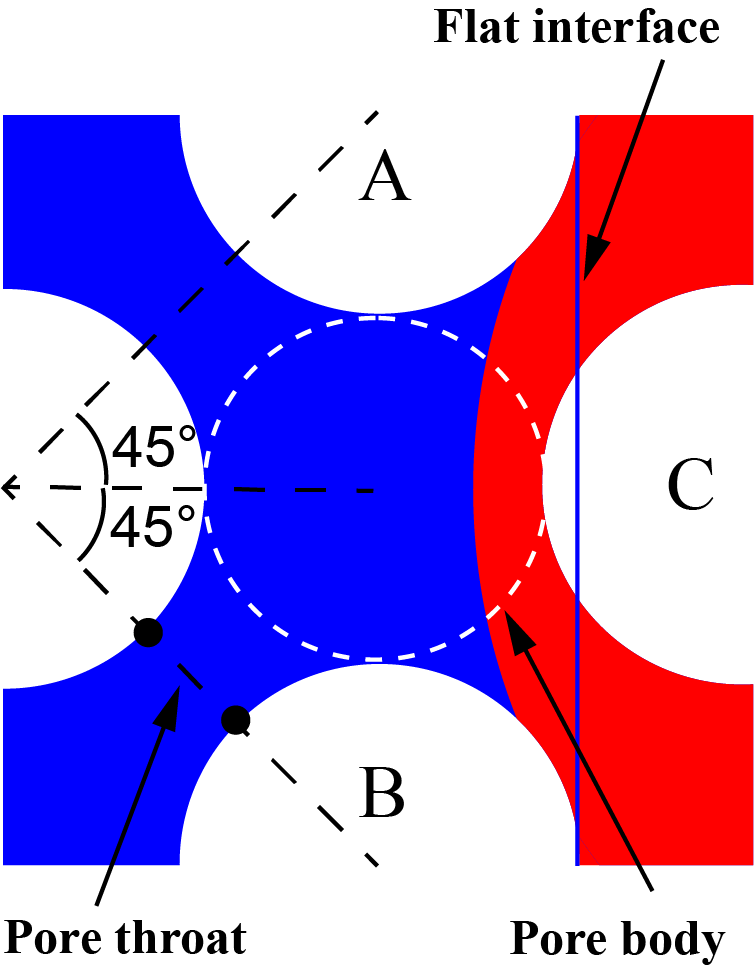}}
  \caption{(Colour online) (a) The initial fluid distribution and setup of the boundary conditions for the imbibition simulations in the dual-permeability porous geometry. The white circles represent the solid grains, while the blue and red regions represent the wetting and non-wetting fluids, respectively. The whole computational domain has a size of $1428\times 1441$ lattices, which consists of an inlet and an outlet section, connected by a pore network. (b) Representation for staggered array of circular grains in the pore network. A pore body is defined by the largest circle fitting locally the pore space. The size of a pore throat is defined by the narrowest width between two nearest solid grains.}
   \label{fig:dualPerm-geometry}
\end{figure}

As shown in figure~\ref{fig:dualPerm-geometry}(a), the porous media geometry used in this study consists of an inlet and an outlet section, connected by a pore network.
The pore network includes two distinct permeability zones with each occupying approximately a half width of the domain. Each homogeneous zone contains a staggered periodic array of uniform circular grains (see figure~\ref{fig:dualPerm-geometry}b). We run the simulations in a $1428\times 1441$ lattice domain, which corresponds to the physical size of $0.714\times 0.721$~$\si{\cm}^2$. The length of the pore network is $1078$ lattices. The diameter of solid grains is 64 lattices in the high permeability zone and 32 lattices in the low permeability zone. The diameter of pore bodies in the high (low) permeability zone is 56 (28) lattices, and the corresponding pore throat width is 20.8 (10.4) lattices. Both permeability zones have equal porosity of 0.55. Initially, the pore network is saturated with the non-wetting (red) fluid, and the wetting (blue) fluid is injected from the left inlet continuously with a constant velocity of $u_\text{in}$, and the outlet pressure is set to a constant. The top and bottom boundaries are no-slip walls. The densities of the two fluids are assumed to be equal since the displacement mainly occurs in the horizontal direction, where the effect of gravity can be negligible. Each simulation is run until the wetting fluid breaks through the right boundary of the pore network.

\begin{figure}
\centering
\subfloat[]{\includegraphics[width=0.33\textwidth]{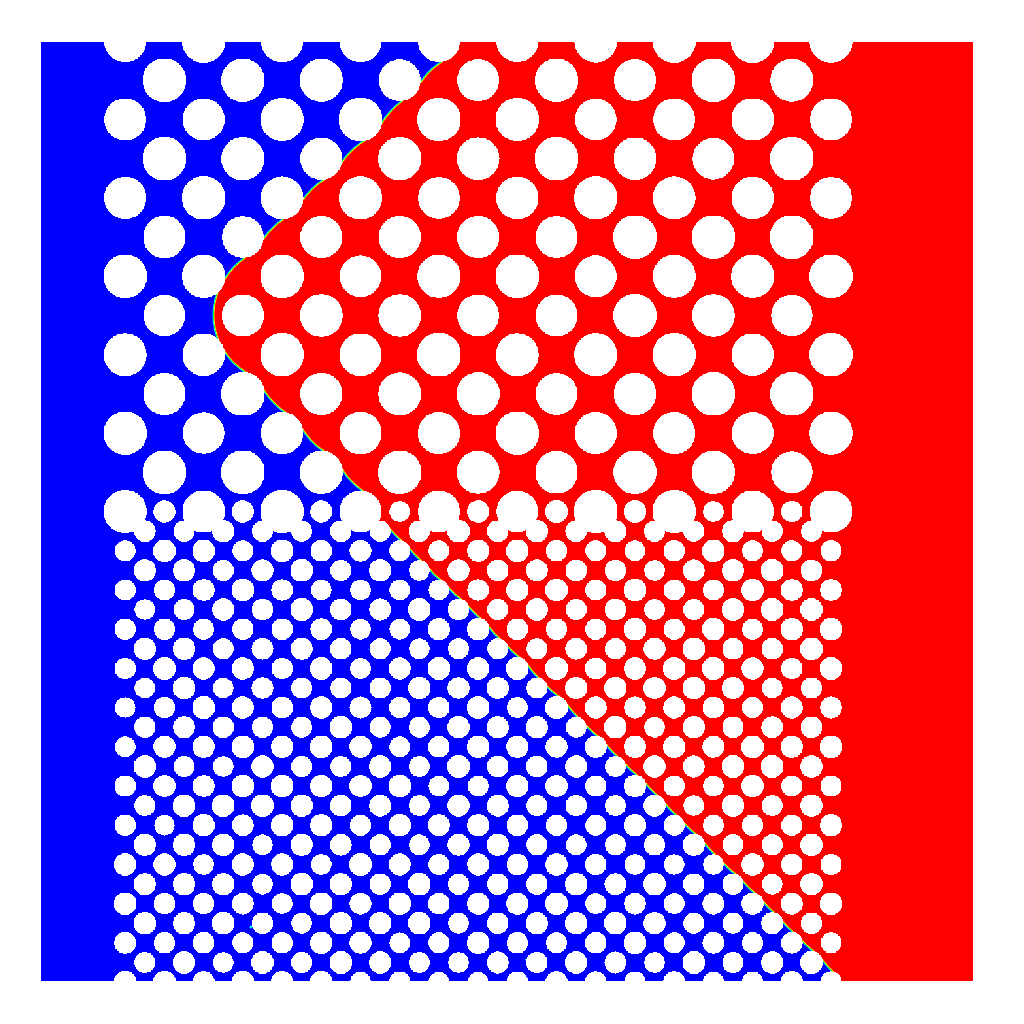}}~
\subfloat[]{\includegraphics[width=0.33\textwidth]{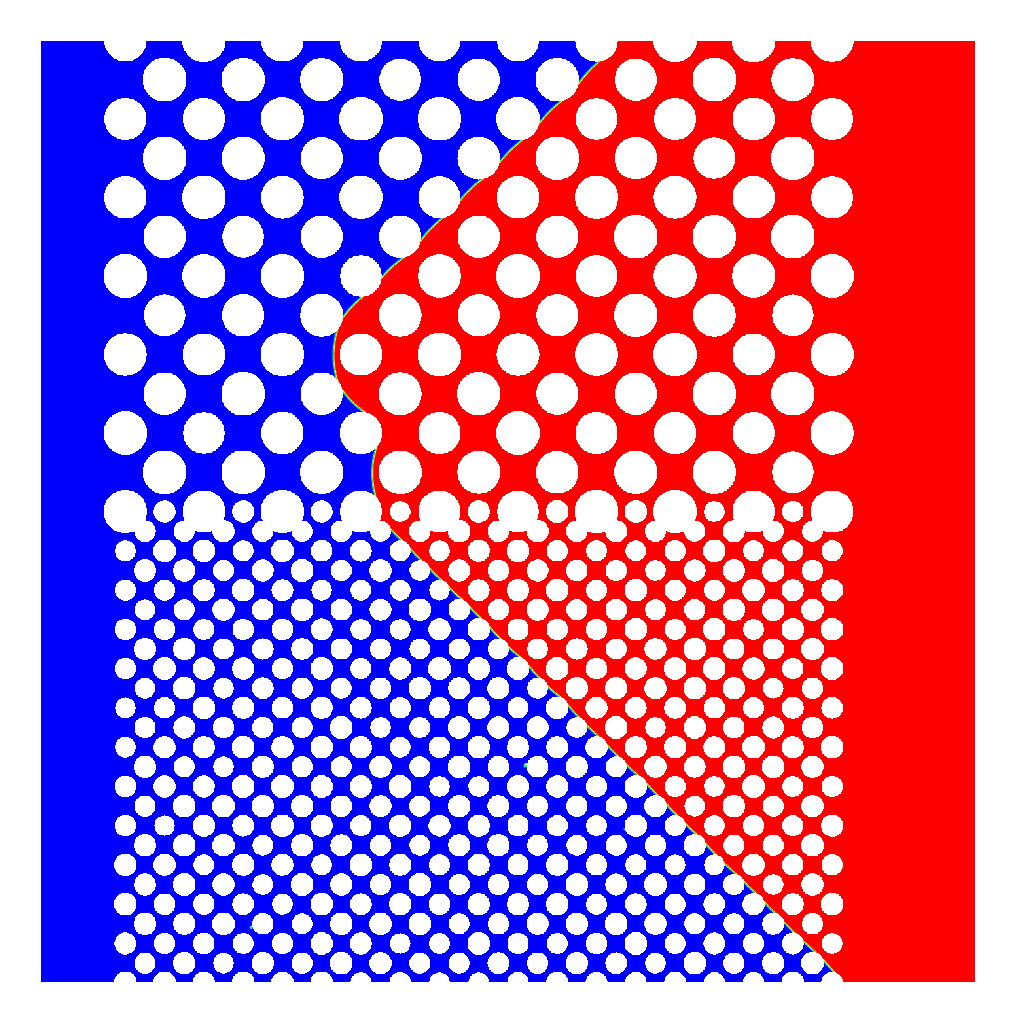}}~
\subfloat[]{\includegraphics[width=0.33\textwidth]{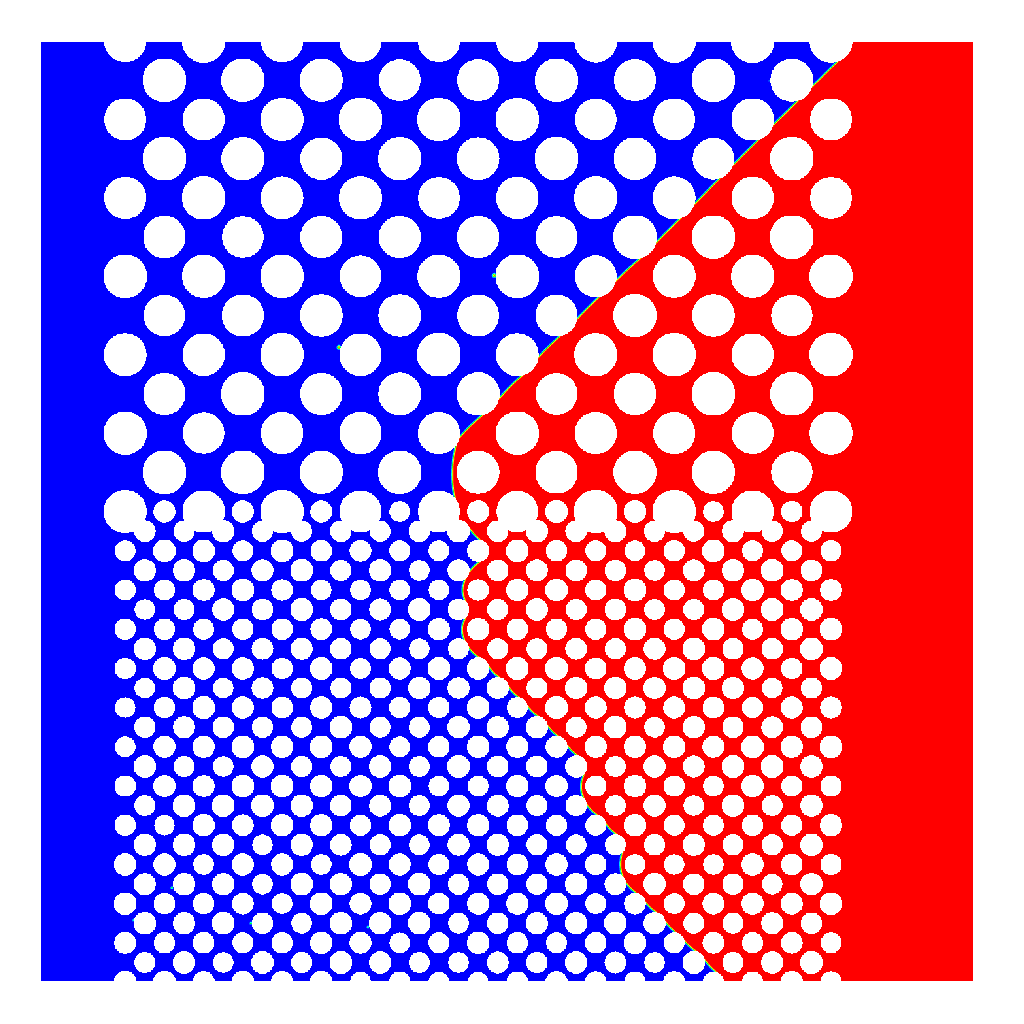}} \\
\subfloat[]{\includegraphics[width=0.33\textwidth]{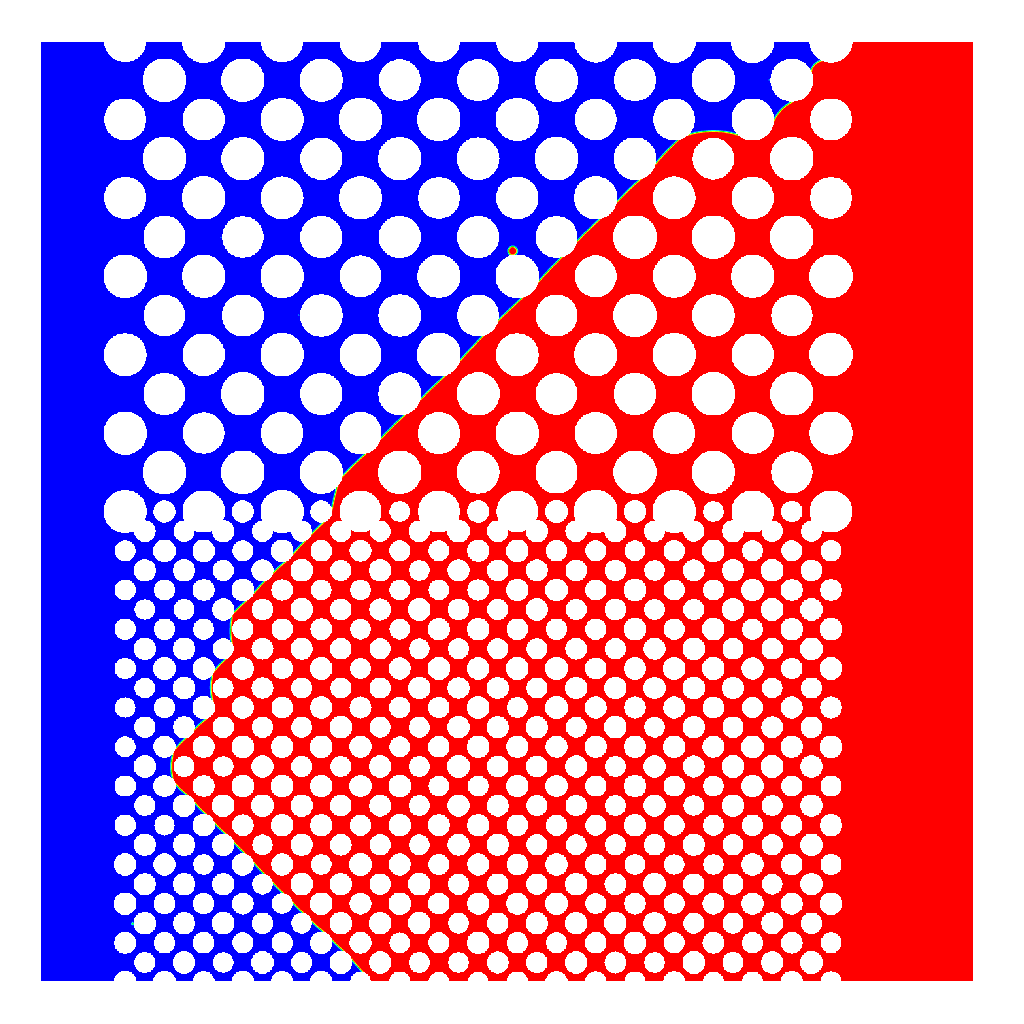}}~
\subfloat[]{\includegraphics[width=0.33\textwidth]{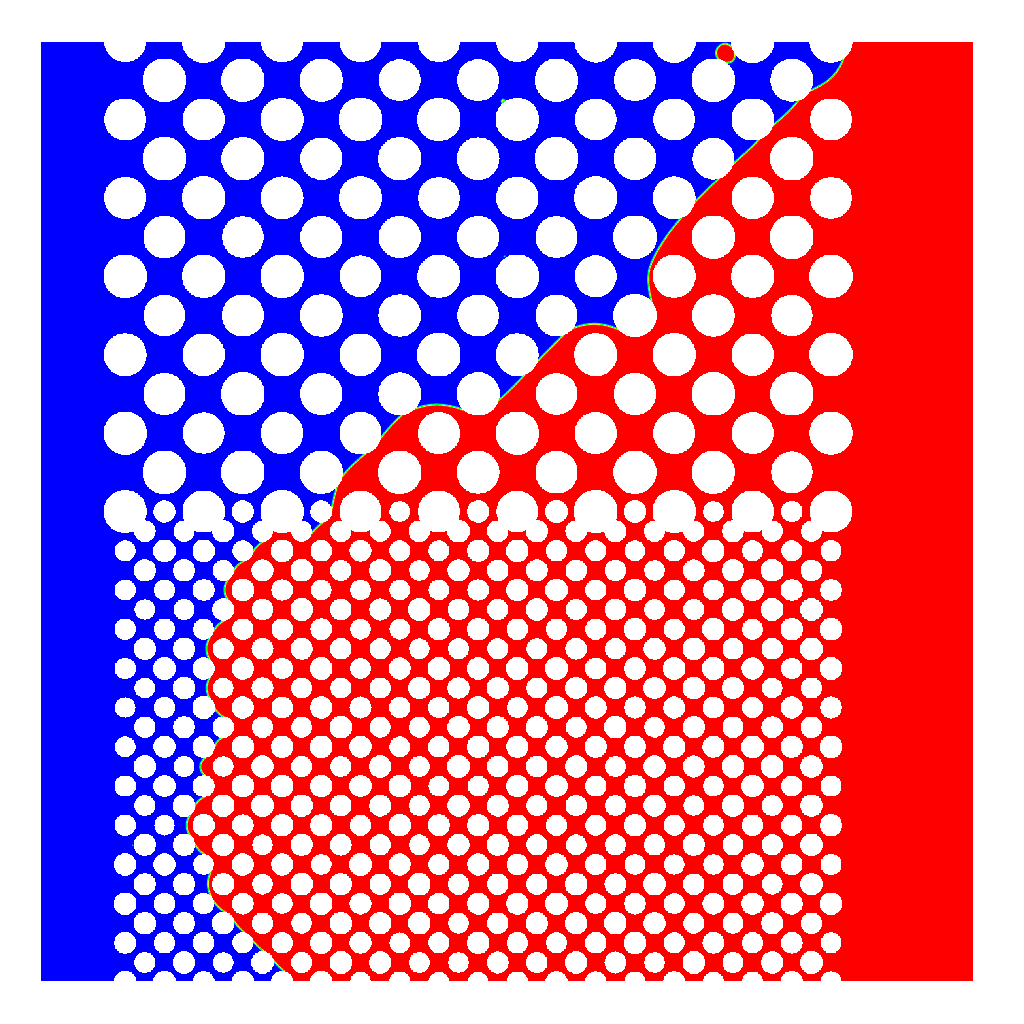}}~
\subfloat[]{\includegraphics[width=0.33\textwidth]{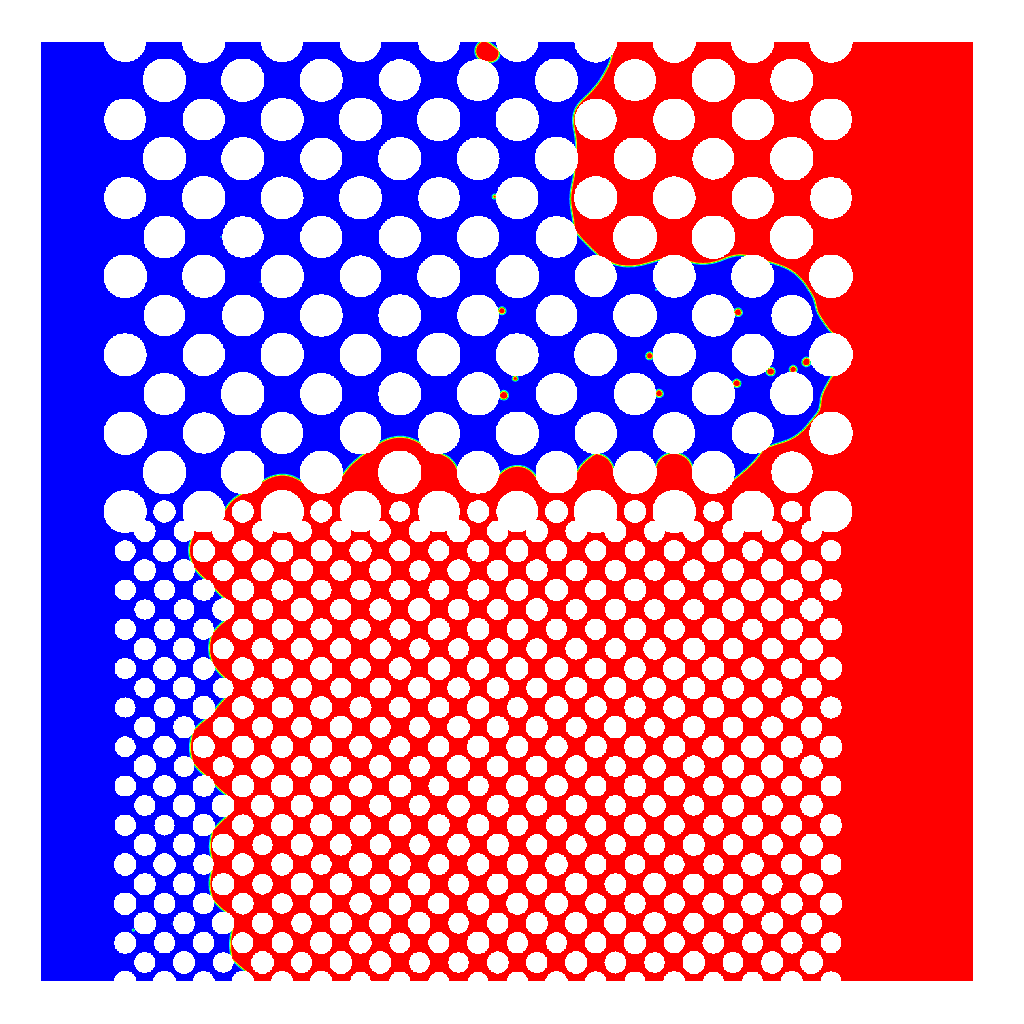}}
\caption{(Colour online) Fluid distributions in the dual-permeability pore network at breakthrough for: (a) $Ca_m=0.5944$, (b) $Ca_m=1.1888$, (c) $Ca_m=1.7832$, (d) $Ca_m=5.9441$, (e) $Ca_m=29.7206$ and (f) $Ca_m=59.4413$. The viscosity ratio of wetting to non-wetting fluids is fixed at 0.1. The non-wetting and wetting fluids are shown in red and blue, respectively.}\label{fig:imbibition-fluid-distributions-visratio01}
\end{figure}

We first consider the viscosity ratio of 0.1 for various values of $Ca_m$, where $Ca_m$ is defined by  $Ca_m=3\eta_n qL/2r_1^2\sigma$ with $r_1$ and $L$ taken as the average of pore body radius and half throat width (see figure~\ref{fig:dualPerm-geometry}b) in the low permeability zone and the length of the pore network. Figure~\ref{fig:imbibition-fluid-distributions-visratio01} shows the corresponding fluid distributions in the dual-permeability pore network at breakthrough. It is found that at low (high) values of $Ca_m$,  the wetting fluid prefers to invade the low (high) permeability zone and the breakthrough first occurs in the low (high) permeability zone, consistent with the previous observations in the pore doublet. In all cases, very few drops of the non-wetting fluid are trapped as the residual phase in the imbibition process as the wetting fluid progresses. For figure~\ref{fig:imbibition-fluid-distributions-visratio01}(a--e) ($Ca_m=0.5944\sim29.7206$), we notice an oblique advancing pattern of the wetting fluid in both high and low permeability zones, but this phenomenon disappears when $Ca_m$ is increased up to $Ca_m=59.4413$ (figure~\ref{fig:imbibition-fluid-distributions-visratio01}f). This suggests that the oblique advancing of the wetting fluid arises from the non-trivial interfacial tension.
\begin{figure}
\centering
\subfloat[]{\includegraphics[height=4.0cm]{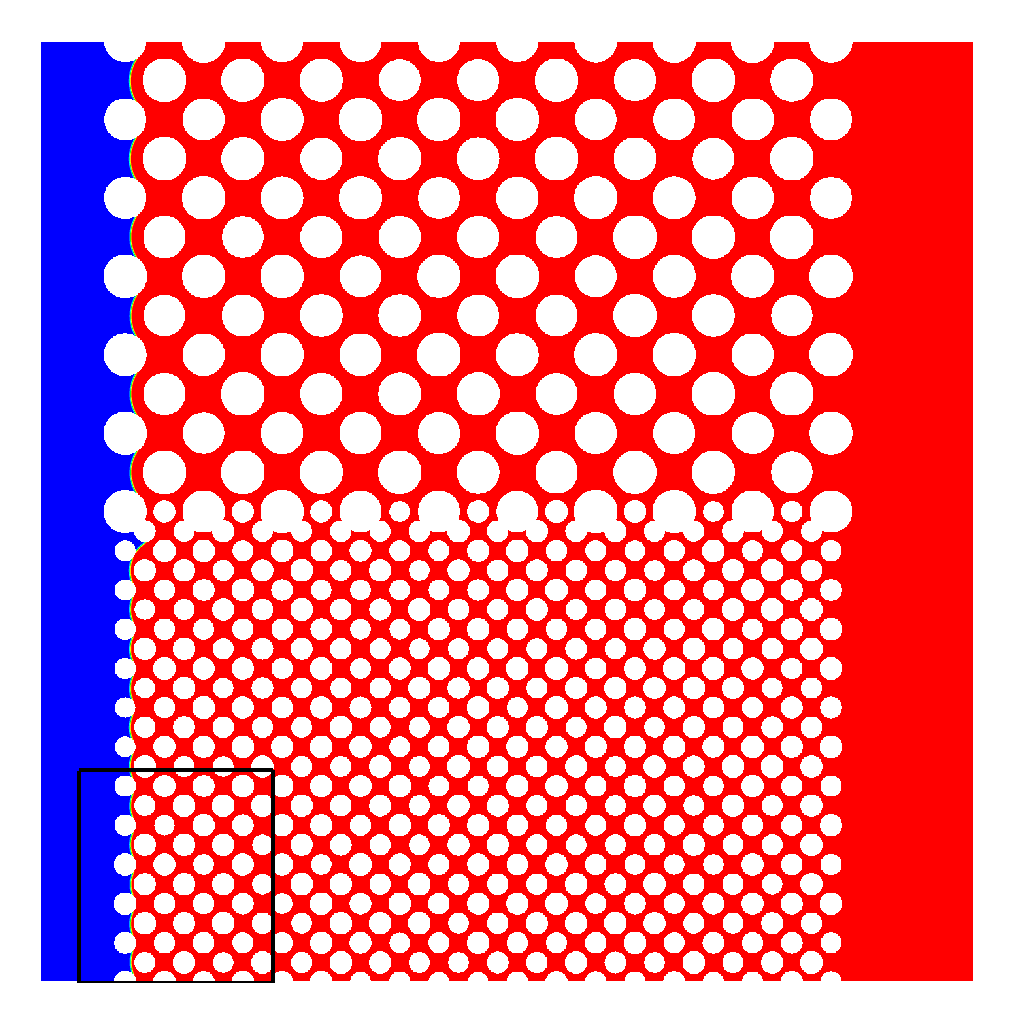}%
    \llap{\raisebox{1.1cm}{
      \includegraphics[height=2.8cm]{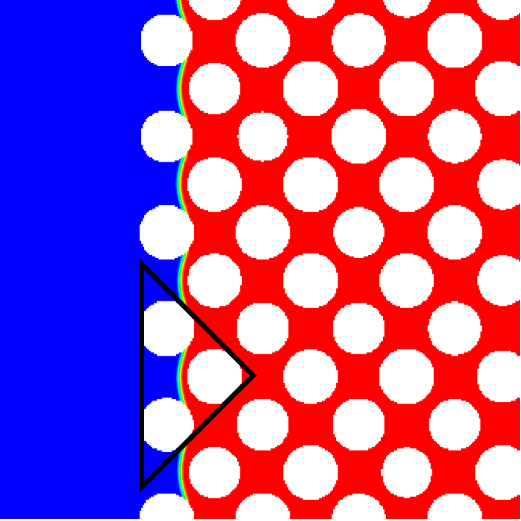}%
    }}
}~
\subfloat[]{\includegraphics[height=4.0cm]{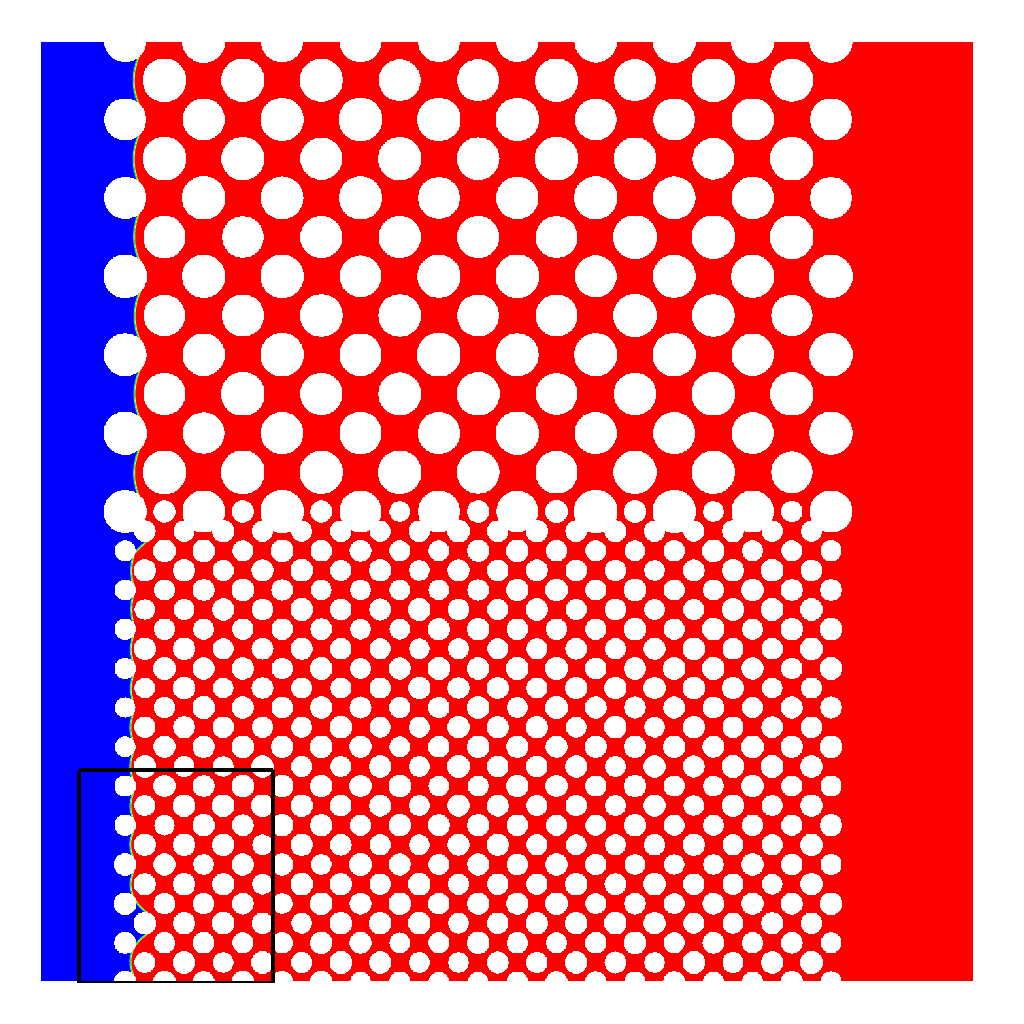}%
    \llap{\raisebox{1.1cm}{
      \includegraphics[height=2.8cm]{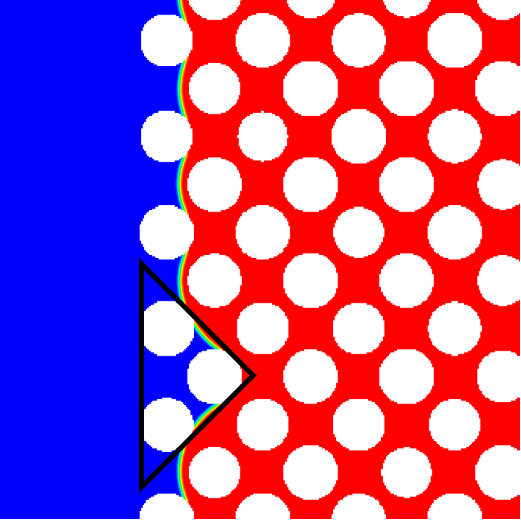}%
    }}
}~
\subfloat[]{\includegraphics[height=4.0cm]{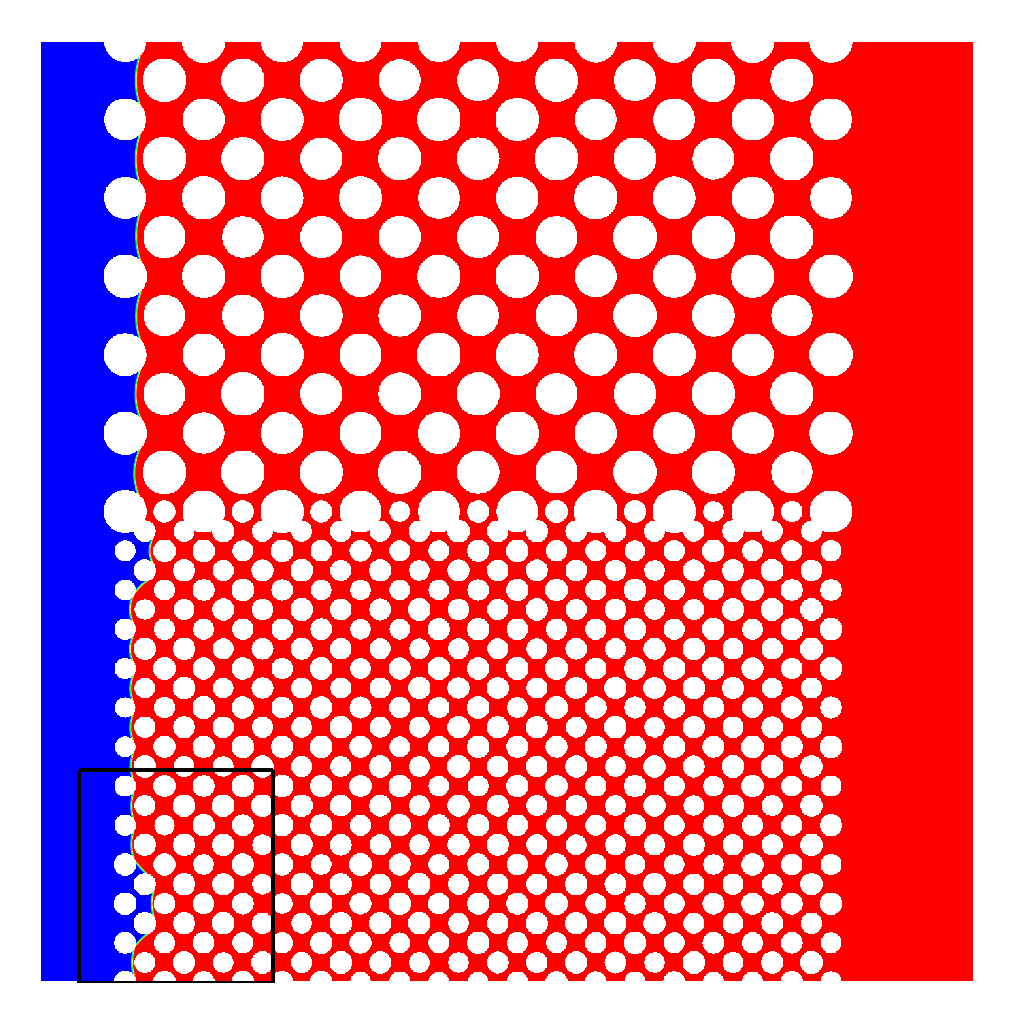}%
    \llap{\raisebox{1.1cm}{
      \includegraphics[height=2.8cm]{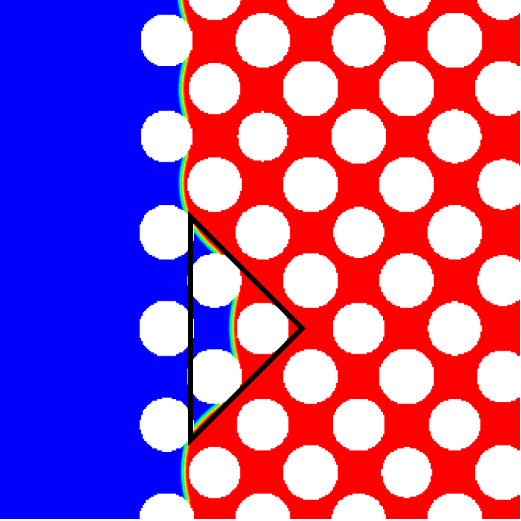}%
    }}
} \\
\subfloat[]{\includegraphics[height=4.0cm]{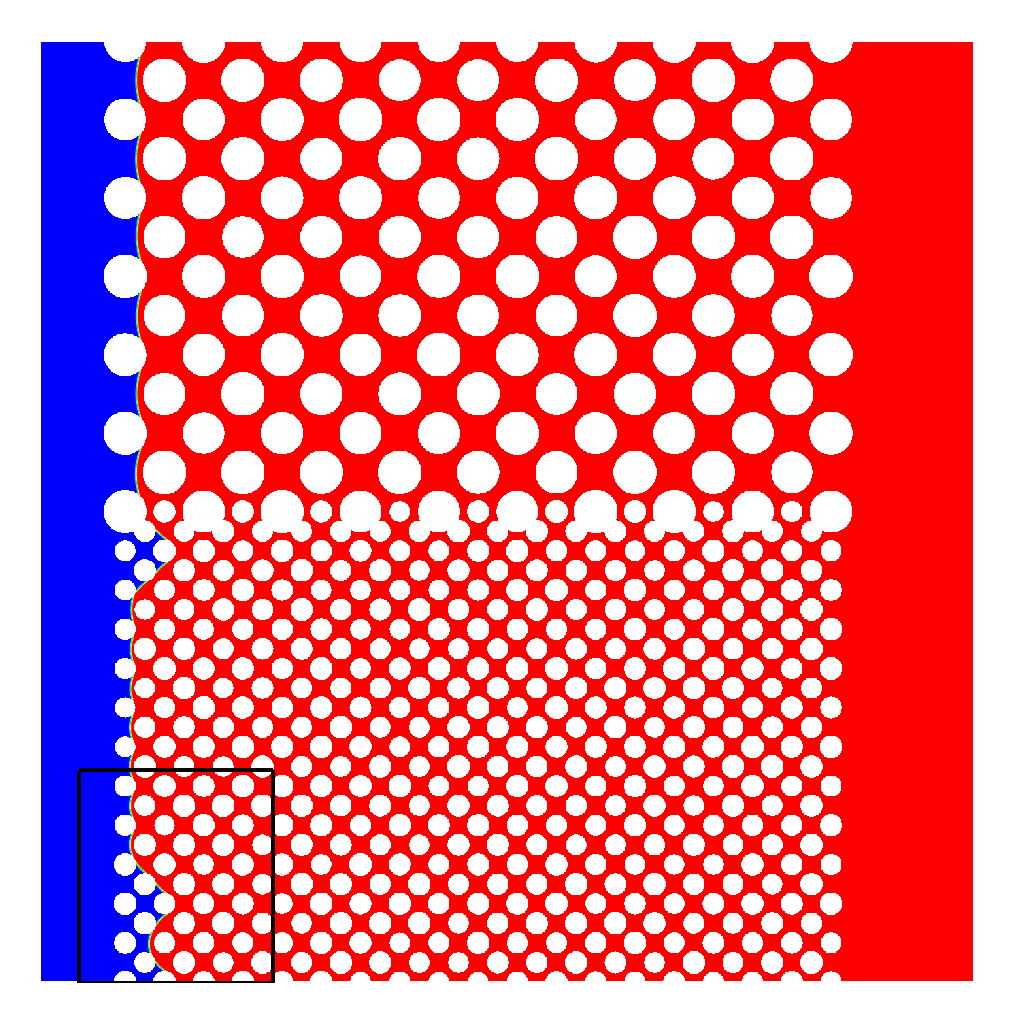}%
    \llap{\raisebox{1.1cm}{
      \includegraphics[height=2.8cm]{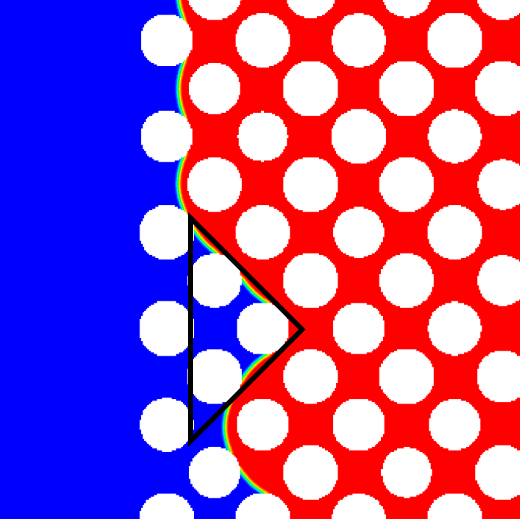}%
    }}
}~
\subfloat[]{\includegraphics[height=4.0cm]{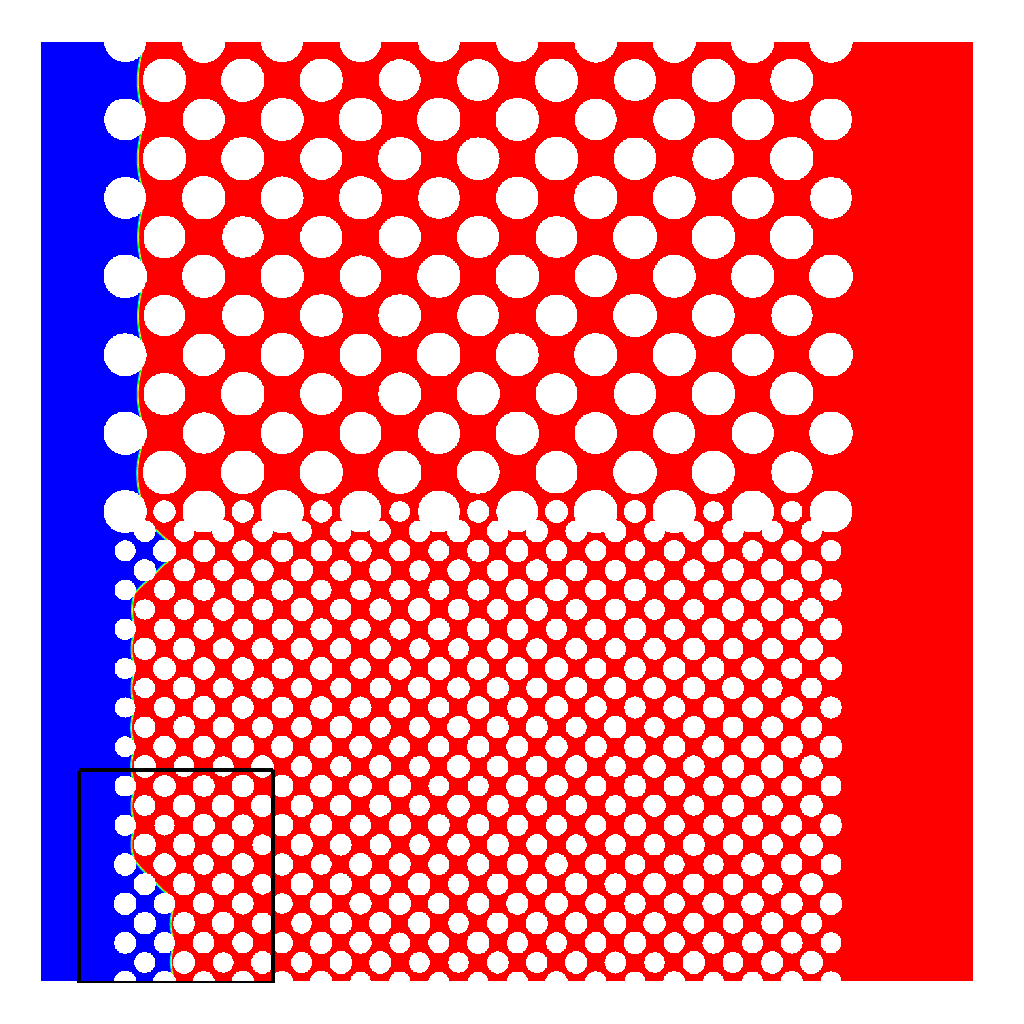}%
    \llap{\raisebox{1.1cm}{
      \includegraphics[height=2.8cm]{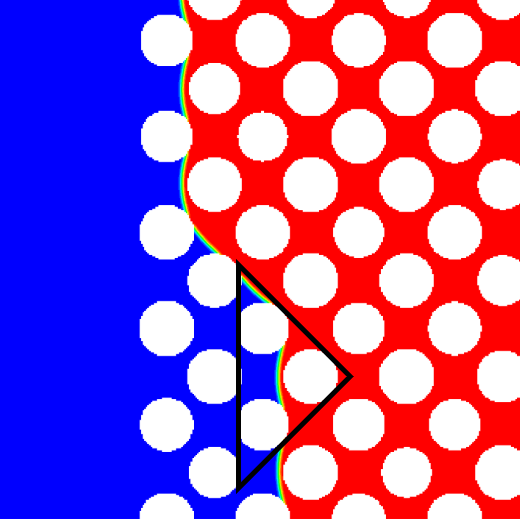}%
    }}
}~
\subfloat[]{\includegraphics[height=4.0cm]{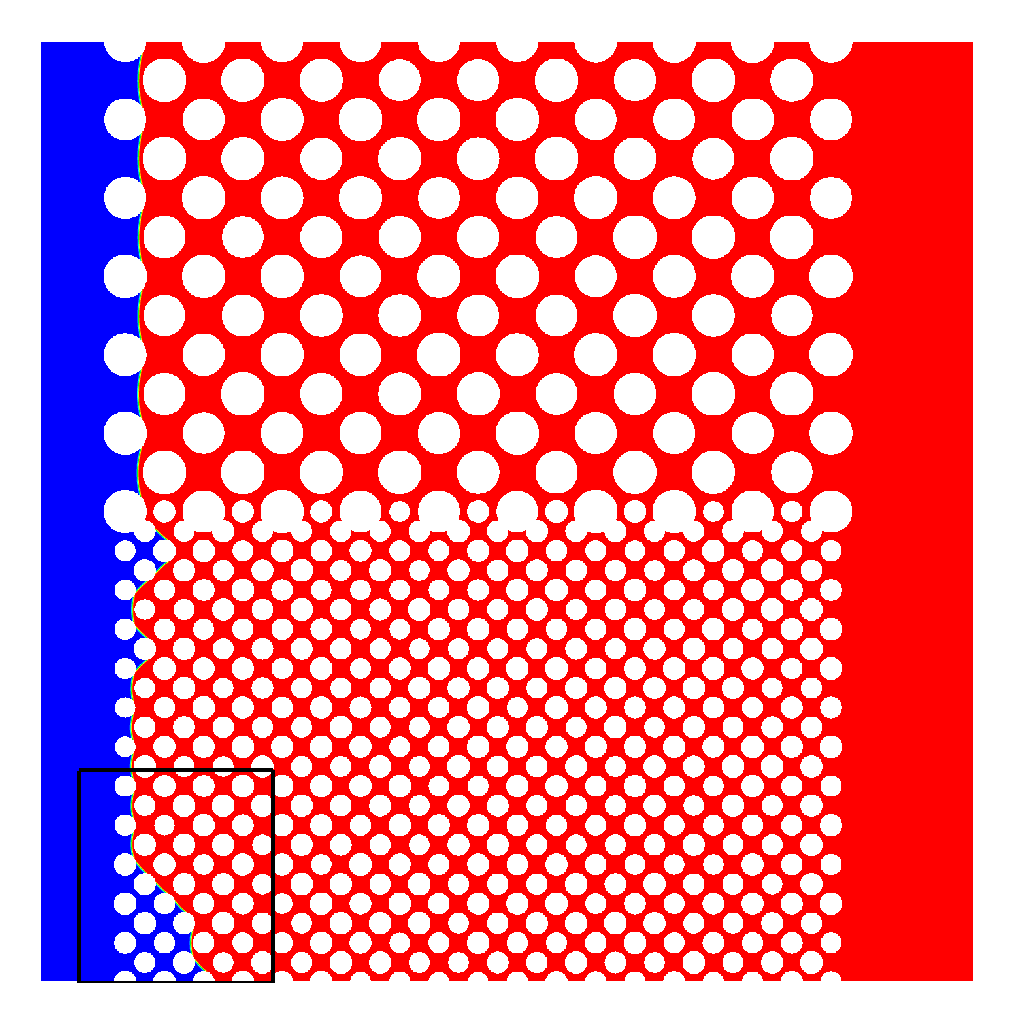}%
    \llap{\raisebox{1.1cm}{
      \includegraphics[height=2.8cm]{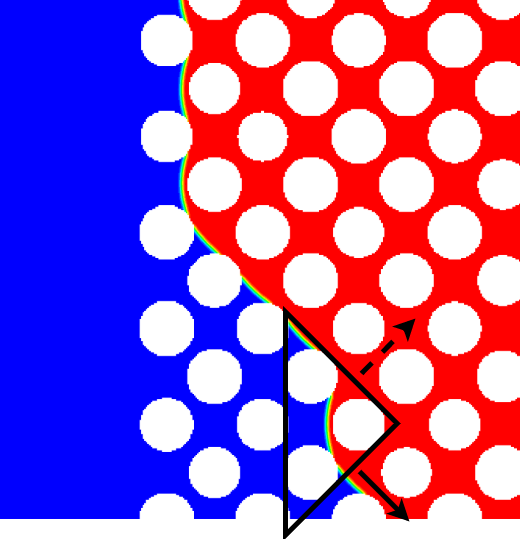}%
    }}
}
\caption{(Colour online) Fluid distributions in the porous media geometry for $Ca_m=0.5944$ and $\lambda=0.1$ at: (a) $\hat t/\hat{t}_B=0.0145$, (b) $\hat t/\hat{t}_B=0.0217$, (c) $\hat t/\hat{t}_B=0.0290$, (d) $\hat t/\hat{t}_B=0.0362$, (e) $\hat t/\hat{t}_B=0.0434$ and (f) $\hat t/\hat{t}_B=0.0507$. The insets are to show the close-up view of the region indicated by the black rectangle box in the lower left corner.}\label{fig:imbibition-oblique-advance-combined}
\end{figure}


\begin{table}
  \begin{center}
\def~{\hphantom{0}}
  \begin{tabular}{lcccc}
      $\lambda$   & ~$Ca_{m}$ ~& $~S_{1}~$   &  ~ $S_{2}$~ & ~$S_w$~ \\[3pt]
       0.02  & $\num{0.5944}$ & 0.7051 & 0.2428 & 0.4651\\
       0.02  & $\num{1.1888}$ & 0.7061 & 0.5267 & 0.5990\\
       0.02  & $\num{1.9814}$ & 0.3357 & 0.7126 & 0.5141\\
       0.02   & $\num{2.9721}$ & 0.2243 & 0.6367 & 0.4192\\
       0.025 & $\num{0.4755}$  & 0.7050 & 0.2427 & 0.4650\\
       0.025 & $\num{1.1888}$ & 0.7401 & 0.5405 & 0.6276\\
       0.025 & $\num{2.3777}$ & 0.2784 & 0.6378 & 0.4458\\
       0.1   & $\num{0.5944}$ & 0.7049 & 0.2415 & 0.4643\\
       0.1   & $\num{1.1888}$ & 0.7053 & 0.4273 & 0.5521\\
       0.1   & $\num{1.7832}$ & 0.6230 & 0.7191 & 0.6572\\
       0.1   & $\num{5.9441}$ & 0.1643 & 0.6261 & 0.3862\\
       0.1   & $\num{29.7206}~$ & 0.1288 & 0.7124 & 0.4086\\
       0.1   & $\num{59.4413}~$ & 0.1229 & 0.7791 & 0.4290\\
       0.25   & $\num{0.4755}$ & 0.7043 & 0.2421 & 0.4644\\
       0.25   & $\num{0.7133}$ & 0.7048 & 0.2847 & 0.4844\\
       0.25   & $\num{1.1888}$ & 0.7060 & 0.5784 & 0.6229\\
       0.25   & $\num{2.3777}$ & 0.5192 & 0.7163 & 0.6075\\
       1.0   &$\num{0.1783}$& 0.7046 & 0.2641 & 0.4746\\
       1.0   & $\num{0.5944}$ & 0.7048 & 0.3980 & 0.5383\\
       1.0   & $\num{1.1888}$ & 0.6846 & 0.7307 & 0.6984\\
       1.0   & $\num{2.9721}$ & 0.4038 & 0.7551 & 0.5734\\
       5.0   &$\num{0.0357}$ & 0.7043 & 0.2417 & 0.4641\\
       5.0   &$\num{0.1189}$ & 0.8228 & 0.5764 & 0.6904\\
       5.0   &$\num{0.2378}$ & 0.8448 & 0.8259 & 0.8253\\
       5.0   & $\num{0.3566}$ & 0.7068 & 0.7589 & 0.7281\\
       5.0   & $\num{0.5944}$ & 0.6977 & 0.8293 & 0.7578\\
       20   &~$\num{0.02972}$ & 0.7023 & 0.3172 & 0.4983\\
       20   &~$\num{0.05944}$ & 0.8886 & 0.7761 & 0.8216\\
       20   &~$\num{0.08916}$ & 0.8455 & 0.8645 & 0.8504\\
       20   &$\num{0.1486}$ & 0.7410 & 0.8785 & 0.8083\\
       50   &~$\num{0.01189}$ & 0.7037 & 0.3447 & 0.5117\\
       50   &~$\num{0.02378}$ & 0.8694 & 0.7959 & 0.8218\\
       50   &~$\num{0.03566}$ & 0.8456 & 0.8645 & 0.8505\\
       50   &~$\num{0.05944}$ & 0.7329 & 0.8602 & 0.7951\\
  \end{tabular}
  \caption{Saturations $S_1$, $S_2$ and $S_w$ at breakthrough for various values of viscosity ratio ($\lambda$) and capillary number ($Ca_m$), where $S_{1}$ and $S_{2}$ are the wetting fluid saturations in the low and high permeability zone, and $S_w$ is the wetting fluid saturation in the whole pore network.}
  \label{tab:Sw-L-H-S-Ca-m-lambda}
  \end{center}
\end{table}

 \begin{figure}
    \centering
    \begin{overpic}[scale=.6]{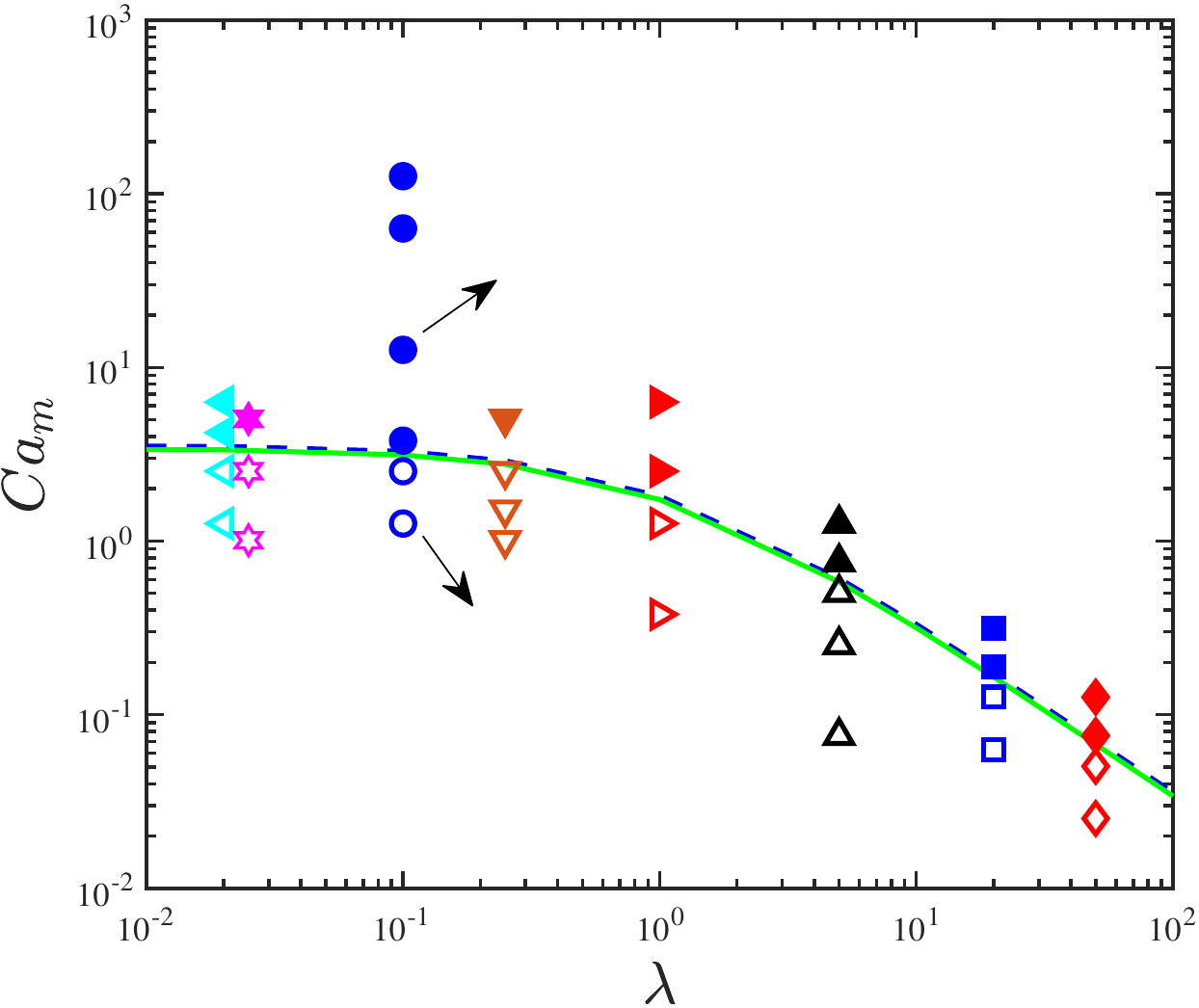}
    \put(25.5,23){\color{black} (\rom{1})}
    \put(31,14){\includegraphics[scale=.04]{figures/imbibition_Ca1E_5.png}}
    \put(39,69){\color{black} (\rom{2})}
    \put(46,58){\includegraphics[scale=.04]{figures/imbibition_Ca1E_4.png}}
    \end{overpic}
	\caption{(Colour online) The $\lambda- Ca_m$ diagram showing preferential imbibition in a dual-permeability pore network. The open symbols represent the cases where $S_{1}>S_{2}$ at breakthrough (\rom{1}), while the filled symbols represent the cases where $S_{1}<S_{2}$ at breakthrough (\rom{2}). Two images of fluid distributions are inserted to show the regions \rom{1} and \rom{2}. The green solid line represents the $Ca_{m,c}$ curve, on which $S_1=S_2$ at breakthrough. The $Ca_{m,c}$ curve (represented by the blue dashed lines) from the pore doublet model is also plotted for comparison.}
	\label{fig:new-phase-diagram-imbibition}
\end{figure}

\begin{figure}
 \centering
    \subfloat[]{\includegraphics[width=0.4\textwidth]{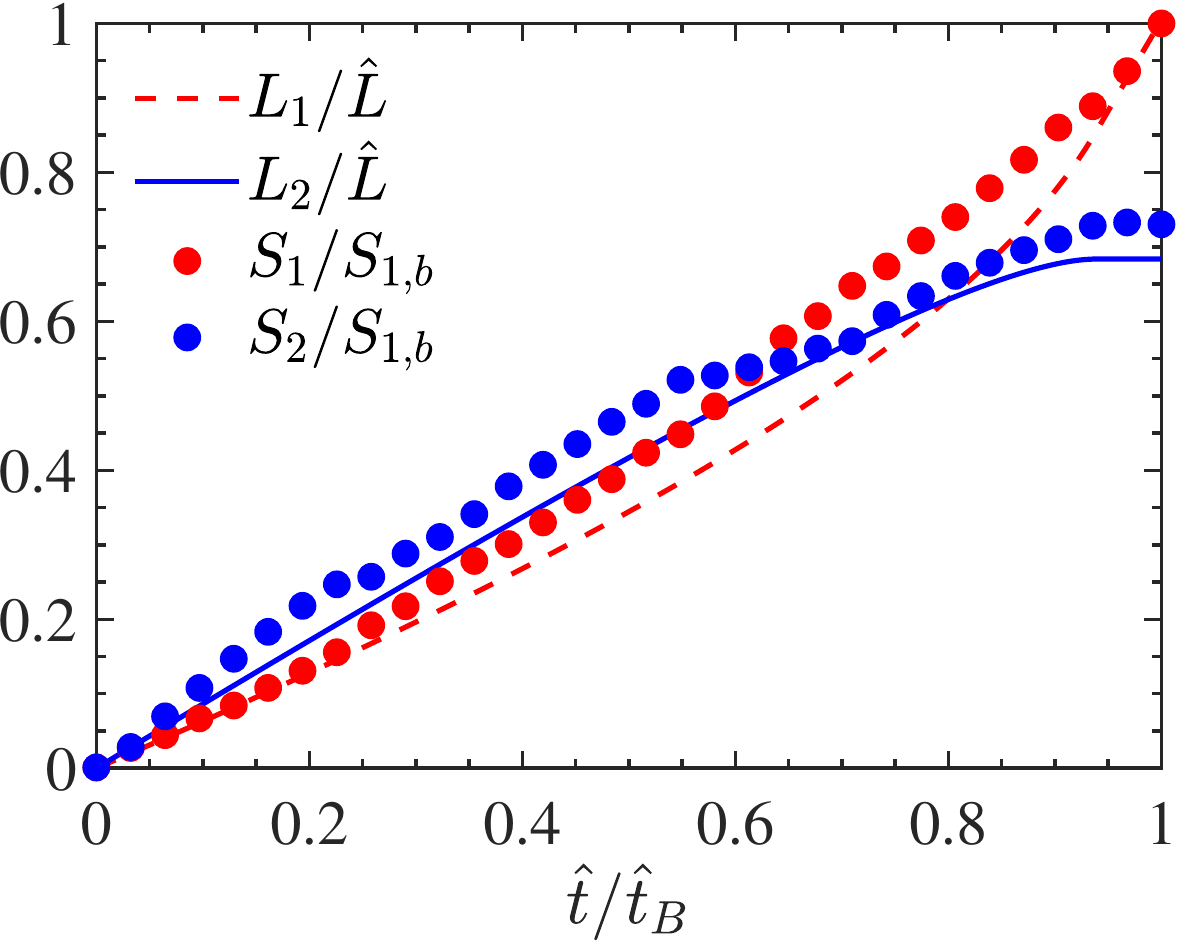}}~
    \subfloat[]{\includegraphics[width=0.4\textwidth]{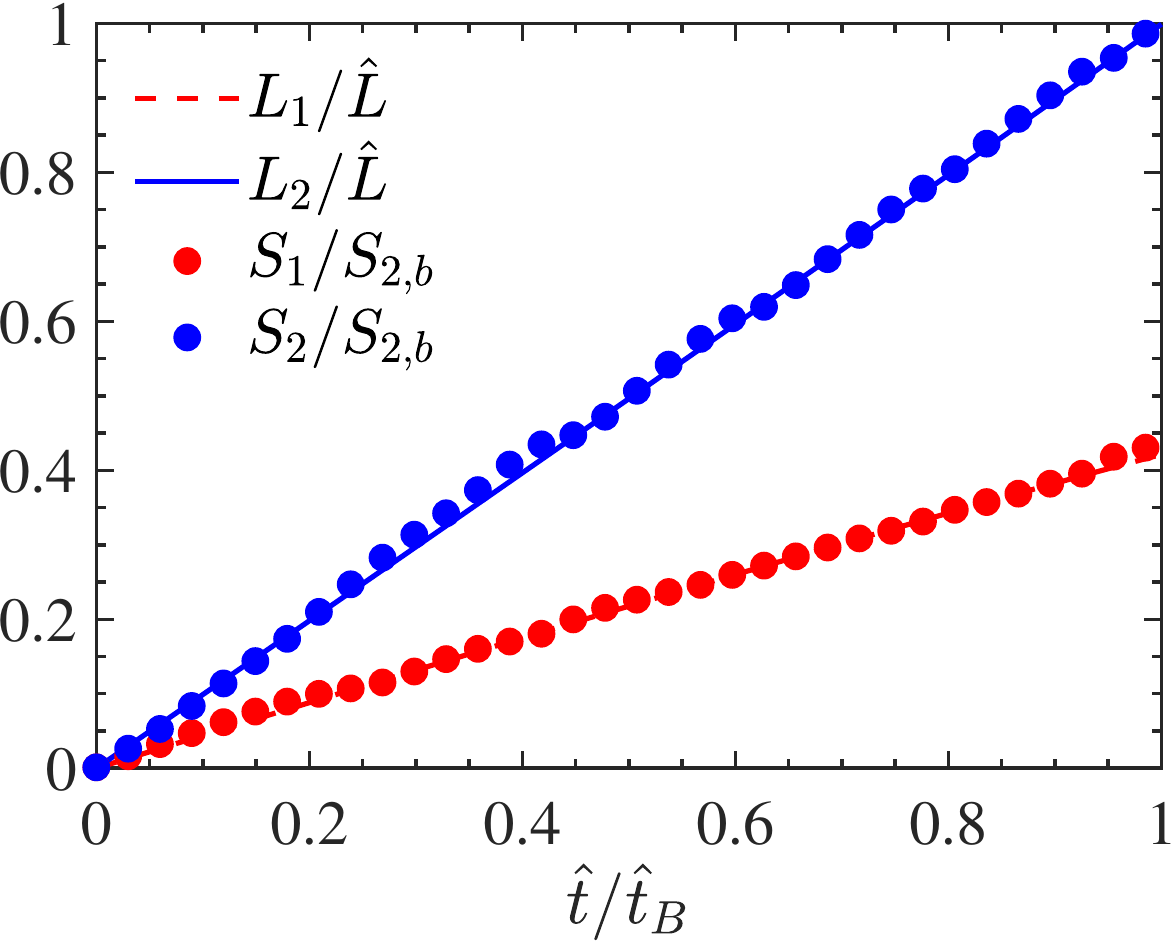}}\\
    \subfloat[]{\includegraphics[width=0.4\textwidth]{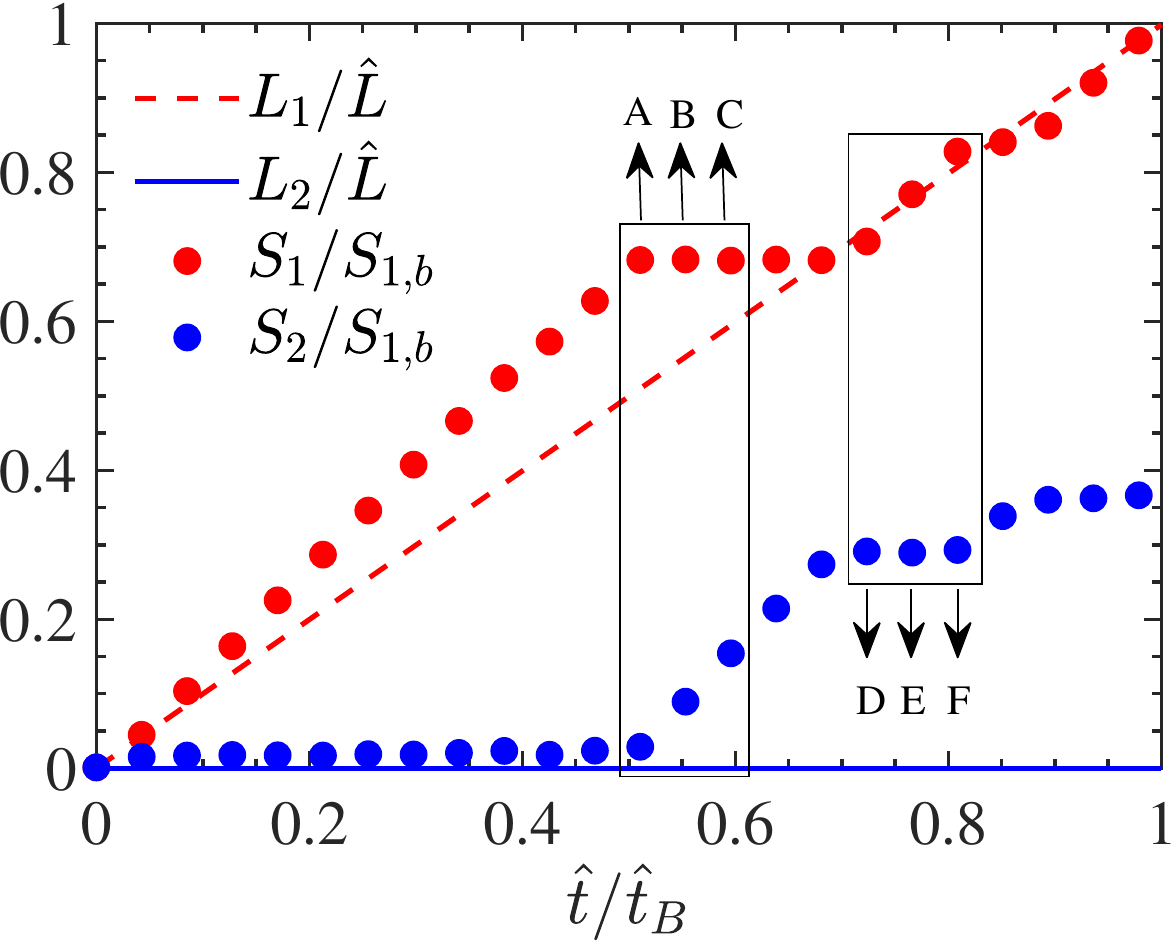}}~
    \subfloat[]{\includegraphics[width=0.4\textwidth]{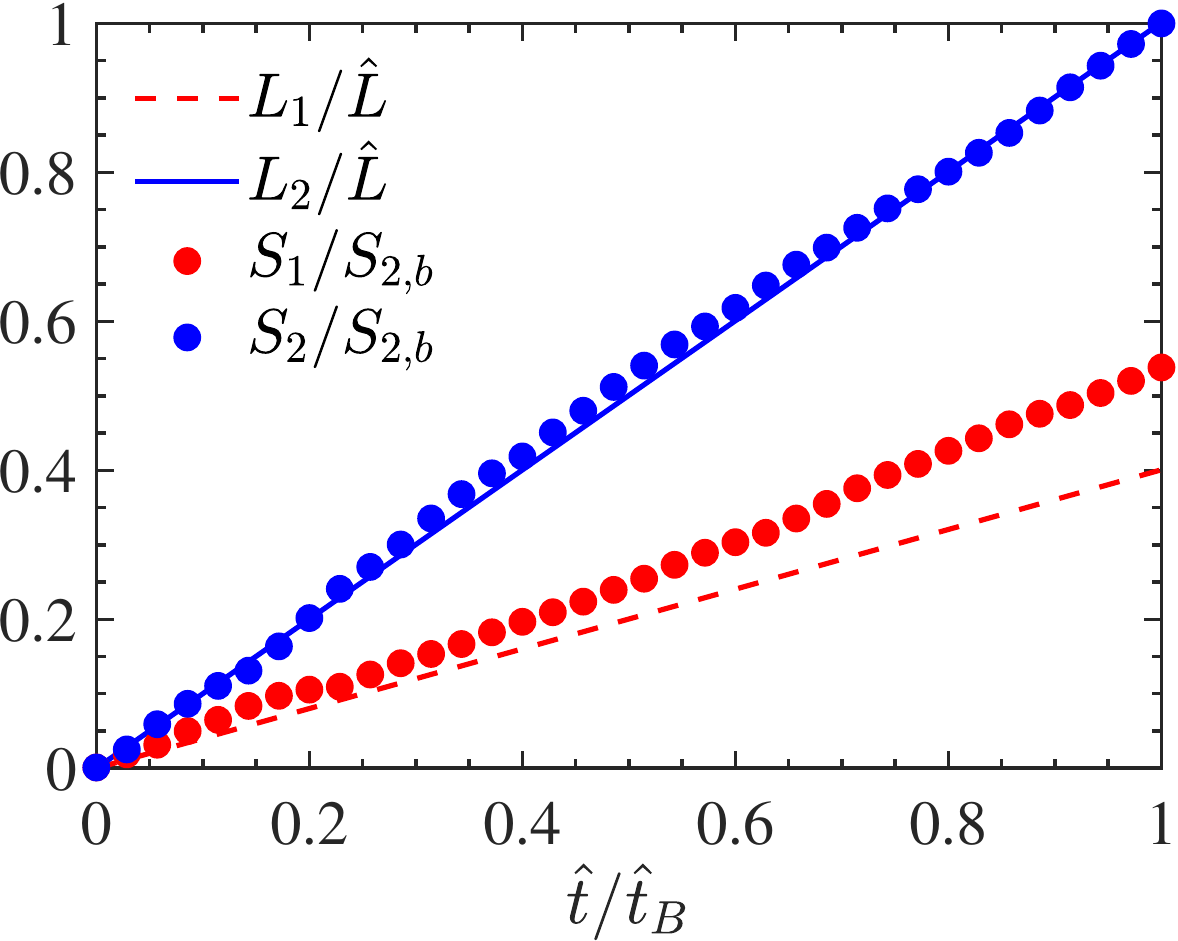}}\\
    \subfloat[]{\includegraphics[width=0.4\textwidth]{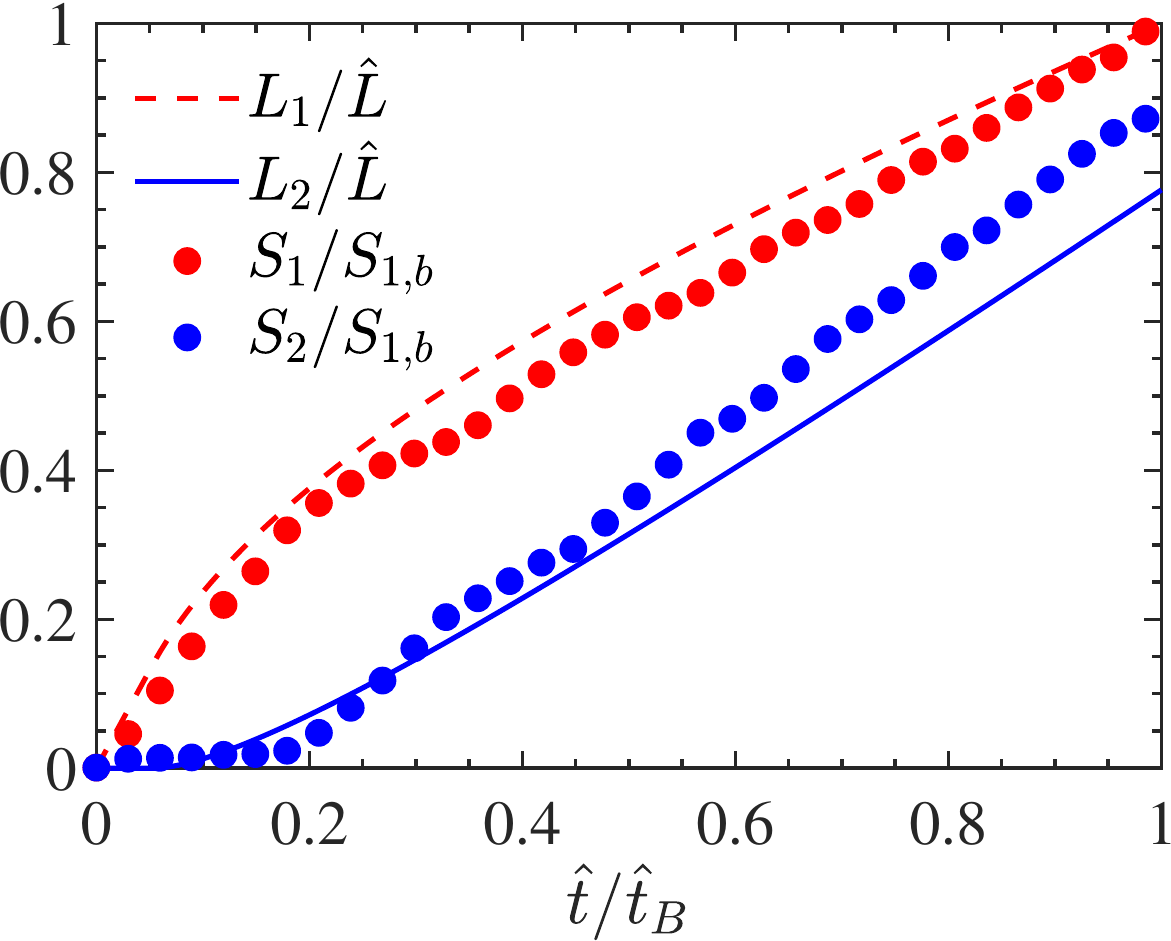}}~
    \subfloat[]{\includegraphics[width=0.4\textwidth]{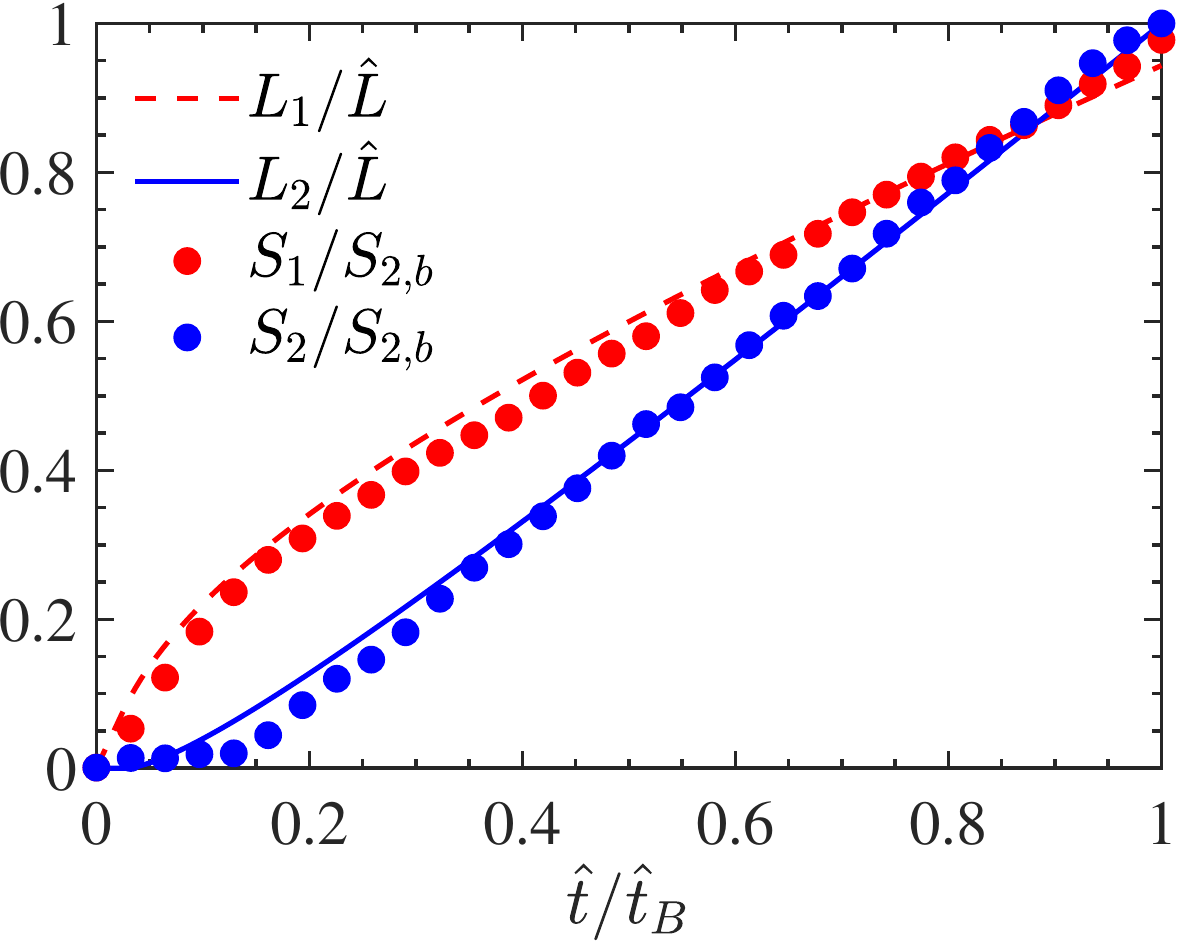}}\\
\caption{(Colour online) The saturations in the low and high permeability regions (normalised by their maximum value at breakthrough) as a function of time in the dual-permeability pore network at $\lambda=0.025$ for (a) $Ca_m=\num{1.1888}$ and (b) $Ca_m=\num{2.3777}$; at $\lambda=1$ for (c) $Ca_m=\num{0.1783}$ and (d) $Ca_m=\num{2.9721}$; at $\lambda=20.0$ for (e) $Ca_m=\num{0.05944}$ and (f) $Ca_m=\num{0.08916}$. The semi-analytical solutions from the pore doublet model at the same values of $\lambda$ and $Ca_m$ are also shown for the comparison.}
   \label{fig:Poredoublet-Dualper-lambda-0025-1-20-combine}
\end{figure}

To better understand the oblique advancing pattern, as an example, we plot the evolution of fluid distributions during the early imbibition at $Ca_m=0.5944$, which is shown in  figure~\ref{fig:imbibition-oblique-advance-combined}. It is known that the dominant capillary pressure is larger in the smaller pores and throats according to the Young-Laplace equation, so the smallest pores and throats are filled first. For the present grain arrangement, let us take a close look at the interface between two vertically aligned solid grains A and B, as shown in figure~\ref{fig:dualPerm-geometry}(b). It is seen that a flat interface (represented by the blue solid line) with zero capillary pressure is able to touch the solid grain C, and thus the advancing meniscus of the wetting fluid always progresses towards the next column of grains through a triangle shape, as marked by the black triangles in figure~\ref{fig:imbibition-oblique-advance-combined}. In addition, as shown in figure~\ref{fig:imbibition-oblique-advance-combined}(f), as the wetting fluid invades the region marked by the black triangle, it cannot infiltrate in the direction highlighted by the dashed arrow due to the requirement of a positive pressure difference between the wetting and non-wetting fluids to overcome the capillary valve resistance~\citep{xuLatticeBoltzmannSimulation2017}, but progress toward the direction highlighted by the solid arrow due to the merging with the neighbouring interface. As a result, the wetting fluid penetrates layer by layer along the direction pointed by the solid arrow, forming an oblique advancing pattern. A similar process occurs in the high permeability zone, but in a direction perpendicular to the invading direction in the low permeability zone. On the other hand, at the highest $Ca_m$ in figure~\ref{fig:imbibition-fluid-distributions-visratio01}(f), the aforementioned pore filling order is disrupted and no longer applicable, as here the viscous force dominates the imbibition behaviour.

We then study the effect of viscosity ratio on the imbibition preference. A wide range of viscosity ratios, varying from $\lambda=0.02$ to 50.0, is considered. For each viscosity ratio, at least three different values of $Ca_m$ are simulated, covering three typical patterns observed at breakthrough. The saturation data at breakthrough for various viscosity ratios and capillary numbers are listed in table~\ref{tab:Sw-L-H-S-Ca-m-lambda}, where $S_1$, $S_2$ and $S_w$ are the wetting fluid saturations in the low permeability zone, the high permeability zone and the entire pore network. Among all the cases considered, the maximum imbibition efficiency is obtained under the conditions of $\lambda=50$ and $Ca_m=0.03566$, where the wetting fluid saturations in both permeability zones are roughly the same (the corresponding values $S_1=0.8456$ and $S_2=0.8645$). In addition, for each viscosity ratio, the highest imbibition efficiency is always achieved when $S_1$ is closest to $S_2$. This implies that the critical capillary numbers $Ca_{m,c}$ are the optimal condition to improve the imbibition efficiency.

\begin{figure}
 \centering
    \subfloat[]{\includegraphics[width=0.33\textwidth]{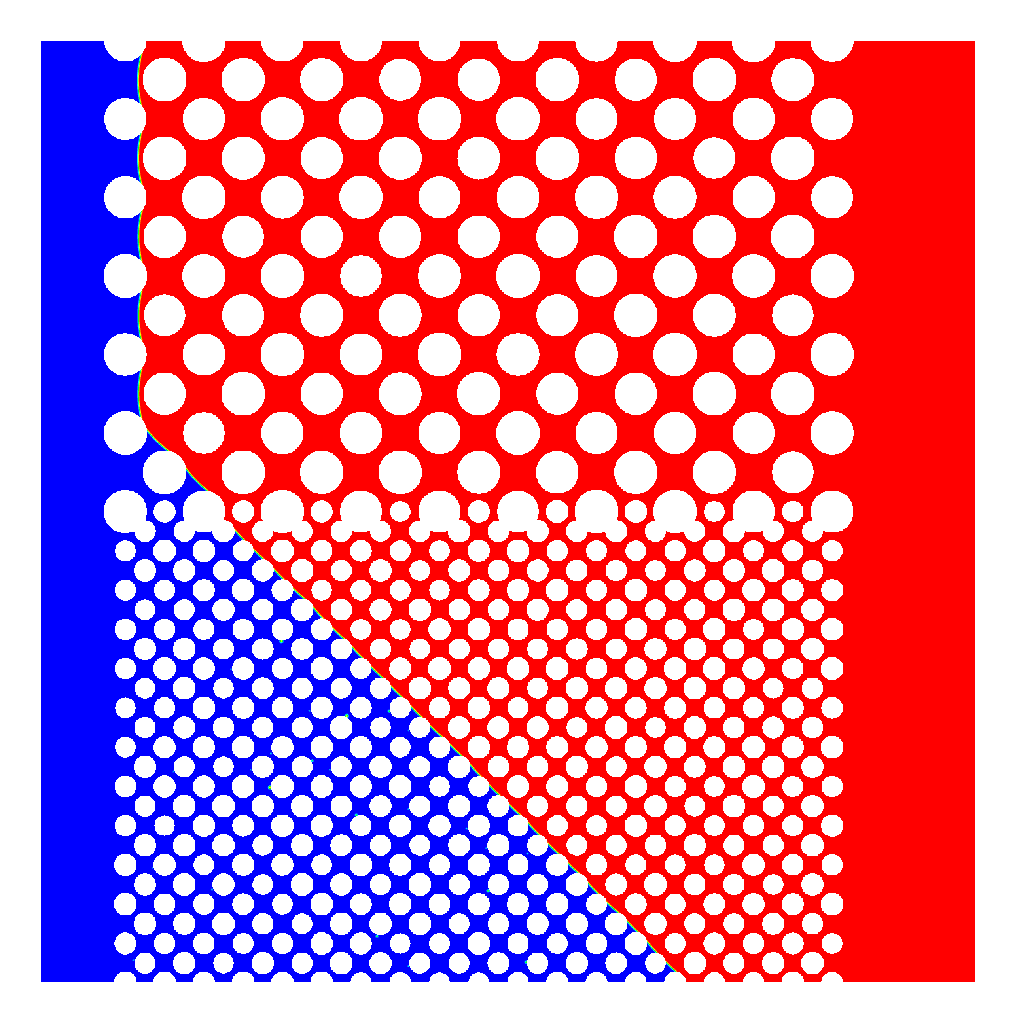}}~
    \subfloat[]{\includegraphics[width=0.33\textwidth]{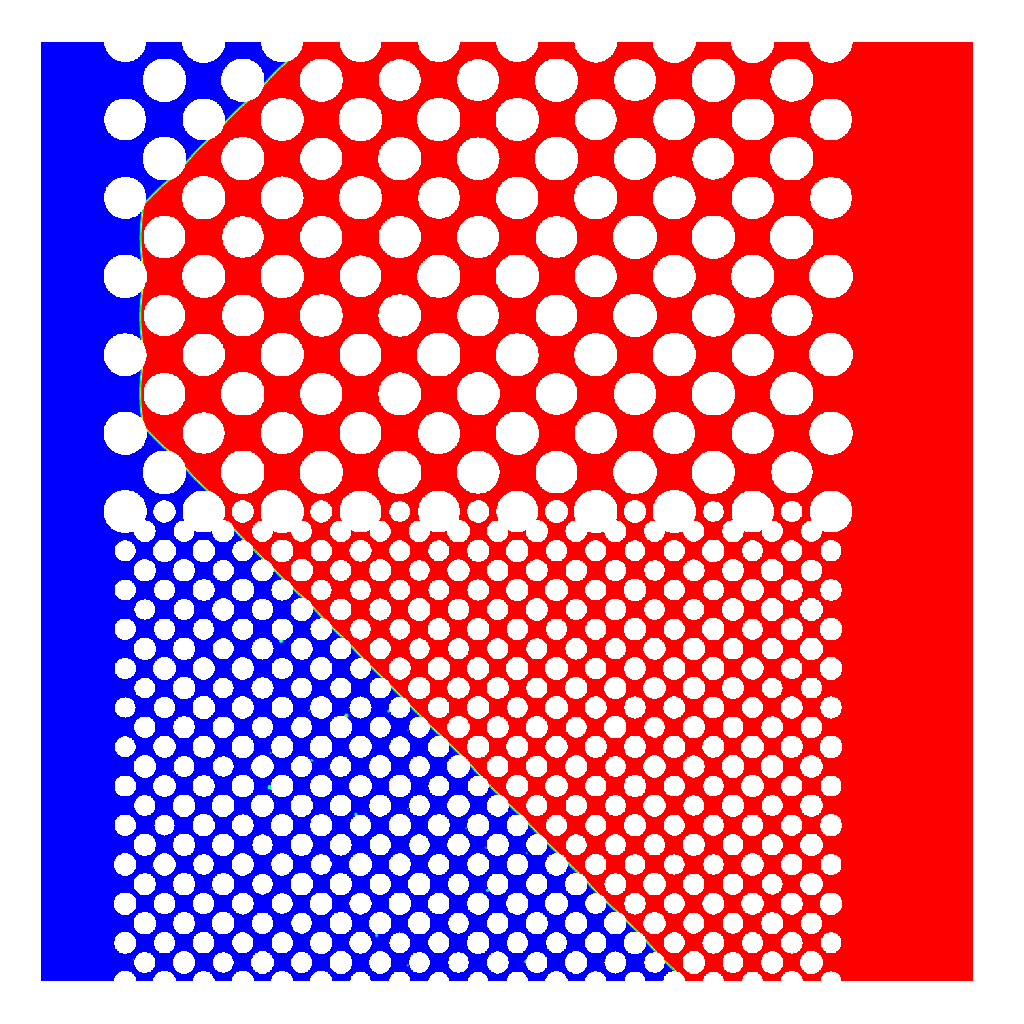}}~
    \subfloat[]{\includegraphics[width=0.33\textwidth]{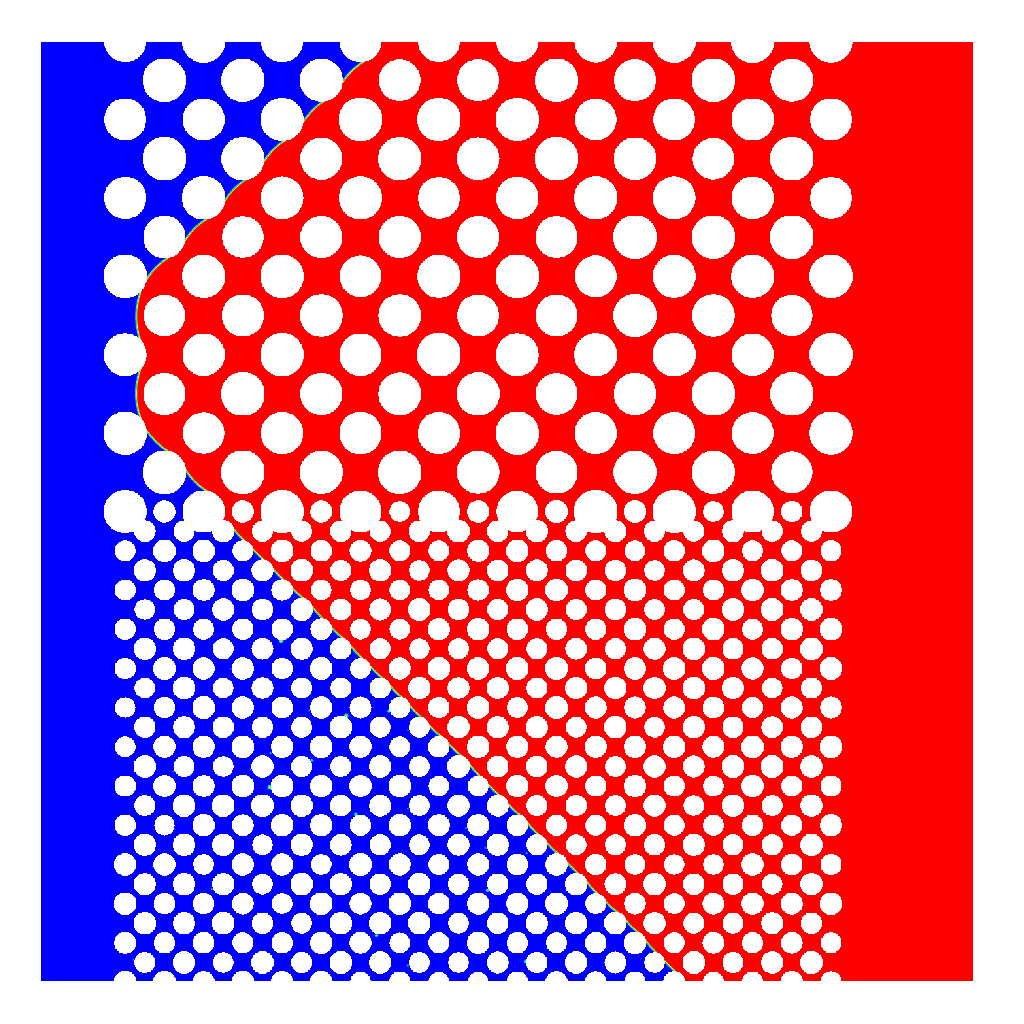}}\\
    \subfloat[]{\includegraphics[width=0.33\textwidth]{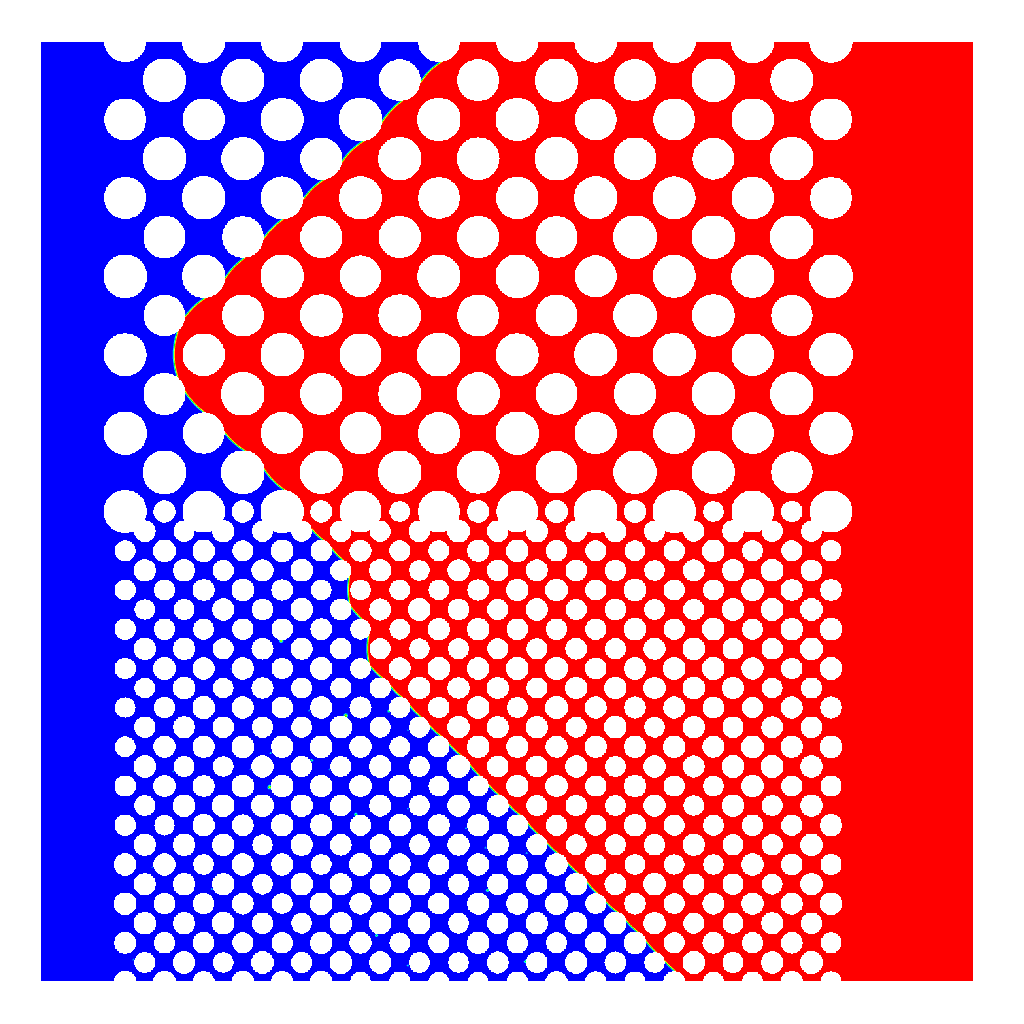}}~
    \subfloat[]{\includegraphics[width=0.33\textwidth]{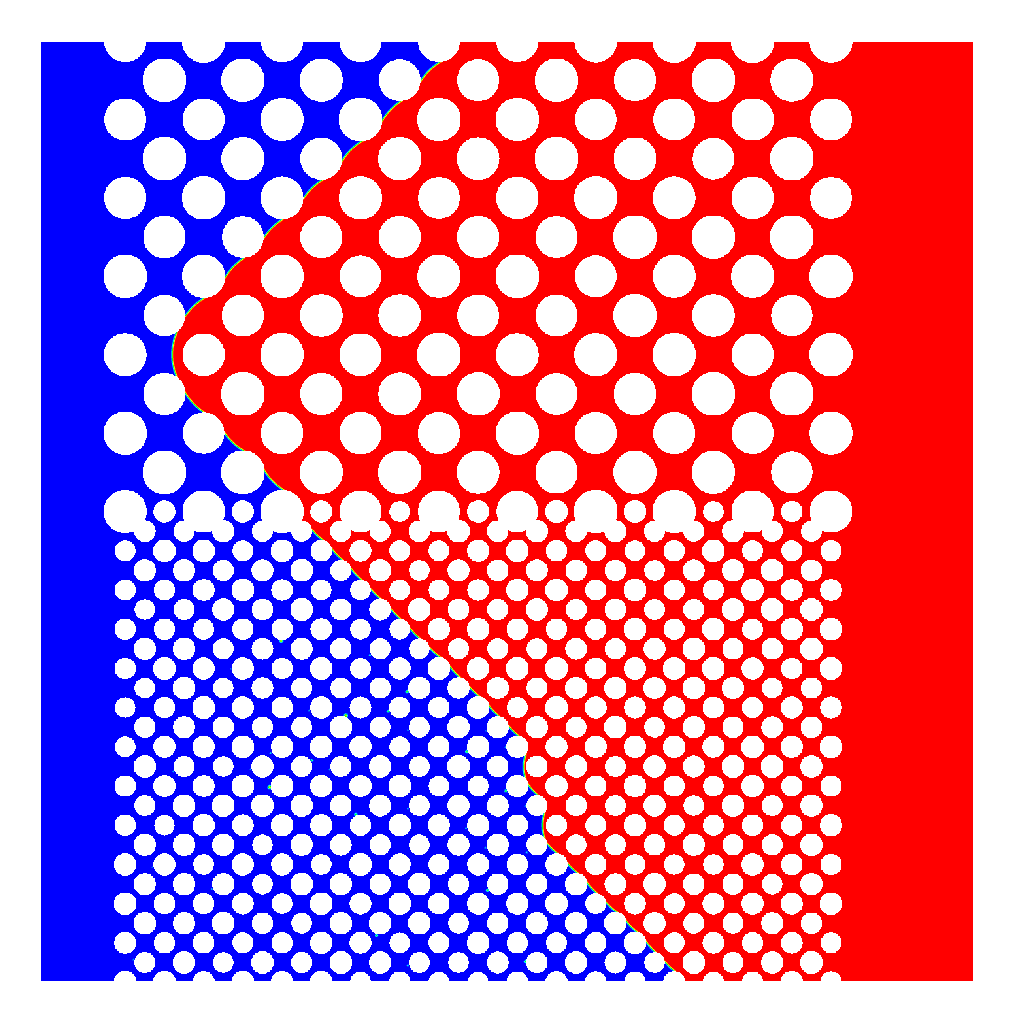}}~
    \subfloat[]{\includegraphics[width=0.33\textwidth]{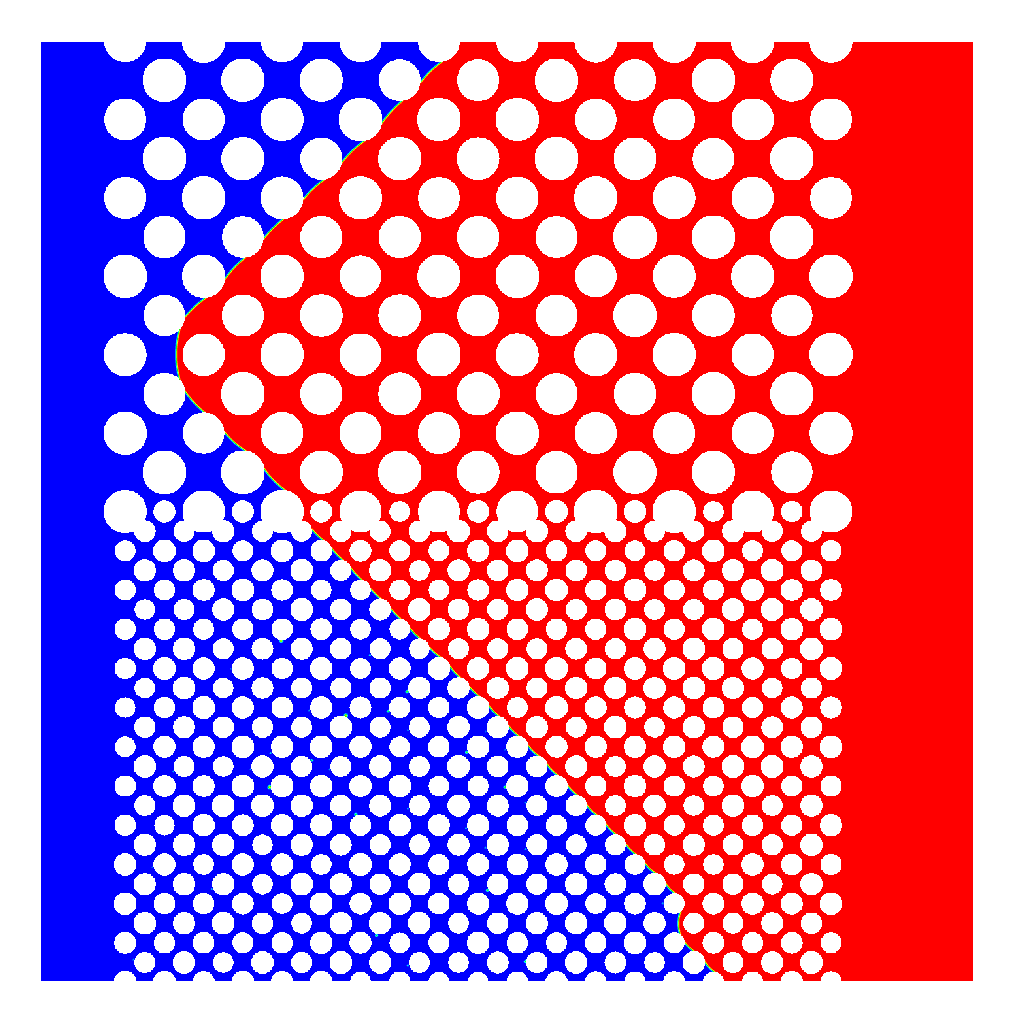}}\\
\caption{(Colour online) Snapshots of the imbibition for $Ca_m=0.1783$ and $\lambda=1$ at: (a) $\hat t/\hat{t}_B=0.5106$, (b) $\hat t/\hat{t}_B=0.5532$, (c) $\hat t/\hat{t}_B=0.5957$, (d) $\hat t/\hat{t}_B=0.7234$, (e) $\hat t/\hat{t}_B=0.7660$ and (f) $\hat t/\hat{t}_B=0.8085$. The snapshots from (a) to (f) correspond to the solid dots marked by A to F in figure~\ref{fig:Poredoublet-Dualper-lambda-0025-1-20-combine}(c).}
   \label{fig:snapshots-Ca3E5Vis1-lambda-1-fluid-distributions}
\end{figure}

To locate the values of $Ca_{m,c}$ for different viscosity ratios, we extract the data regarding the imbibition preference from table~\ref{tab:Sw-L-H-S-Ca-m-lambda} and plot them in the $\lambda- Ca_{m}$ diagram, as shown in figure~\ref{fig:new-phase-diagram-imbibition}. In this figure, the hollow symbols represent the cases where $S_1>S_2$, while the filled symbols represent the cases where $S_1<S_2$. This means that for each value of $\lambda$, the critical capillary number $Ca_{m,c}$ lies between two nearest hollow and filled symbols. As such, the $Ca_{m,c}$ curve can be approximately obtained, which are represented by the green solid lines. For the sake of comparison, figure~\ref{fig:new-phase-diagram-imbibition} also plots the $Ca_{m,c}$ curve from the pore doublet model (represented by the blue dashed lines and directly taken from figure~\ref{fig:poredoublet-cri-Ca-Tendency}). It is clear that the present $Ca_{m,c}$ curve overlaps well with the one from the pore doublet model. This suggests that the simplified pore doublet model can provide insights into the physics of immiscible displacement in the more complex dual-permeability pore network.

Although the pore doublet model can predict the variation of $Ca_{m,c}$ with $\lambda$ in a dual-permeability pore network, it is not clear whether the transient imbibition behaviour in the dual-permeability pore network can be correctly captured by the pore doublet model. In order to clarify this, we plot the time evolution of $S_1$ and $S_2$ (normalized by their maximum value at breakthrough) at three typical viscosity ratios in figure~\ref{fig:Poredoublet-Dualper-lambda-0025-1-20-combine}, where the semi-analytical solutions $L_1$ and $L_2$ (normalized by $\hat{L}$), obtained from (\ref{eq:coupled-ode-3}) and (\ref{eq:coupled-ode-L2t}) with the dimensionless numbers $Ca_m$ and $\lambda$ identical to those in the pore network, are also shown for comparison. For each viscosity ratio, the agreement between the LBM results and the semi-analytical solutions is generally better at higher $Ca_m$ where $S_1<S_2$, but worse when $S_1>S_2$ where the interfacial tension is dominant. The larger discrepancy when $S_1>S_2$ (see figure~\ref{fig:Poredoublet-Dualper-lambda-0025-1-20-combine}a,c and e) is attributed to the fact that in the dual-permeability pore network, the interface varies and thus the capillary pressure varies when the meniscus moves from the throat to the pore body or from the pore body to the  throat, while the capillary pressure remains a constant in the pore doublet. In addition, we interestingly notice in figure~\ref{fig:Poredoublet-Dualper-lambda-0025-1-20-combine}(c) that after $\hat t/t_B = 0.5$, the wetting fluid infiltrates into the high and low permeability zones alternately. Figure~\ref{fig:snapshots-Ca3E5Vis1-lambda-1-fluid-distributions} shows the corresponding snapshots, from which it is seen that the wetting fluid only invades into the high permeability zone in (a-c) but only into the low permeability zone in (d-f).

\section{Conclusions}
\label{conclusion}
We have studied the imbibition behaviour of two immiscible fluids in a dual-permeability pore network by a combination of pore-scale LBM simulation and mathematical modeling. First, we establish a mathematical model of the forced imbibition in a pore doublet, consisting of two branch channels with different widths, and find that the imbibition dynamics can be fully described by the viscosity ratio and the capillary number $Ca_m$, which additionally incorporates the influence of channel width and length. By solving the mathematical model, a phase diagram of $\lambda$ versus $Ca_m$ is proposed to characterise the imbibition preference in the pore doublet. Then, the colour-gradient LBM is used to simulate the imbibition process in the pore doublet and its capability and accuracy are validated against the semi-analytical solutions of mathematical model. Finally, the lattice Boltzmann simulations are used for the imbibition dynamics in a dual-permeability pore network. For each viscosity ratio, it is observed at breakthrough that, the imbibition is preferred to occur in low permeability zone at low values of $Ca_m$ but in high permeability zone at high values of $Ca_m$, which is attributed to the competition between capillary and viscous forces. When the capillary effects cannot be ignored, the wetting fluid is found to progress layer by layer in an oblique manner. In addition, for each viscosity ratio, there exists a critical capillary number $Ca_{m,c}$ at which the wetting fluid saturations are equal in both permeability zones, and $Ca_{m,c}$ is the optimal condition to improve the imbibition efficiency. By comparing the phase diagram obtained in the dual-permeability pore network with that from the pore doublet model, we demonstrate for the first time that, the pore doublet model can predict the variation of $Ca_{m,c}$ with the viscosity ratio in a dual-permeability pore network. Nevertheless, the pore doublet model cannot describe all features of the imbibition process in the dual-permeability pore network, especially when the imbibition is preferred to occur in low permeability zone. The present study not only facilitates fundamental understanding of the imbibition mechanism within the dual-permeability porous media, but also provides operational guidelines to improve the oil recovery in practice.

\section*{Acknowledgments}
This work is supported by the National Natural Science Foundation of China (No.51876170) and the Natural Science Basic Research Plan in Shaanxi Province of China (No. 2019JM-343).

\section*{Declaration of interests}
The authors declare no conflict of interest.

\bibliographystyle{jfm}
\bibliography{P_dualpermeability}

\end{document}